\documentclass[acmsmall, screen]{acmart}

\usepackage{graphicx} 
\usepackage{amsmath}  
\usepackage{amsthm}   
\usepackage{bbold}
\usepackage{bbm}
\usepackage{bussproofs}
\usepackage{lipsum}
\usepackage{mathtools}
\usepackage{cancel}
\usepackage{enumitem}
\usepackage{pifont}
\newcommand{\xmark}{\ding{55}}%
\newcommand{\cmark}{\ding{51}}%

\usepackage{epigraph}  
\makeatletter
\renewenvironment{proof}[1][\proofname]{%
  \par\pushQED{\qed}%
  \normalfont
  \topsep6\p@\@plus6\p@\relax
  \trivlist
  \item[\hskip\labelsep\sffamily \ul{#1}.]\ignorespaces
}{%
  \popQED\endtrivlist\@endpefalse
}
\makeatother

\usepackage{adjustbox}

\usepackage{quiver}
\usepackage{array}
\usepackage{tikz}
\usepackage{tikz-cd}
\usetikzlibrary{shadows}

\usepackage[most]{tcolorbox}
\usepackage[T1]{fontenc}
\usepackage[utf8]{inputenc}

\newcommand{\tul}[1]{%
  \leavevmode
  \sbox0{#1}
  \dimen0=\wd0
  {#1}%
  \llap{\rule[-3.8pt]{\dimen0}{.8pt}}
}

\tcbset{
    bthmbox/.style={
        enhanced,
        unbreakable,
        sharp corners=all,
        fonttitle=\bfseries\sffamily\normalsize,
        fontupper=\normalsize,
        top=0mm,
        bottom=0mm,
        right=0mm,
        borderline west={3.35mm}{0pt}{black},
        borderline west={3.0mm}{0.35mm}{bmidnight},
        colback=white,
        colframe=white,
        colbacktitle=white,
        coltitle=black,
        attach boxed title to top right,
        boxed title style={empty, size=minimal, bottom=1.5mm},
        overlay unbroken={
            \draw[line width=.8pt]
                (frame.north west) -- (frame.north east);
        },
        overlay first={
            \draw[line width=.8pt] (title.south west)--(title.south east); 
            \draw[line width=.8pt] ([xshift=3.5mm]frame.north west)--([xshift=3.5mm]frame.south west);
        },
        overlay middle={
            \draw[line width=.8pt] ([xshift=3.5mm]frame.north west)--([xshift=3.5mm]frame.south west);
        },
        overlay last={
        },
    },
}

\tcbset{
    blembox/.style={
        enhanced,
        unbreakable,
        sharp corners=all,
        fonttitle=\bfseries\sffamily\normalsize,
        fontupper=\normalsize,
        top=0mm,
        bottom=0mm,
        right=0mm,
        borderline west={3.35mm}{0pt}{black},
        borderline west={3.0mm}{0.35mm}{cayenne},
        colback=white,
        colframe=white,
        colbacktitle=white,
        coltitle=black,
        left skip=.2\textwidth,
        width=.8\textwidth,
        attach boxed title to top right,
        boxed title style={empty, size=minimal, bottom=1.5mm},
        overlay unbroken={
            \draw[line width=.8pt]
                (frame.north west) -- (frame.north east);
        },
        overlay first={
            \draw[line width=.8pt] (title.south west)--(title.south east); 
            \draw[line width=.8pt] ([xshift=3.5mm]frame.north west)--([xshift=3.5mm]frame.south west);
        },
        overlay middle={
            \draw[line width=.8pt] ([xshift=3.5mm]frame.north west)--([xshift=3.5mm]frame.south west);
        },
        overlay last={
        },
    },
}

\tcbset{
    rthmbox/.style={
        enhanced,
        unbreakable,
        sharp corners=all,
        fonttitle=\sffamily\normalsize,
        fontupper=\normalsize,
        top=0mm,
        bottom=0mm,
        right=0mm,
        borderline west={3.35mm}{0pt}{black},
        borderline west={3.0mm}{0.35mm}{cayenne},
        colback=white,
        colframe=white,
        colbacktitle=white,
        coltitle=black,
        attach boxed title to top left,
        boxed title style={empty, size=minimal, bottom=1.5mm},
        overlay unbroken ={
            \draw[line width=.8pt] (title.south west)--(title.south east);
        },
        overlay first={
            \draw[line width=.8pt] (title.south west)--(title.south east); 
            \draw[line width=.8pt] ([xshift=3.5mm]frame.north west)--([xshift=3.5mm]frame.south west);
        },
        overlay middle={
            \draw[line width=.8pt] ([xshift=3.5mm]frame.north west)--([xshift=3.5mm]frame.south west);
        },
        overlay last={
        },
    },
}

\newtcbtheorem{bthm}{\textbf{\tul{Theorem}}}{bthmbox}{thm}
\newtcbtheorem{blem}{\textbf{\tul{Lemma}}}{blembox}{lemm}
\newtcbtheorem{rlem}{\textbf{\tul{Lemma}}}{rthmbox}{lemm}
\newtcbtheorem{rthm}{Theorem}{rthmbox}{thm}
\newtcbtheorem{bex}{Example}{bthmbox}{ex}
\newtcbtheorem{rex}{Example}{rthmbox}{ex}
\newtcbtheorem{bdef}{Definition}{bthmbox}{ex}
\newtcbtheorem{rdef}{Definition}{rthmbox}{ex}

\usepackage{tabularray}
\usepackage{mdframed}
\usepackage{subcaption}
\usepackage{wrapfig}

\newcommand{\myhrulefill}{\leavevmode\leaders\hrule height .8pt\hfill\kern-.8pt}
\newcommand{\rulediv}[2]{\raisebox{.5ex}{\makebox[\linewidth]{\rule{\dimexpr50pt-#1\relax}{.8pt}\ \raisebox{-.412ex}{\makebox[2\dimexpr#1\relax]{\textsf{\small{#2}}}}\hspace{.7ex}\myhrulefill}}}

\usepackage{tcolorbox}
\tcbuselibrary{skins}

\definecolor{hscodebg}{rgb}{0.98,0.98,0.98}
\definecolor{cayenne}{rgb}{0.58,0.067,0}
\definecolor{midnight}{rgb}{0.008, 0.08, 0.47}
\definecolor{offwhite}{rgb}{0.98,0.98,0.98}

\usepackage[skip=2pt]{caption} 

\usepackage{preamble}
\newcommand{\Cj}{$\cj${\scriptsize $(\nb{\Bool},\nb{\lto})$}}

\begin{document}

\title{Compiling to Recurrent Neurons}

\author{Joey Velez-Ginorio}
\email{joeyv@seas.upenn.edu}
\orcid{0009-0004-6451-5107}
\affiliation{
  \institution{University of Pennsylvania}
  \country{USA}
}

\author{Nada Amin}
\email{namin@seas.harvard.edu}
\orcid{0000-0002-0830-7248}
\affiliation{
  \institution{Harvard University}
  \country{USA}
}

\author{Konrad Kording}
\email{kording@upenn.edu}
\orcid{0000-0001-8408-4499}
\affiliation{
  \institution{University of Pennsylvania}
  \country{USA}
}

\author{Steve Zdancewic}
\email{stevez@seas.upenn.edu}
\orcid{0000-0002-3516-1512}
\affiliation{
  \institution{University of Pennsylvania}
  \country{USA}
}


\newif\ifcomments\commentstrue   
\newif\ifaftersubmission \aftersubmissionfalse 
\newif\ifplentyofspace \plentyofspacefalse 
\ifcomments
  \newcommand{\proposecut}[1]{\ifcomments{\color{gray}\sout{#1}}\fi}
  \newcommand{\sz}[1]{\textcolor{brown}{{[SZ:~#1]}}}
  \newcommand{\kk}[1]{\textcolor{blue}{{[KK:~#1]}}}
  \newcommand{\na}[1]{\textcolor{orange}{{[NA:~#1]}}}
  \newcommand{\jvg}[1]{\textcolor{ForestGreen}{{[JVG:~#1]}}}
  
\else
  \newcommand{\proposecut}[1]{}
  \newcommand{\todo}[1]{}
\fi

\def\Cj{$\cj${\scriptsize $(\nb{\Bool},\nb{\lto})$}}

\setcopyright{cc}
\setcctype{by}
\acmJournal{PACMPL}
\acmYear{2026} \acmVolume{10} \acmNumber{PLDI} \acmArticle{267}
\acmMonth{6} \acmDOI{10.1145/3808345}

\begin{abstract}

Discrete structures are currently second-class in differentiable programming. Since functions over discrete structures lack overt derivatives, differentiable programs do not differentiate through them and limit where they can be used. For example, when programming a neural network, conditionals and iteration cannot be used everywhere; they can break the derivatives necessary for gradient-based learning to work. This limits the class of differentiable algorithms we can directly express, imposing restraints on how we build neural networks and differentiable programs more generally. However, these restraints are not fundamental.
Recent work shows conditionals can be first-class, by compiling them into differentiable form as linear neurons.
Similarly, this work shows iteration can be first-class---by compiling to linear \textit{recurrent} neurons.
We present a minimal typed, higher-order and linear programming language with iteration called $\cj{\scriptstyle(\nb{\lto},\, \nb{\Bool},\, \nb{\mathbb{N}})}$. We prove its programs compile correctly to recurrent neurons, allowing discrete algorithms to be expressed in a differentiable form compatible with gradient-based learning. With our implementation, we conduct two experiments where we link these recurrent neurons against a neural network solving an iterative image transformation task. This determines part of its function prior to learning. As a result, the network learns faster and with greater data-efficiency relative to a neural network programmed without first-class iteration.
A key lesson is that recurrent neurons enable a rich interplay between learning and the discrete structures of ordinary programming.
\end{abstract}


\maketitle

\section{Introduction}

\begin{wrapfigure}{h!}{0.33\textwidth}
    \centering
    \vspace{-4.5em}
    \includegraphics[width=.33\textwidth]{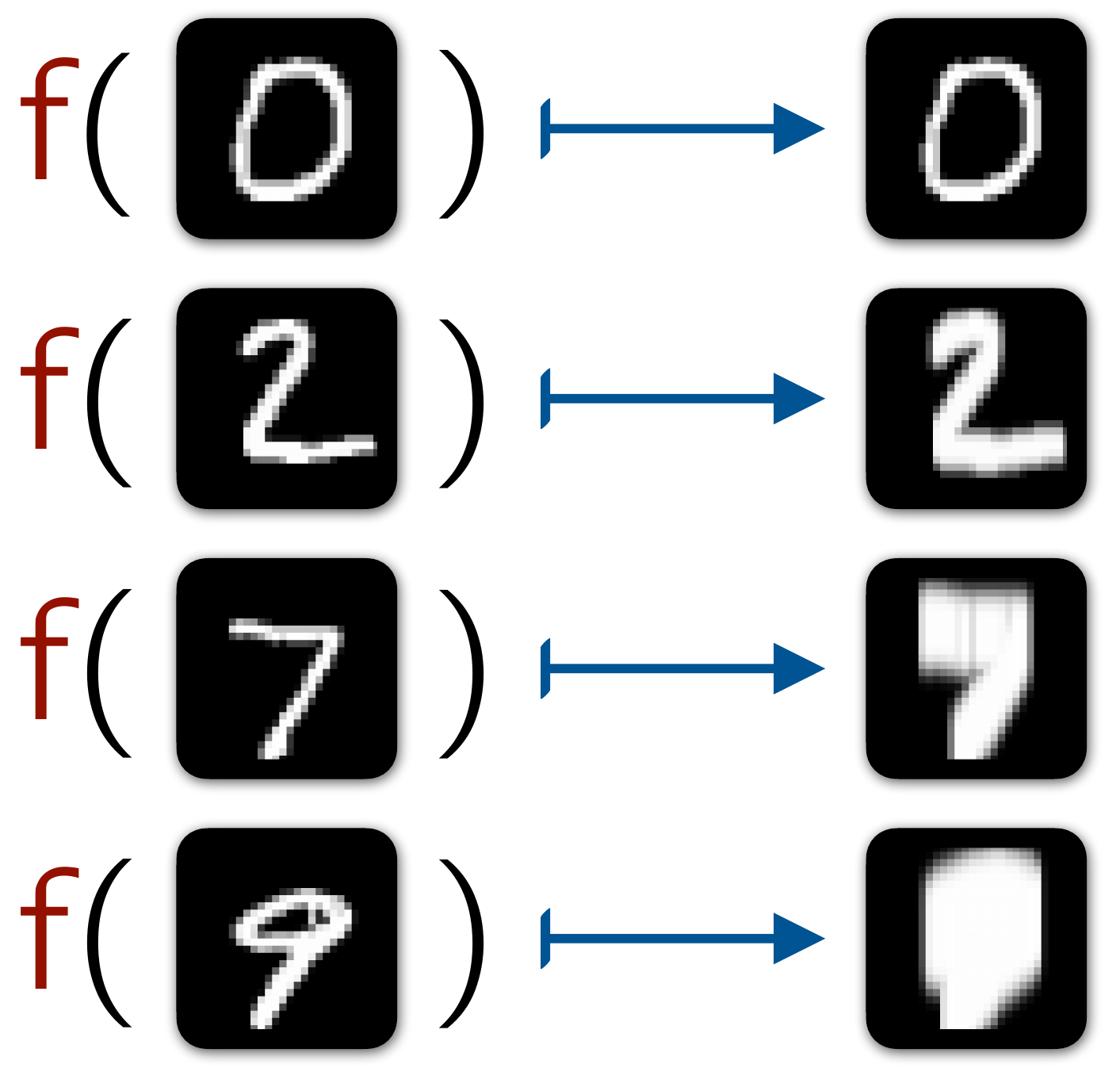}
    \vspace{-.5em}
  \caption{Iterative image transform}
  \label{fig:task}
\end{wrapfigure}

Functions over discrete structures lack overt derivatives. Nevertheless, they could be differentiated if compiled to differentiable form, enabling discrete structure to appear anywhere in a differentiable program. But it is not yet clear how to build these compilers. For conditionals, recent work shows it is possible by compiling programs to linear neurons\footnote{Linear neurons are essentially linear maps.} \cite{jvg26linear}. For iteration over natural numbers, we build on those efforts to show it is also possible---by compiling programs to linear recurrent neurons.\footnote{Linear recurrent neurons are essentially linear dynamical systems.} This allows us to directly express differentiable algorithms with discrete iterative structure. For example, consider the following image transformation in Fig. \ref{fig:task}.

\setlist[itemize]{topsep=8pt, itemsep=1pt, parsep=2pt, leftmargin=1.5em, rightmargin=1.5em}
\begin{itemize}[label=$\triangleright$]
\item If $\nb{x}$ is the digit $\nb{0}$, return $\nb{x}$
\item If $\nb{x}$ is the digit $\nb{n}$, apply vertical blur $\nb{n}$ times to $\nb{x}$
\end{itemize}

Recognizing whether a digit is $\nb{0}$ or some $\nb{n}$ is a classic differentiable algorithm that a neural network can learn \cite{lecun1989handwritten}. However, applying vertical blur $\nb{n}$ times is not overtly differentiable with respect to the natural number $\nb{n}$. Therefore it seems a neural network could not be programmed with this algorithm, as the derivatives necessary for gradient-based learning would be absent for $\nb{n}$. 

\begin{figure}[h]
  \centering
  \begin{minipage}[b]{0.5\textwidth}
      \centering
    \includegraphics[width=\linewidth]{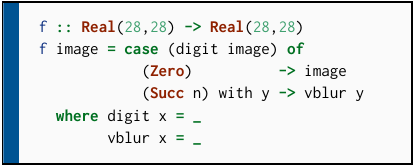}
    \subcaption{Programming an iterative image transform}
    \label{fig:program-iteration}
    \end{minipage}
  \hfill
  \begin{minipage}[b]{0.49\textwidth}
    \centering
    \includegraphics[width=.79\textwidth]{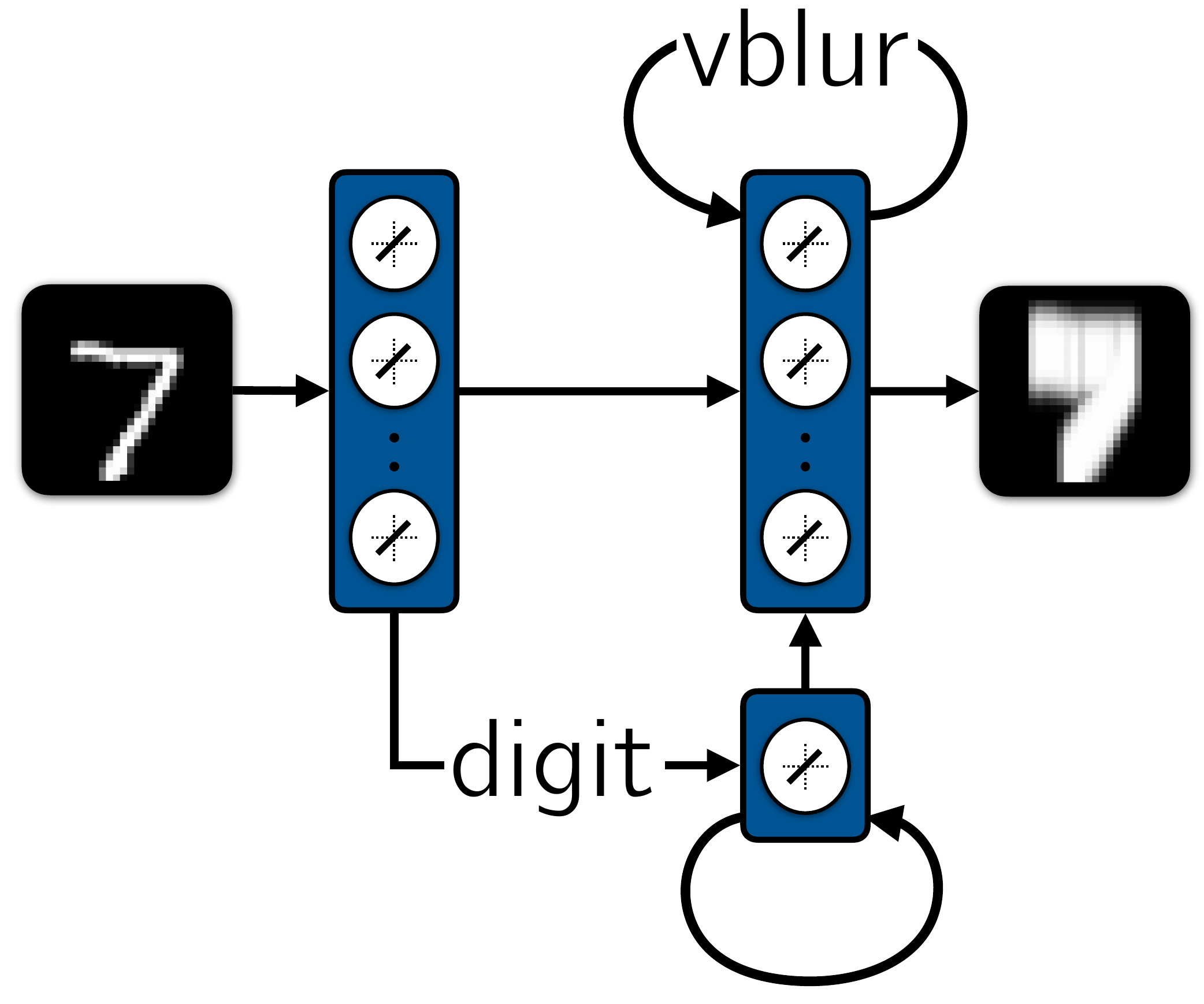} 
    \hfill
    \subcaption{Compiling iteration into differentiable form}
    \label{fig:compiling-iteration}

    \end{minipage}
  \vspace{.3em}
  \caption{Programming and compiling an iterative image transform}
  \label{fig:program-style}
\end{figure}

But it is possible. With $\cj{\scriptstyle(\nb{\lto},\, \nb{\Bool},\, \nb{\mathbb{N}})}$, a neural network can be programmed, as in Fig. \ref{fig:program-iteration}, to learn the iterative transformation. The program pattern matches on the output of {\small \texttt{(digit image)}}, a neural network which checks whether {\small \texttt{image}} is the digit $\nb{0}$ or some $\nb{n}$. If it is $\nb{0}$, the program returns the original {\small \texttt{image}}. Otherwise, it will use {\small \texttt{vblur}} to apply vertical blur $\nb{n}$ times to {\small \texttt{image}}. Both {\small \texttt{digit}} and {\small \texttt{vblur}} are left undefined; a neural network will learn them. Though for this to work, $\cj{\scriptstyle(\nb{\lto},\, \nb{\Bool},\, \nb{\mathbb{N}})}$ must compile iteration into differentiable form.

Prior work identifies linear programming languages as promising source languages for such compilers \cite{jvg26linear}. Their semantic connection to linear algebra opens a route towards differentiating discrete structure, by compiling linear programs to linear maps between $\rd{\R}$-vector spaces. However, while the case of conditionals is clear, the differentiable status of iteration is more subtle. In prior work, the compiled linear maps were between \textit{finite}-dimensional $\rd{\R}$-vector spaces, which are unambiguously differentiable \cite{spivak2018calculus}. For iteration, the compiled linear maps can be between \textit{infinite}-dimensional $\rd{\R}$-vector spaces---and without additional assumption, these linear maps may not be differentiable \cite{kreyszig1991introductory}.

Fortunately, an assumption is common when programming recurrent neurons: you differentiate with respect to a finite-dimensional \textit{subspace}. For example, we show that the program in Fig. \ref{fig:program-iteration} compiles to the recurrent neural network in Fig. \ref{fig:compiling-iteration}, which can be implemented as a differentiable program (Section \ref{sec:experiments}). The critical observation is that the recurrence \textit{if observed finitely}, renders a linear recurrent neuron equivalent to a linear map between finite-dimensional $\rd{\R}$-vector spaces. 

With $\cj{\scriptstyle(\nb{\lto},\, \nb{\Bool},\, \nb{\mathbb{N}})}$, our aim is to present these essential issues in plain terms, using a minimal calculus whose formal specification and metatheory are tractable and pleasing to explore. We are interested in an empirical \textit{and} mathematical method for giving discrete structure first-class status in differentiable programming. Specifically, our contributions are:

\setlist[itemize]{topsep=4pt, itemsep=2pt, parsep=2pt, leftmargin=1.5em, rightmargin=1.5em}
\begin{itemize}[label=$\triangleright$]
\item We specify the syntax, evaluation, and typing of $\cj{\scriptstyle(\nb{\lto},\, \nb{\Bool},\, \nb{\mathbb{N}})}$ programs (Section \ref{sec:defining}). 
\item We prove that $\cj{\scriptstyle(\nb{\lto},\, \nb{\Bool},\, \nb{\mathbb{N}})}$ programs compile correctly to recurrent neurons (Section \ref{sec:compiling}).
\item We implement $\cj{\scriptstyle(\nb{\lto},\, \nb{\Bool},\, \nb{\mathbb{N}})}$, programming a neural network to solve the iterative transform task (Fig. \ref{fig:task}) and a conditional variant using first-class conditionals and iteration (Section \ref{sec:experiments}).
\end{itemize}

These findings suggest that more complex forms of recurrence in neural networks could implement richer forms of differentiable primitive recursion---over lists, trees, and arbitrary algebraic data (Section \ref{sec:discussion}).

\section{Linear Recurrent Neurons}
\label{sec:lin-neuron}

Linear recurrent neurons are discrete-time linear dynamical systems \cite{antsaklis2006linear}. They transform vectors using linear maps, but are additionally parameterized by \textit{time}. For example, consider the following system whose input is a timestep $\nb{n} \in \nb{\N}$, and whose output is a vector $\rd{v} \in \rd{\R^2}$. 

$$
    \rd{f}(\nb{0}) = \rd{\begin{bmatrix}1\\0\end{bmatrix}}
    \quad\quad
    \rd{f}(\nb{n+1}) = \rd{\begin{bmatrix}0 & 2\\3 & 0\end{bmatrix}} \Bigl(\rd{f}(\nb{n})\Bigr) 
$$
\vspace{.5em}

\noindent We can imagine this system evolving over time by enumerating in sequence its \textit{state} at each $\nb{n}$.

$$
\Biggl(
\rd{\begin{bmatrix}1\\0\end{bmatrix}},
\rd{\begin{bmatrix}0\\3\end{bmatrix}},
\rd{\begin{bmatrix}6\\0\end{bmatrix}},
\rd{\begin{bmatrix}0\\18\end{bmatrix}},
\cdots
\Biggr)
$$
\vspace{0.1em}

\noindent It is a \textit{linear} system because each transition is determined by a matrix multiplication; the next state is a linear combination of its previous state. And it is a \textit{discrete-time} system because time is parameterized by the natural numbers.

Equivalently, we may view the system as a linear recurrent neural network with two recurrent neurons, shown in Fig. \ref{fig:folded-neuron}. The input is the initial state determined by $\rd{f}(\nb{0})$. The recurrent connections between the two neurons $\rd{h_0}$ and $\rd{h_1}$ are determined by the coefficients of the matrix in $\rd{f}(\nb{n+1})$. Its output at any timestep $\nb{n}$ is given by $\rd{y}=\rd{f}(\nb{n})$. Importantly, when unfolding the dynamics up to a particular timestep $\nb{n}$, as in Fig. \ref{fig:unfolded-neuron}, recurrent neurons become equivalent to an ordinary linear neural network. They are linear and differentiable maps in the initial state $\rd{f}(\nb{0})$.

\begin{figure}[h]
  \centering
  \begin{minipage}[b]{0.35\textwidth}
        \centering
        \begin{tikzcd}[
          every arrow/.append style={line width=.7pt},
        ]
          \rd{1} & 
          {\tikz[baseline=(x.base)] \node[drop shadow, fill=white, circle, draw, line width=1.pt, minimum size=.8cm, inner sep=0pt] (x) {$\rd{h_0}$};} & 
          \rd{y_0} \\
          \rd{0} & 
          {\tikz[baseline=(x.base)] \node[drop shadow, fill=white, circle, draw, line width=1.pt, minimum size=.8cm, inner sep=0pt] (x) {$\rd{h_1}$};} &
          \rd{y_1} \\
          \arrow[from=1-1, to=1-2]
          \arrow["\rd{0}"{description, font=\normalsize}, dotted, tail reversed, no head, from=1-2, to=1-2, loop, in=55, out=125, distance=10mm]
          \arrow[from=1-2, to=1-3]
          \arrow["\rd{3}"{description, font=\normalsize}, curve={height=30pt}, from=1-2, to=2-2]
          \arrow["\rd{2}"{description, font=\normalsize}, curve={height=30pt}, from=2-2, to=1-2]
          \arrow[from=2-1, to=2-2]
          \arrow["\rd{0}"{description, font=\normalsize}, dotted, from=2-2, to=2-2, loop, in=305, out=235, distance=10mm]
          \arrow[from=2-2, to=2-3]
        \end{tikzcd}
        \subcaption{Linear recurrent neurons}
        \label{fig:folded-neuron}
    \end{minipage}
  \begin{minipage}[b]{0.62\textwidth}
        \centering
        \begin{tikzcd}[
          every arrow/.append style={line width=.7pt},
        ]
        	\rd{1} & 
            {\tikz[baseline=(x.base)] \node[drop shadow, fill=white, circle, draw, line width=1.pt, minimum size=.8cm, inner sep=0pt] (x) {$\rd{1}$};} & 
            {\tikz[baseline=(x.base)] \node[drop shadow, fill=white, circle, draw, line width=1.pt, minimum size=.8cm, inner sep=0pt] (x) {$\rd{0}$};} & 
            {\tikz[baseline=(x.base)] \node[drop shadow, fill=white, circle, draw, line width=1.pt, minimum size=.8cm, inner sep=0pt] (x) {$\rd{6}$};} & 
            \rd{6}
            \\
        	\rd{0} & 
            {\tikz[baseline=(x.base)] \node[drop shadow, fill=white, circle, draw, line width=1.pt, minimum size=.8cm, inner sep=0pt] (x) {$\rd{0}$};} & 
            {\tikz[baseline=(x.base)] \node[drop shadow, fill=white, circle, draw, line width=1.pt, minimum size=.8cm, inner sep=0pt] (x) {$\rd{3}$};} & 
            {\tikz[baseline=(x.base)] \node[drop shadow, fill=white, circle, draw, line width=1.pt, minimum size=.8cm, inner sep=0pt] (x) {$\rd{0}$};} & 
            \rd{0}
            \\
        	\arrow[from=1-1, to=1-2]
        	\arrow[dotted, from=1-2, to=1-3]
        	\arrow["\rd{3}"{description, pos=0.15, font=\normalsize}, shorten <=-6pt, shorten >=-6pt, from=1-2, to=2-3]
        	\arrow[dotted, from=1-3, to=1-4]
        	\arrow["\rd{3}"{description, pos=0.15, font=\normalsize}, shorten <=-6pt, shorten >=-6pt, from=1-3, to=2-4]
        	\arrow[from=1-4, to=1-5]
        	\arrow[from=2-1, to=2-2]
        	\arrow["\rd{2}"{description, pos=0.15, font=\normalsize}, shorten <=-6pt, shorten >=-6pt, from=2-2, to=1-3]
        	\arrow[dotted, from=2-2, to=2-3]
        	\arrow["\rd{2}"{description, pos=0.15, font=\normalsize}, shorten <=-6pt, shorten >=-6pt, from=2-3, to=1-4]
        	\arrow[dotted, from=2-3, to=2-4]
        	\arrow[from=2-4, to=2-5]
        \end{tikzcd}

        \subcaption{Unfolding linear recurrent neurons for $\nb{n}=\nb{2}$}
        \label{fig:unfolded-neuron}
    \end{minipage}
  \vspace{1.3em}
  \caption{Linear recurrent neurons and their dynamics}
  \label{fig:rec-neurons}
\end{figure}

However, observe that $\rd{f}$ is not overtly differentiable in $\nb{n}$. The derivative should describe how small changes to $\nb{n}$ impact the behavior of $\rd{f}$. But natural numbers, as in other discrete structure, lack an adequate notion of \textit{small changes}. As a result, there is no overt derivative with respect to $\nb{n}$. Yet there is recourse. Consider the following function $\rd{g}$ whose input is an infinite sequence of real numbers $\rd{x_i}$ with finite support, and whose output is a vector $\rd{v} \in \rd{\R^2}$.

$$
\rd{g}(\rd{x}) = \rd{\sum_{\bl{\nb{n} \in \nb{\N}}}} \;\rd{x}_\nb{n} \;\rd{\.}\; \rd{\begin{bmatrix}0 & 2\\3 & 0\end{bmatrix}^\nb{n}} \Bigl(\rd{\begin{bmatrix}1\\0\end{bmatrix}}\Bigr)
$$
\vspace{.1em}

First, observe that $\rd{g}$ is linear in $\rd{x}$.

$$
\;\;\rd{3} \;\rd{\cdot}\; \rd{g}(\rd{\begin{bmatrix}1 \!\!&\!\! 2 \!\!&\!\! 0 \!\!&\!\! \cdots\end{bmatrix}}) \;\;\rd{+}\;\; \rd{4} \;\rd{\cdot}\; \rd{g}(\rd{\begin{bmatrix}0 \!\!&\!\! 3 \!\!&\!\! 0 \!\!&\!\! \cdots\end{bmatrix}})
= \rd{g}(\rd{3 \cdot \begin{bmatrix}1 \!\!&\!\! 2 \!\!&\!\! 0 \!\!&\!\! \cdots\end{bmatrix} \;+\; 4 \cdot \begin{bmatrix}0 \!\!&\!\! 3 \!\!&\!\! 0 \!\!&\!\! \cdots\end{bmatrix}})
$$
\vspace{.01em}

\noindent Next, if we restrict $\rd{x}$ to be a \textit{finite}\footnote{The finiteness assumption is necessary for the notion of differentiation to be canonical.} sequence, then $\rd{g}$ is also differentiable in $\rd{x}$. For example, consider the restriction of $\rd{g}$ to sequences of length $\rd{3}$.

$$
\rd{g}_{\mid\rd{3}}(\rd{\begin{bmatrix}7 \!\!&\!\! 8 \!\!&\!\! 9\end{bmatrix}}) = \rd{7} \;\rd{\cdot}\;\rd{\begin{bmatrix}0 & 2\\3 & 0\end{bmatrix}^\nb{0}} \Bigl(\rd{\begin{bmatrix}1\\0\end{bmatrix}}\Bigr) 
\;\rd{+}\; 
 \rd{8} \;\rd{\cdot}\;\rd{\begin{bmatrix}0 & 2\\3 & 0\end{bmatrix}^\nb{1}} \Bigl(\rd{\begin{bmatrix}1\\0\end{bmatrix}}\Bigr)
 \;\rd{+}\; 
  \rd{9} \;\rd{\cdot}\;\rd{\begin{bmatrix}0 & 2\\3 & 0\end{bmatrix}^\nb{2}} \Bigl(\rd{\begin{bmatrix}1\\0\end{bmatrix}}\Bigr)
  = \rd{\begin{bmatrix}61\\24\end{bmatrix}}
$$
\vspace{.5em}

\noindent From $\rd{g}$ we obtain a linear map $\rd{g}_{\mid\rd{3}}:\rd{\R^3 \to \R^2}$. The meaning of a small change to its input is standard. From the above example, you can imagine how these changes would impact its behavior. But the most remarkable property of $\rd{g}_{\mid\rd{3}}$ is that, with appropriate choice of $\rd{x}$, it \textit{exactly} implements the system $\rd{f}$ up to its first three states as a linear and differentiable map.

$$
\begin{aligned}
\rd{g}_{\mid\rd{3}}(\rd{\begin{bmatrix}1 \!\!&\!\! 0 \!\!&\!\! 0\end{bmatrix}}) &= \rd{1} \;\rd{\cdot}\;\rd{\begin{bmatrix}0 & 2\\3 & 0\end{bmatrix}^\nb{0}} \Bigl(\rd{\begin{bmatrix}1\\0\end{bmatrix}}\Bigr) 
\;\rd{+}\; 
 \rd{0} \;\rd{\cdot}\;\rd{\begin{bmatrix}0 & 2\\3 & 0\end{bmatrix}^\nb{1}} \Bigl(\rd{\begin{bmatrix}1\\0\end{bmatrix}}\Bigr)
 \;\rd{+}\; 
  \rd{0} \;\rd{\cdot}\;\rd{\begin{bmatrix}0 & 2\\3 & 0\end{bmatrix}^\nb{2}} \Bigl(\rd{\begin{bmatrix}1\\0\end{bmatrix}}\Bigr)
  = \rd{\begin{bmatrix}1\\0\end{bmatrix}} = \rd{f}(\nb{0})
\\
\rd{g}_{\mid\rd{3}}(\rd{\begin{bmatrix}0 \!\!&\!\! 1 \!\!&\!\! 0\end{bmatrix}}) &= \rd{0} \;\rd{\cdot}\;\rd{\begin{bmatrix}0 & 2\\3 & 0\end{bmatrix}^\nb{0}} \Bigl(\rd{\begin{bmatrix}1\\0\end{bmatrix}}\Bigr) 
\;\rd{+}\; 
 \rd{1} \;\rd{\cdot}\;\rd{\begin{bmatrix}0 & 2\\3 & 0\end{bmatrix}^\nb{1}} \Bigl(\rd{\begin{bmatrix}1\\0\end{bmatrix}}\Bigr)
 \;\rd{+}\; 
  \rd{0} \;\rd{\cdot}\;\rd{\begin{bmatrix}0 & 2\\3 & 0\end{bmatrix}^\nb{2}} \Bigl(\rd{\begin{bmatrix}1\\0\end{bmatrix}}\Bigr)
  = \rd{\begin{bmatrix}0\\3\end{bmatrix}} = \rd{f}(\nb{1})
\\
\rd{g}_{\mid\rd{3}}(\rd{\begin{bmatrix}0 \!\!&\!\! 0 \!\!&\!\! 1\end{bmatrix}}) &= \rd{0} \;\rd{\cdot}\;\rd{\begin{bmatrix}0 & 2\\3 & 0\end{bmatrix}^\nb{0}} \Bigl(\rd{\begin{bmatrix}1\\0\end{bmatrix}}\Bigr) 
\;\rd{+}\; 
 \rd{0} \;\rd{\cdot}\;\rd{\begin{bmatrix}0 & 2\\3 & 0\end{bmatrix}^\nb{1}} \Bigl(\rd{\begin{bmatrix}1\\0\end{bmatrix}}\Bigr)
 \;\rd{+}\; 
  \rd{1} \;\rd{\cdot}\;\rd{\begin{bmatrix}0 & 2\\3 & 0\end{bmatrix}^\nb{2}} \Bigl(\rd{\begin{bmatrix}1\\0\end{bmatrix}}\Bigr)
  = \rd{\begin{bmatrix}6\\0\end{bmatrix}} = \rd{f}(\nb{2})
\end{aligned}
$$
\vspace{.8em}

This is the essential theory underlying how we compile iteration in $\cj{\scriptstyle(\nb{\lto},\, \nb{\Bool},\, \nb{\mathbb{N}})}$. In fact, $\rd{g}_{\mid\rd{3}}$ implements (modulo rescaling) the logical operator $\nb{\texttt{not}}$ iterated over $\nb{\texttt{true}}$ (Section \ref{sec:cajal}).


\section{Linear Programs}
\label{sec:lin-prog}
$\cj{\scriptstyle(\nb{\lto},\, \nb{\Bool},\, \nb{\mathbb{N}})}$ is a linearly typed programming language. Linear types restrict how variables are used in a program \cite{walker2005substructural}. They carefully permit the duplication of variables and forbid the discarding of variables. Curiously, these restrictions are the basis for their connection to linear algebra \cite{mellies2009categorical}. It is what links \textit{linear} programs and \textit{linear} maps.

\vspace{.3em}
\begin{figure}[h]
  \centering
  \begin{minipage}[b]{0.42\textwidth}
    \centering
    \includegraphics[width=\linewidth]{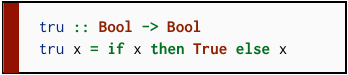}
    \vspace{-.5em}
    \subcaption{Programming nonlinearly}
    \label{fig:nonlinear-dup}
  \end{minipage}
  \hfill
  \begin{minipage}[b]{0.45\textwidth}
      \centering
    \includegraphics[width=\linewidth]{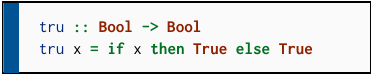}
      \vspace{-0.5em}
      \subcaption{Programming linearly}
      \label{fig:linear-dup}
    \end{minipage}
  \vspace{1.0em}
  \caption{Programming with and without duplicating/discarding \texttt{x}}
  \label{fig:prog-duplication}
\end{figure}

Consider the linear and nonlinear version of \nb{\texttt{tru}} in Fig. \ref{fig:prog-duplication}. Both implement a constant function which always return true. However, observe that the nonlinear version in Fig. \ref{fig:nonlinear-dup} duplicates \texttt{x} in the else branch of the conditional. If applied to an argument, $\nb{\texttt{tru}}$ uses \texttt{x} when checking the condition \textit{and} to return a value in the else branch. But the linear version in Fig. \ref{fig:linear-dup} only uses \texttt{x} once when checking the condition. Additionally, observe that the nonlinear version in Fig. \ref{fig:nonlinear-dup} does not use \texttt{x}, which is in scope, when returning true in the if branch. But the linear version in Fig. \ref{fig:linear-dup} uses everything in scope, as \texttt{x} is not in scope in either branch. 
 
\begin{figure}[h]
  \centering
  \begin{minipage}[b]{0.45\textwidth}
    \centering

        \scalebox{.8}{
        \def\extraVskip{4pt}
        \def\defaultHypSeparation{\hskip .35in}
        \bottomAlignProof
        \AxiomC{}
        \UnaryInfC{$\p{\textcolor{red}{\bind{x}{\Bool}}} {\bl{x}} {\bl{\Bool}}$}
        \AxiomC{}
        \UnaryInfC{$\p{\textcolor{red}{\bind{x}{\Bool}}} {\bl{\t{tt}}} {\bl{\Bool}}$}
        \AxiomC{}
        \UnaryInfC{$\p{\textcolor{red}{\bind{x}{\Bool}}} {\bl{x}} {\bl{\Bool}}$}
        \LeftLabel{\textcolor{red}{\huge \xmark}}
        \TrinaryInfC{$\p{\textcolor{red}{\bind{x}{\Bool}}} {\bl{\ite{\ul{x}}{\ul{\t{tt}}}{\ul{x}}}} {\bl{\Bool}}$}
        \UnaryInfC{$\p{\bl{\emp}} {\bl{\lam{x}{\ite{\ul{x}}{\ul{\t{tt}}}{\ul{x}}}}} {\bl{\Bool \to \Bool}}$}
        \DisplayProof}
    \vspace{.5em}
    \subcaption{Typing nonlinearly}
    \label{fig:typ-nonlinear-dup}
  \end{minipage}
  \hfill
  \begin{minipage}[b]{0.45\textwidth}
      \centering

        \scalebox{.8}{
        \def\extraVskip{4pt}
        \def\defaultHypSeparation{\hskip .35in}
        \bottomAlignProof
        \AxiomC{}
        \UnaryInfC{$\p{\textcolor{ForestGreen}{\bind{x}{\Bool}}} {\bl{x}} {\bl{\Bool}}$}
        \AxiomC{}
        \UnaryInfC{$\p{\bl{\emp}} {\bl{\t{tt}}} {\bl{\Bool}}$}
        \AxiomC{}
        \UnaryInfC{$\p{\bl{\emp}} {\bl{\t{tt}}} {\bl{\Bool}}$}
        \LeftLabel{\textcolor{ForestGreen}{\huge \cmark}}
        \TrinaryInfC{$\p{\textcolor{ForestGreen}{\bind{x}{\Bool}}} {\bl{\ite{\ul{x}}{\ul{\t{tt}}}{\ul{\t{tt}}}}}{\bl{\Bool}}$}
        \UnaryInfC{$\p{\bl{\emp}} {\bl{\lam{x}{\ite{\ul{x}}{\ul{\t{tt}}}{\ul{\t{tt}}}}}} {\bl{\Bool \lto \Bool}}$}
        \DisplayProof}

      \vspace{.5em}
      \subcaption{Typing linearly}
      \label{fig:typ-linear-dup}
    \end{minipage}
  \vspace{1.0em}
  \caption{Typing with and without duplicating/discarding \texttt{x}}
  \label{fig:typ-duplication}
\end{figure}

Typing can detect these distinctions. The nonlinear typing in Fig. \ref{fig:typ-nonlinear-dup} duplicates $\textcolor{red}{\bind{x}{\Bool}}$ when typing the conditional; it duplicates its context across each subprogram. But the linear typing in Fig. \ref{fig:typ-linear-dup} does not duplicate $\textcolor{ForestGreen}{\bind{x}{\Bool}}$ when typing the conditional. Its context is \textit{split} across each subprogram. Additionally, the nonlinear typing in Fig. \ref{fig:typ-nonlinear-dup} discards the variable $\textcolor{red}{\bind{x}{\Bool}}$ when typing $\t{tt}$ in the then branch of the conditional. However, by splitting the context, the linear typing in Fig. \ref{fig:typ-linear-dup} does not discard anything.

Generally, linear typing must ensure the structural rules of \textit{contraction} and \textit{weakening} are not permissible \cite{walker2005substructural}. Their impermissibility establish an important link between linear programs and linear maps, a connection we can use to differentiate through discrete structure.

\newcommand{\ttvm}{\rd{\begin{bmatrix} 1 \\ 0\end{bmatrix}}}
\newcommand{\ffvm}{\rd{\begin{bmatrix} 0 \\ 1\end{bmatrix}}}
\newcommand{\bvm}[2]{\rd{\begin{bmatrix} #1 \\ #2\end{bmatrix}}}

\section{\texorpdfstring{Cajal{\small$(\nb{\lto},\nb{\Bool},\nb{\N})$}}{Cajal}}
\label{sec:cajal}

\begin{figure}[h]
\[\begin{tikzcd}
	&&&& {\text{\large \nb{$\iter{\tt}{x}{\t{not}\,x}{2}$}}} \\
	{\text{\large \nb{$\mathsf{ff}$}}} &&&&&&&& {\text{\large \nb{$\mathsf{tt}$}}} \\
	\\
	&&&& \begin{array}{c} 
    \rd{\displaystyle\sum\limits_{\bl{\nb{n} \in \nb{\N}}}} \;\;\rd{\begin{bmatrix}0 \\0 \\ 1 \\ 0 \\ \vdots\end{bmatrix}_\nb{n}} \rd{\.}\;\;\rd{\begin{bmatrix}0 \!\!\!&\!\!\! 1\\1 \!\!\!&\!\!\! 0\end{bmatrix}^\nb{n}} \Bigl(\rd{\begin{bmatrix}1\\0\end{bmatrix}}\Bigr)
    \end{array} \\
	\begin{array}{c} \rd{\begin{bmatrix}0\\1\end{bmatrix}} \end{array} &&&&&&&& \begin{array}{c} \rd{\begin{bmatrix}1\\0\end{bmatrix}} \end{array}
	\arrow["{\scalebox{2.0}{\nb{$\not\bstep$}}}"{description}, curve={height=10pt}, shorten >=2pt, dotted, from=1-5, to=2-1]
	\arrow["{\scalebox{2.0}{\nb{$\bstep$}}}"{description},shorten >=2pt, curve={height=-10pt}, from=1-5, to=2-9]
	\arrow["{\scalebox{1.4}{{$\c{\nb{-}}$}}}"{description}, Rightarrow, from=1-5, to=4-5]
	\arrow["{\scalebox{1.4}{{$\c{\nb{-}}$}}}"{description}, Rightarrow, from=2-1, to=5-1]
	\arrow["{\scalebox{1.4}{{$\c{\nb{-}}$}}}"{description}, Rightarrow, from=2-9, to=5-9]
	\arrow["{\scalebox{2.0}{\rd{$\neq$}}}"{description}, shorten >=-5pt, shorten <= -5pt, curve={height=10pt}, dotted, from=4-5, to=5-1]
	\arrow["{\scalebox{2.0}{\rd{$=$}}}"{description}, shorten >=-5pt, shorten <= -5pt,curve={height=-10pt}, from=4-5, to=5-9]
\end{tikzcd}\]

\caption{Correctly compiling to recurrent neurons}
\label{fig:ccc1}
\end{figure}

To enable first-class iteration in differentiable programming, a \textit{correct} compiler should establish a correspondence between evaluation $\nb{\bstep}$ of $\cj{\scriptstyle(\nb{\lto},\, \nb{\Bool},\, \nb{\mathbb{N}})}$ programs and equations underlying the dynamics of linear recurrent neurons. Fig. \ref{fig:ccc1} details this notion of correctness, equivalent to establishing the \textit{soundness} and \textit{adequacy} of a denotational semantics \cite{winskel1993formal}. Establishing this correspondence is the central focus of this section. It sets the stage for differentiable programming with first-class iteration, where we program neural networks with discrete iterative structure in a differentiable way (Section \ref{sec:experiments}).

\subsection{Defining}
\label{sec:defining}

\subsubsection{Syntax} $\cj{\scriptstyle(\nb{\lto},\, \nb{\Bool},\, \nb{\mathbb{N}})}$ is a minimal typed, higher-order and linear programming language. Fig. \ref{fig:syntax} details its syntax, similar to Gödel's \textit{System T} of higher-order primitive recursion \cite{harper2016practical}. The main differences are the use of iterators (rather than recursors) for natural numbers and inclusion of a boolean type. These distinctions are inessential to the underlying theory, but simplify the presentation of certain ideas. When convenient, instead of explicit successors, we use an ordinary syntax for natural numbers. For example, using $\nb{2}$ instead of $\suc{\suc{\z}}$.

\begin{figure}[h]
  \begin{mdframed}[topline=false,innertopmargin=-1.07ex,innerleftmargin=-0.1ex,innerrightmargin=0ex,linewidth=.8pt,skipbelow=2ex]
    \rulediv{18pt}{\large \textsf{Syntax}}
    \vspace{.1em}
    \[  
    \begin{tabular}{c c}
        \scalebox{.8}{
        $\begin{aligned}
          &\;\;\;\;\;\;\;\;\;\nb{e} \in \nb{\t{Expression}}\\[3mm]
          \nb{e} \,\,\,\coloneqq 
	       \ & \ \ \ \nb{x} \in \t{\nb{Var}} &&\t{variable}\\
            	 &\mid \tt &&\t{true}\\
            	 &\mid \ff &&\t{false}\\
            	 &\mid \z &&\t{zero}\\
            	 &\mid \suc{e} &&\t{successor}\\
            	 &\mid \nb{\iter{e_1}{y}{e_2}{e_3}} &&\t{iterator}\\
            	 &\mid \nb{\lam{x}{e}} &&\t{linear map}\\
            	 &\mid \nb{e_1e_2}&&\t{linear application}\\
        \end{aligned}$}
    &
        \scalebox{.8}{
        $\begin{aligned}
          &\;\;\;\;\;\;\;\;\;\nb{\tau} \in \nb{\t{Type}}\\[3mm]
          \nb{\tau} \,\,\,\coloneqq
	       &\ \ \ \nb{\mathbb{2}} &&\t{boolean}\\
        	 &\mid \nb{\N} &&\t{natural number}\\
        	 &\mid \nb{\tau_1 \lto \tau_2} &&\t{linear map}\\
        \end{aligned}$}
    \\[22mm]
        \scalebox{.8}{
        $\begin{aligned}
          &\;\;\;\;\;\;\;\;\;\nb{v} \in \nb{\t{Value}}\\[3mm]
          \nb{v} \,\,\,\coloneqq 
	       &\ \ \ \nb{\tt} && \t{true}\\
        	 &\mid \ff && \t{false}\\
        	 &\mid \z && \t{zero}\\
        	 &\mid \suc{v} && \t{successor}\\
        	 &\mid \nb{\lam{x}{e}} && \t{linear map}
        \end{aligned}$}
    &
        \scalebox{.8}{
        $\begin{aligned} 
          &\;\;\;\;\;\;\;\;\;\nb{\D} \in \nb{\t{Context}}\\[3mm]
          \nb{\D} \,\,\,\coloneqq
	       &\ \ \ \nb{\emp} &&\t{empty}\\
	       &\mid \nb{\D,\bind{x}{\ty}} &&\t{binder}\\
        \end{aligned}$}
    \\[15mm]
    \end{tabular}
    \]
  \end{mdframed}
  \vspace{.8em}
  \caption{Syntax of $\cj${\scriptsize $(\nb{\lto}, \nb{\Bool}, \nb{\N})$}}
  \label{fig:syntax}
\end{figure}

\begin{figure}[h]
  \begin{mdframed}[topline=false,innertopmargin=-1.15ex,innerleftmargin=-0.1ex,innerrightmargin=0ex,linewidth=.8pt,skipbelow=2ex]
    \rulediv{25pt}{\large \textsf{Evaluating}}
    \vspace{.1em}
    \[  
    \begin{tabular}{c}
        \scalebox{.88}{
        $\step {e} {v} \iff \t{expression }\nb{e} \t{ evaluates to }\nb{v}$}
    \\[5mm]
        \scalebox{.8}{
        \def\extraVskip{3pt}
        \def\defaultHypSeparation{\hskip .05in}
        \bottomAlignProof
        \AxiomC{}
        \UnaryInfC{$\step{v}{v}$}
        \DisplayProof}
    \quad
        \scalebox{.8}{
        \def\extraVskip{3pt}
        \def\defaultHypSeparation{\hskip .05in}
        \bottomAlignProof
        \AxiomC{$\step{e}{v}$}
        \UnaryInfC{$\step{\suc{e}}{\suc{v}}$}
        \DisplayProof}
    \quad
        \scalebox{.8}{
        \def\extraVskip{3pt}
        \def\defaultHypSeparation{\hskip .05in}
        \bottomAlignProof
        \AxiomC{$\step {e_1} {\lam{x}{e}}$}
        \AxiomC{$\step {e_2} {v_2}$}
        \AxiomC{$\step {\{x \map v_2\}(e)} {v}$}
        \TrinaryInfC{$\step {e_1e_2} {v}$}
        \DisplayProof}
    \\[5mm]
        \scalebox{.8}{
        \def\extraVskip{3pt}
        \def\defaultHypSeparation{\hskip .05in}
        \bottomAlignProof
        \AxiomC{$\step{e_1}{\tt}$}
        \AxiomC{$\step {e_2} {v_2}$}
        \BinaryInfC{$\step  {\ite{e_1}{e_2}{e_3}}  {v_2}$}
        \DisplayProof}
    \quad
        \scalebox{.8}{
        \def\extraVskip{3pt}
        \def\defaultHypSeparation{\hskip .05in}
        \bottomAlignProof
        \AxiomC{$\step{e_1}{\ff}$}
        \AxiomC{$\step {e_3} {v_3}$}
        \BinaryInfC{$\step  {\ite{e_1}{e_2}{e_3}}  {v_3}$}
        \DisplayProof}
    \\[5mm]
        \scalebox{.8}{
        \def\extraVskip{3pt}
        \def\defaultHypSeparation{\hskip .05in}
        \bottomAlignProof
        \AxiomC{$\step{e_1}{v_1}$}
        \AxiomC{$\step{e_3}{\z}$}
        \BinaryInfC{$\step{\iter{e_1}{y}{e_2}{e_3}}{v_1}$}
        \DisplayProof}
    \quad
        \scalebox{.8}{
        \def\extraVskip{3pt}
        \def\defaultHypSeparation{\hskip .05in}
        \bottomAlignProof
        \AxiomC{$\step{\iter{e_1}{y}{e_2}{v_3}}{v_\t{n}}$}
        \AxiomC{$\step{\{y \map v_\t{n}\}(e_2)}{v}$}
        \AxiomC{$\step{e_3}{\suc{v_\t{3}}}$}
        \TrinaryInfC{$\step{\iter{e_1}{y}{e_2}{e_3}}{v}$}
        \DisplayProof}
    \\[5mm]
    \end{tabular}
    \]
  \end{mdframed}
  \vspace{.8em}
  \caption{Evaluating $\cj${\scriptsize $(\nb{\lto}, \nb{\Bool}, \nb{\N})$} programs}
  \label{fig:evaluating}
\end{figure}

\subsubsection{Evaluating} 

$\cj{\scriptstyle(\nb{\lto},\, \nb{\Bool},\, \nb{\mathbb{N}})}$ expressions evaluate according to the \textit{big-step} evaluation relation in Fig. \ref{fig:evaluating}; it specifies a call-by-value interpreter for the language. The notation $\nb{\{x \map v\}(e)}$ describes a capture-avoiding substitution, where $\nb{x}$ is replaced by $\nb{v}$ in the expression $\nb{e}$. In general, evaluation is similar to a simply-typed lambda calculus, with iteration as an exception. To clarify, consider the following iterator which doubles $\nb{1}$ to $\nb{2}$.

\vspace{-.8em}
\[
        \scalebox{.8}{
        \def\extraVskip{4pt}
        \def\defaultHypSeparation{\hskip .35in}
        \bottomAlignProof
        \AxiomC{$\step{\z}{\z}$}
        \AxiomC{$\step{\z}{\z}$}
        \BinaryInfC{$\step{\iter{\z}{y}{\suc{\suc{y}}}{\z}}{\z}$}
        \AxiomC{$\step{\z}{\z}$}
        \UnaryInfC{$\step{\suc{\z}}{\suc{\z}}$}
        \UnaryInfC{$\step{\{y \map \z\}(\suc{\suc{y}})}{\suc{\suc{\z}}}$}
        \AxiomC{$\step{\z}{\z}$}
        \UnaryInfC{$\step{\suc{\z}}{\suc{\z}}$}
        \TrinaryInfC{$\step{\iter{\z}{y}{\suc{\suc{y}}}{\suc{\z}}}{\suc{\suc{\z}}}$}
        \DisplayProof}
\]
\vspace{.1em}

The iterator dispatches on the value of $\nb{e_3}$. If it is $\nb{0}$, it returns the value of the base case $\nb{e_1}$. Otherwise, if it is $\nb{n}$ then the recurrence $(\nb{y \hookrightarrow e_2})$ will apply $\nb{n}$ times over the base case.

\begin{figure}[h]
  \begin{mdframed}[topline=false,innertopmargin=-1.08ex,innerleftmargin=-0.1ex,innerrightmargin=0ex,linewidth=.8pt,skipbelow=2ex]
    \rulediv{18pt}{\large \textsf{Typing}}
    \vspace{0.1em}
    \[  
    \begin{tabular}{c}
        \scalebox{.88}{
        $\p{\D} {e} {\ty} \iff \t{program }\nb{e} \t{ has type }\nb{\ty}$}
    \\[5mm]
        \scalebox{.8}{
        \def\extraVskip{3pt}
        \def\defaultHypSeparation{\hskip .05in}
        \bottomAlignProof
        \Axiom$\fCenter$
        \UnaryInf$\fCenter \p{\bind{x}{\ty}} {x} {\ty}$
        \DisplayProof}
    \quad
        \scalebox{.8}{
        \def\extraVskip{3pt}
        \def\defaultHypSeparation{\hskip .05in}
        \bottomAlignProof
        \Axiom$\fCenter$
        \UnaryInf$\fCenter \p{\emp} {\tt} {\Bool}$
        \DisplayProof}
    \quad
        \scalebox{.8}{
        \def\extraVskip{3pt}
        \def\defaultHypSeparation{\hskip .05in}
        \bottomAlignProof
        \Axiom$\fCenter$
        \UnaryInf$\fCenter \p{\emp} {\ff} {\Bool}$
        \DisplayProof}
    \quad
        \scalebox{.8}{
        \def\extraVskip{3pt}
        \def\defaultHypSeparation{\hskip .05in}
        \bottomAlignProof
        \AxiomC{}
        \UnaryInfC{$\p{\emp} {\z} {\N}$}
        \DisplayProof}
    \quad
        \scalebox{.8}{
        \def\extraVskip{3pt}
        \def\defaultHypSeparation{\hskip .05in}
        \bottomAlignProof
        \AxiomC{$\p{\D}{e}{\N}$}
        \UnaryInfC{$\p{\D} {\suc{e}} {\N}$}
        \DisplayProof}
    \\[5mm]
        \scalebox{.8}{
        \def\extraVskip{3pt}
        \def\defaultHypSeparation{\hskip .05in}
        \bottomAlignProof
        \AxiomC{$\p{ \D,\bind{x}{\ty_1}} {e} {\ty_2}$}
        \UnaryInfC{$\p{\D} {\lam{x}{e}} {\ty_1 \lto \ty_2}$}
        \DisplayProof}
    \quad
        \scalebox{.8}{
        \def\extraVskip{3pt}
        \def\defaultHypSeparation{\hskip .05in}
        \bottomAlignProof
        \AxiomC{$\p{\D_1} {e_1} {\ty_1 \lto \ty_2}$}
        \AxiomC{$\p{\D_2} {e_2} {\ty_1}$}
        \BinaryInfC{$\p{\D_1\o\D_2} {e_1e_2} {\ty_2}$}
        \DisplayProof}
    \\[5mm]
        \scalebox{.8}{
        \def\extraVskip{3pt}
        \def\defaultHypSeparation{\hskip .05in}
        \bottomAlignProof
        \AxiomC{$\p{\D_1} {e_1} {\Bool}$}
        \AxiomC{$\p{\D_2} {e_2} {\ty}$}
        \AxiomC{$\p{\D_2} {e_3} {\ty}$}
        \TrinaryInfC{$\p{\D_1\o\D_2} {\ite{e_1}{e_2}{e_3}} {\ty}$}
        \DisplayProof}
    \quad
        \scalebox{.8}{
        \def\extraVskip{3pt}
        \def\defaultHypSeparation{\hskip .05in}
        \bottomAlignProof
        \AxiomC{$\p{\D_1} {e_1} {\ty}$}
        \AxiomC{$\p{\bind{y}{\ty}} {e_2} {\ty}$}
        \AxiomC{$\p{\D_3} {e_3} {\N}$}
        \TrinaryInfC{$\p{\D_1 \o \D_3} {\iter{e_1}{y}{e_2}{e_3}} {\ty}$}
        \DisplayProof}
    \\[5mm]
    \end{tabular}
    \]
  \end{mdframed}
  \vspace{.8em}
  \caption{Typing $\cj${\scriptsize $(\nb{\lto}, \nb{\Bool}, \nb{\N})$} programs}
  \label{fig:typing}
\end{figure}

\subsubsection{Typing} 

$\cj{\scriptstyle(\nb{\lto},\, \nb{\Bool},\, \nb{\mathbb{N}})}$ is linearly typed. The presence of context splitting in typing application, conditionals, and iteration ensures that the structural rule \textit{contraction} is not permissible. This relation relates two contexts if their variable names are disjoint. For example, \scalebox{1.0}{$\nb{(\bind{x}{\Bool}, \bind{y}{\Bool}) \o (\bind{z}{\Bool})}$} inhabits the relation, but \scalebox{1.0}{$\nb{(\bind{x}{\Bool}, \bind{y}{\Bool}) \o (\bind{y}{\Bool})}$} does not (Appendix A.1). Separately, the demand of singleton or empty contexts in typing variables or constants ensures that the structural rule \textit{weakening} is not permissible.


There is now a key property we must prove about typing---well typed programs must have behavior. Otherwise, we cannot establish a correspondence with equations underlying the dynamics of recurrent neurons. Later, it helps us prove our analog of \textit{adequacy} for a denotational semantics.
\newcounter{thm_ctr}
\newcounter{lem_ctr}
\refstepcounter{thm_ctr}\label{thm:type-sound}
\vspace{.8em}
\begin{bthm}{\textsf{(Programs evaluate to values)}}{}
\vspace{.25em}
\t{If }$\p{\emp}{e}{\ty}$ \t{, then } $\exists \nb{v}, \step{e}{v}$  
\end{bthm}

\textit{Proof sketch.} The argument is similar to \textit{logical relations} proofs of termination for typed lambda calculi. Appendix B.1 contains a detailed proof. In short, the argument shows that $\cj{\scriptstyle(\nb{\lto},\, \nb{\Bool},\, \nb{\mathbb{N}})}$ programs inhabit the logical relation shown in Fig. \ref{fig:logical-relation}, and that programs in this logical relation terminate.

\vspace{.4em}
\begin{figure}[h]
  \begin{mdframed}[topline=false,innertopmargin=-1.15ex,innerleftmargin=-0.1ex,innerrightmargin=0ex,linewidth=.8pt,skipbelow=2ex]
    \rulediv{38pt}{\large \textsf{Logical Relation}}
    \vspace{.1em}
    \[  
    \begin{tabular}{c c}
        \scalebox{.8}{
        $\begin{aligned}
          &\;\;\;\V{\_}: \nb{\t{Type}} \to \Big\{\nb{\t{Value}}\Big\}\\[3mm]
            \V{\Bool}&=\Big\{\;\tt,\ff\;\Big\}\\
            \V{\N}&=\Big\{\begin{array}{l}
            \nb{\z}
            \end{array}\Big\} \cup
            \Big\{\begin{array}{l}
            \nb{\suc{v}} \mid \nb{v} \in \V{\N}
            \end{array}\Big\}
            \\
            \V{\ty_1 \lto \ty_2} &=
            \Big\{\begin{array}{l}
            \nb{\lam{x}{e}} \mid \forall\nb{v}\in \V{\ty_1},\nb{(\lam{x}{e})v} \in \E{\ty_2}\\
            \end{array}
            \Big\}
        \end{aligned}$}
    &
        \scalebox{.8}{
        $\begin{aligned} 
          \Nv{\_}:& \; \nb{\t{Context}} \to \Big\{\nb{\t{Environment}} \Big\}\\[3mm]
            \Nv{\emp} &= \Big\{\nb{\emp} \Big\}\\
            \Nv{\D,\bind{x}{\ty}} &= 
            \Big\{\begin{array}{l}
             \nb{\d\{x \map v\}}\mid \nb{\d} \in \Nv{\D}, \nb{v} \in \V{\ty}
            \end{array}
            \Big\}
        \end{aligned}$}
    \\[18mm]
        \scalebox{.8}{
        $\begin{aligned} 
          \E{\_}:& \; \nb{\t{Type}} \to \Big\{\nb{\t{Expression}} \Big\}\\[3mm]
            \E{\ty} &= \left\{
            \nb{e} \;\middle|\;
            \begin{array}{l}
            \nb{e} \bstep \nb{v},\\
            
            \nb{v}\in \V{\ty}\\
            
            \end{array}
            \right\}
        \end{aligned}$}
    &
        \scalebox{.8}{
        $
        \begin{aligned}
        \ul{\t{Logical Typing}} &\\[2mm]
        \lp{\D}{e}{\ty} &\iff \forall \nb{\d} \in \Nv{\D},\, \nb{\d(e)}\in \E{\ty}
        \end{aligned}
        $}
    \\[11mm]
    \end{tabular}
    \]
  \end{mdframed}
  \vspace{.8em}
  \caption{Logical Relation for $\cj${\scriptsize $(\nb{\lto}, \nb{\Bool}, \nb{\N})$}}
  \label{fig:logical-relation}
\end{figure}

At each type, the logical relation $\E{\ty}$ returns a set of terminating $\cj{\scriptstyle(\nb{\lto},\, \nb{\Bool},\, \nb{\mathbb{N}})}$ expressions. It includes expressions which are \textit{not} syntactically well-typed. For example, $\nb{(\ite{\ul{\tt}}{\ul{\tt}}{\ul{0}})} \in \E{\Bool}$. We say these expression are \textit{logically} well-typed, $\lp{\emp}{\ite{\ul{\tt}}{\ul{\tt}}{\ul{{0}}}}{\Bool}$. To show termination for $\cj{\scriptstyle(\nb{\lto},\, \nb{\Bool},\, \nb{\mathbb{N}})}$, we first prove by induction on typing derivations that syntactically well-typed programs can be logically typed.

\vspace{.5em}
\begin{blem}{\textsf{(Programs can be logically typed)}}{fundamental-lemma}
\vspace{.25em}
\t{If }$\p{\D}{e}{\ty} $ \t{, then } $\lp{\D}{e}{\ty}$  
\end{blem}
\vspace{.5em}

The bulk of the effort is proving that programs can be logically typed. Afterwards, it is immediate from the definition of the logical relation that logically typed programs terminate.

\vspace{.5em}
\begin{blem}{\textsf{(Logically typed programs evaluate to values)}}{logical-termination}
\vspace{.25em}
\t{If }$\lp{\D}{e}{\ty}  $ \t{,  then } $\exists \nb{v},\step{\d(e)}{v}$  
\end{blem}
\vspace{.5em}

From these lemmas, it follows that syntactically well-typed programs in $\cj{\scriptstyle(\nb{\lto},\, \nb{\Bool},\, \nb{\mathbb{N}})}$ evaluate to values. In other words, they have behavior. Our efforts may now shift toward implementing these behaviors with the dynamics of recurrent neurons.

\subsection{Compiling}
\label{sec:compiling}

\subsubsection{Types}
The first step in compiling $\cj{\scriptstyle(\nb{\lto},\, \nb{\Bool},\, \nb{\mathbb{N}})}$ is to compile its types to $\rd{\R}$-vector spaces. To implement a type with $\nb{n}$ distinct\footnote{Distinct meaning not contextually equivalent. For example, $\lam{x}{x}$ and $\lam{y}{y}$ are contextually equivalent.} values, an $\nb{n}$-dimensional space generally suffices. For finite data like booleans, the choice of $\nb{n}$-dimensional space is simple. $\rd{\R^2}$ suffices to implement boolean programs as linear maps, shown in Fig. \ref{fig:compiling-types}. Note that if the target vector space has too few dimensions, then a linear map that adequately implements a program \textit{may not exist}.

\begin{figure}[h]
  \begin{mdframed}[topline=false,innertopmargin=-1.15ex,innerleftmargin=-0.1ex,innerrightmargin=0ex,linewidth=.8pt,skipbelow=2ex]
    \rulediv{28pt}{\large \textsf{Compiling}\scalebox{1.3}{$\hspace{.03em}_\nb{\ty}$}}
    \vspace{0.1em}
    \[  
    \begin{tabular}{c}
        \scalebox{.88}{
        $\begin{aligned}
        \c{\nb{\_}}:&\; \nb{\t{Type}} \to \rd{\t{Vector Space}}\\[.8em]
            \c{\nb{\Bool}} &= \rd{\R^2}\\[1mm]
            \c{\nb{\N}} &= \bigl\{\rd{x}:\rd{\N \to \R} \mid \exists \nb{n},\, \rd{\t{supp}}(\rd{x})=\nb{n}\bigr\}\\[1mm]
            \c{\nb{\ty_1 \lto \ty_2}} &= \rd{\t{Lin}}(\c{\nb{\ty_1}},\c{\nb{\ty_2}})\\
            \end{aligned}$
        }
    \\[12mm]
    \end{tabular}
    \]
  \end{mdframed}
  \vspace{.8em}
  \caption{Compiling $\cj${\scriptsize $(\nb{\lto}, \nb{\Bool}, \nb{\N})$} types}
  \label{fig:compiling-types}
\end{figure}

For natural numbers, the choice of vector space is more subtle. In Fig. \ref{fig:compiling-types}, natural numbers compile to an infinite-dimensional sequence space. Each vector is an infinite-length sequence, represented as a function which yields the value of the sequence at any index $\nb{n \in \N}$. The support of these functions is finite---only a finite number of entries in a sequence may be non-zero. This assumption is critical because our strategy for compiling iteration involves a summation over sequences, and this summation is not well-defined without finite support. To form a vector space, vector addition and scalar multiplication are defined pointwise.

For functions, we compile to the vector space of linear maps between the compiled input and output types. To form a vector space, vector addition and scalar multiplication are defined pointwise. Our approach differs from prior work studying first-class conditionals in differentiable programming, where functions compiled to matrix spaces \cite{jvg26linear}. As we developed our metatheory and implementation, we found it much simpler to represent function types abstractly as linear maps. If necessary, the matrix of a linear map can be determined in a separate compiler pass.

\subsubsection{Contexts}

Programs with free variables require us to compile contexts, shown in Fig. \ref{fig:compiling-contexts}. These compiled contexts specify which vectors we can link compiled neurons against. Generally, a context with $\nb{k}$ variables compiles to a $\nb{k}$-tuple of vectors.\footnote{By convention we denote elements of a compiled $\nb{1}$-tuple using $\xvec$, and elements of a compiled $\nb{k}$-tuple as $\svec$.}

\begin{figure}[h]
  \begin{mdframed}[topline=false,innertopmargin=-1.15ex,innerleftmargin=-0.1ex,innerrightmargin=0ex,linewidth=.8pt,skipbelow=2ex]
    \rulediv{28pt}{\large \textsf{Compiling}\scalebox{1.2}{$\hspace{.03em}_\nb{\D}$}}
    \vspace{0.1em}
    \[
    \begin{tabular}{c}
        \scalebox{.88}{
        $\begin{aligned}
        \c{\nb{\_}}: \nb{\t{Context}}& \to \rd{\t{Tuple}}(\rd{\t{Vector}})\\[.8em]
            \c{\nb{\emp}} &= \rd{\{0\}}\\[1mm]
            \c{\nb{\D,\bind{x}{\ty}}} &= \c{\nb{\D}}\;\rd{\x}\;\c{\nb{\ty}}\\ 
            \end{aligned}$
        }
    \\[9mm]
    \end{tabular}
    \]
  \end{mdframed}
  \vspace{.8em}
  \caption{Compiling $\cj${\scriptsize $(\nb{\lto}, \nb{\Bool}, \nb{\N})$} contexts}
  \label{fig:compiling-contexts}
\end{figure}

\subsubsection{Programs}

Using a compiler for contexts and types, we compile typing judgments $\p{\D}{e}{\ty}$ to multilinear maps from $\cm{\D}$ to $\cm{\ty}$. Generally, a typing judgment with $\nb{k}$ variables in the context will compile to a $\nb{k}$-linear map. The compiler is a function defined recursively on the structure of typing derivations, shown in Fig. \ref{fig:compiling-programs}. Prior work on first-class conditionals used a relational specification. However, we find a functional specification leads to simpler proofs and a simpler implementation.

\begin{figure}[h]
  \begin{mdframed}[topline=false,innertopmargin=-1.15ex,innerleftmargin=-0.1ex,innerrightmargin=0ex,linewidth=.8pt,skipbelow=2ex]
    \rulediv{28pt}{\large \textsf{Compiling}\scalebox{1.3}{$\hspace{.03em}_\nb{e}$}}
    \vspace{0.1em}
    \[
    \begin{tabular}{c}
        \scalebox{.88}{
        $\begin{aligned}
        &\hspace{13em}\cm{-} :\nb{\t{Program}}\to \rd{\t{Multilinear map}}\\[.8em]
            &\c{\p{\bind{x}{\ty}}{x}{\ty}}({\xvec}) = \xvec\\\\
            &\c{\p{\emp}{\tt}{\Bool}}(\rd{0}) = \ttv\\\\
            &\c{\p{\emp}{\ff}{\Bool}}(\rd{0}) = \ffv\\\\
            &\c{\p{\emp}{\z}{\N}}(\rd{0}) = 
             {\rd{n} \map 
              \begin{cases}
                 \rd{1} & \t{if }\rd{n=0}\\
                 \rd{0} & \t{otherwise}
              \end{cases}}\\\\
            &\c{\p{\D}{\suc{e}}{\N}}(\sv{}) = 
             {\rd{n} \map 
              \begin{cases}
                 \rd{0} & \t{if }\rd{n=0}\\
                 \imp{\nb{e}}{\svec}(\rd{n-1}) & \t{otherwise}
              \end{cases}}\\\\
            &\c{\p{\D}{\lam{x}{e}}{\ty_1 \lto \ty_2}}(\svec) = \xvec \map \cm{\nb{e}}{(\sv{},\xvec)}\\\\
            &\c{\p{\D_1\o\D_2}{e_1e_2}{\ty_2}}(\svec) = \cm{e_1}(\svec_\Ds1)(\cm{e_2}(\svec_\Ds2)) \\\\
            &\c{\p{\D_1\o\D_2}{\ite{e_1}{e_2}{e_3}}{\ty_2}}(\svec) = \rd{\pi_1}(\cm{e_1}(\svec_\Ds1))\rt \cm{e_2}(\svec_\Ds2) \rp \rd{\pi_2}(\cm{e_1}(\svec_\Ds1))\rt \cm{e_3}(\svec_\Ds2)\\\\
            &\c{\p{\D_1\o\D_2\o\D_3}{\iter{e_1}{y}{e_2}{e_3}}{\ty}}(\svec) = \rd{\sum_{n \in \N}}\,\c{\nb{e_3}}(\svec_\Ds3)(\rd{n})\,\rd{\.}\,(\c{\nb{e_2}})^\rd{n}(\cm{\nb{e_1}}(\svec_\Ds1))\\\\
            \end{aligned}$
        }
    \\[9mm]
    \end{tabular}
    \]
  \end{mdframed}
  \vspace{.4em}
  \caption{Compiling $\cj${\scriptsize $(\nb{\lto}, \nb{\Bool}, \nb{\N})$} programs}
  \label{fig:compiling-programs}
\end{figure}

Compiling variables is straight-forward. The variable rule maps to identity at each type.
\vspace{.1em}
\[
\cm{\p{\bind{x}{\Bool}}{x}{\Bool}} \mapsto (\xvec \in \rd{\R^2} \mapsto \xvec)
\]

\vspace{.6em}
\noindent Our specification in Fig. \ref{fig:compiling-programs} uses an equivalent, less verbose notation. Moreover, we often omit specifying the domain of these multilinear maps. The domain is always $\cm{\D}$. 

The constants \nb{\tt}, \nb{\ff}, and \nb{\z} compile to constant $\nb{0}$-linear maps. Their domain is $\{\rd{0}\}$ and they are vacuously linear in $\nb{0}$ arguments. These $\nb{0}$-linear maps generally map $\rd{0}$ to orthonormal bases in $\cm{\ty}$. These are so-called \textit{one-hot} encodings. Observe that for natural numbers these encodings occur \textit{in time}.
\vspace{.6em}
\[
\begin{aligned}
\c{\p{\emp}{\nb{\z}}{\N}}(\rd{0}) &= \rd{\begin{bmatrix}1 \!\!&\!\! 0 \!\!&\!\! 0 \!\!&\!\! \cdots\end{bmatrix}}\\
\c{\p{\emp}{\nb{\suc{\z}}}{\N}}(\rd{0}) &= \rd{\begin{bmatrix}0 \!\!&\!\! 1 \!\!&\!\! 0 \!\!&\!\! \cdots\end{bmatrix}}\\
\c{\p{\emp}{\nb{\suc{\suc{\z}}}}{\N}}(\rd{0}) &= \rd{\begin{bmatrix}0 \!\!&\!\! 0 \!\!&\!\! 1 \!\!&\!\! \cdots\end{bmatrix}}\\
\end{aligned}
\]
\vspace{.3em}

\noindent Using time, a \textit{single} recurrent neuron can implement natural numbers.

The successor $\p{\bind{x}{\N}}{\suc{x}}{\N}$ compiles to a $\nb{1}$-linear \textit{right-shift} operator on sequences. Since $\nb{\suc{x}}$ contains a free variable $\nb{x}$, this compiled program can be \textit{linked} with an \textit{environment}. An environment is any element of $\cm{\bind{x}{\N}}$, and linking is any application of the $\nb{1}$-linear right-shift operator to these elements, shown in the following example. 
\vspace{.3em}
\[
\begin{aligned}
\c{\p{\bind{x}{\N}}{\nb{\suc{x}}}{\N}}(\rd{\begin{bmatrix}10 \!\!&\!\! 20 \!\!&\!\! 30 \!\!&\!\! \cdots\end{bmatrix}}) &= \rd{\begin{bmatrix}0 \!\!&\!\! 10 \!\!&\!\! 20 \!\!&\!\! \cdots\end{bmatrix}}
\end{aligned}
\]
\vspace{-.4em}

\noindent If we bind $\nb{x}$ using $\nb{\lambda}$, this compiles instead to a $\nb{0}$-linear map which \textit{returns a linear map}.
\vspace{.5em}
\[
\begin{aligned}
\c{\p{\emp}{\lam{x}{\suc{x}}}{\N \lto \N}}(\rd{0}) &= \bigl(\xvec \in \cm{\N} \mapsto \rd{\begin{bmatrix}0 \!\!&\!\! \vec{x_0} \!\!&\!\! \vec{x_1} \!\!&\!\! \cdots\end{bmatrix}}\bigr)
\end{aligned}
\]

\noindent Note that our specification in Fig. \ref{fig:compiling-programs} omits typing information when it is clear from context. For example, $\cm{e}$ is used in the function case. But the compiler is defined on typing derivations, not expressions. This is shorthand for $\c{\p{\D,\bind{x}{\ty_1}}{e}{\ty_2}}$; the convention is \textit{very} useful in simplifying definitions and proofs.

An application compiles to function application. However, context splitting introduces some subtlety. Observe how the environment $\yvec$ propagates only to specific subprograms in this example.
\vspace{.4em}
\[
\begin{aligned}
& \;\;\;\;\c{\p{\bind{y}{\N}}{(\lam{x}{\suc{x}})y}{\N}}(\yvec)\\[.8mm]
& = \cm{\lam{x}{\suc{x}}}(\yvec_\nb{\emp})(\cm{y}(\yvec_\nb{y:\N})) && \t{By compiling} \\[.8mm]
& = \cm{\lam{x}{\suc{x}}}(\rd{0})(\cm{y}(\yvec)) && \t{By restricting the environment} \\[.6mm]
& = \bigl(\xvec \in \cm{\N} \mapsto \rd{\begin{bmatrix}0 \!\!&\!\! \vec{x_0} \!\!&\!\! \vec{x_1} \!\!&\!\! \cdots\end{bmatrix}}\bigr)\bigl( \rd{\begin{bmatrix} \vec{y_0} \!\!&\!\! \vec{y_1} \!\!&\!\! \vec{y_2} \!\!&\!\! \cdots\end{bmatrix}}\bigr) && \t{By compiling}\\[.6mm]
& = \rd{\begin{bmatrix} 0 \!\!&\!\! \vec{y_0} \!\!&\!\! \vec{y_1} \!\!&\!\! \cdots\end{bmatrix}} && \t{By function application}
\end{aligned}
\]
\vspace{.4em}

\noindent Essentially, the environment must be split to mirror context splitting. In Fig. \ref{fig:compiling-programs}, the notation $\svec_\nb{\D}$ handles this. It defines the restriction of an environment $\svec$ according to a context $\nb{\D}$. Appendix A.7 contains a detailed specification.

Conditionals compile to a \textit{soft-branching} conditional. Both branches evaluate, returning a linear combination of their values. It implements ordinary \textit{hard-branching} conditionals, but in a fully differentiable way. Observe how the following example is differentiable in the environment, enabling first-class differentiable programming with conditionals. It is a differentiable $\nb{\texttt{not}}$ operator.
\vspace{.7em}
\[
\begin{aligned}
& \;\;\;\;\c{\p{\bind{x}{\Bool}}{\ite{\ul{x}}{\ul{\ff}}{\ul{\tt}}}{\Bool}}\bigl(\bv{\a1}{\a2}\bigr)\\[.8mm]
& = \rd{\pi_1}\bigl(\cm{x}\bigl(\bv{\a1}{\a2}\bigr)\bigr) \rt \cm{\ff}(\rd{0}) \rp \rd{\pi_1}\bigl(\cm{x}\bigl(\bv{\a1}{\a2}\bigr)\bigr) \rt \cm{\tt}(\rd{0}) && \t{By compiling}\\[.8mm]
& = \rd{\pi_1}\bigl(\bv{\a1}{\a2}\bigr) \rt \ffv \rp \rd{\pi_2}\bigl(\bv{\a1}{\a2}\bigr) \rt \ttv && \t{By compiling}\\[.8mm]
& = \a1 \rt \ffv \rp \a2 \rt \ttv && \t{By projections }\rd{\pi_i}\\[.8mm]
\end{aligned}
\]
\vspace{.3em}

\noindent The functions $\rd{\pi_1}$ and $\rd{\pi_2}$ used in Fig. \ref{fig:compiling-programs} are the first and second projection for $\rd{v} \in \rd{\R^2}$.

Iteration compiles to equations underlying the dynamics of linear recurrent neurons, as in Fig \ref{fig:ccc1}. It computes a linear combination of states in a linear dynamical system. Recall that with particular inputs, these equations \textit{exactly} implement the dynamics of linear recurrent neurons (Section 2). This example shows how to compile the iterated \nb{\texttt{not}} operator from Fig. \ref{fig:ccc1}. 
\vspace{.7em}
\[
\begin{aligned}
& \;\;\;\;\c{\p{\emp}{\iter{\tt}{x}{\ite{\ul{x}}{\ul{\ff}}{\ul{\tt}}}{2}}{\Bool}}\bigl(\rd{0})\\[1.2mm]
& = \rd{\sum_{n \in \N}} \;(\cm{2}(\rd{0}))(\rd{n}) \rt (\cm{\ite{\ul{x}}{\ul{\ff}}{\ul{\tt}}}(\rd{0}))^\rd{n}(\cm{\tt}(\rd{0})) && \t{By compiling}\\[.2mm]
& = \rd{\sum_{n \in \N}} \; \rd{\begin{bmatrix} 0 \!\!&\!\! 0 \!\!&\!\! 1\!\!&\!\!  0 \!\!&\!\!  \cdots\end{bmatrix}_n} \rt \bigl(\cm{\ite{\ul{x}}{\ul{\ff}}{\ul{\tt}}}\bigl(\rd{0}\bigr)\bigr)^\rd{n}\bigl(\ttv\bigr) && \t{By compiling}\\[.2mm]
& = \rd{\sum_{n \in \N}} \; \rd{\begin{bmatrix} 0 \!\!&\!\! 0 \!\!&\!\! 1\!\!&\!\!  0 \!\!&\!\!  \cdots\end{bmatrix}_n} \rt \bigl(\xvec \mapsto \cm{\ite{\ul{x}}{\ul{\ff}}{\ul{\tt}}}\bigl(\xvec\bigr)\bigr)^\rd{n}\bigl(\ttv\bigr) && \t{By partial application}\\[.2mm]
& \simeq \rd{\sum_{n \in \N}} \; \rd{\begin{bmatrix} 0 \!\!&\!\! 0 \!\!&\!\! 1\!\!&\!\!  0 \!\!&\!\!  \cdots\end{bmatrix}_n} \rt \rd{\begin{bmatrix}0 & 1\\1 & 0\end{bmatrix}^\rd{n}}\bigl(\ttv\bigr) && \t{By matrix of a linear map}\\[.2mm]
\end{aligned}
\]
\vspace{.3em}

Observe that this example satisfies the desired correspondence from Fig. \ref{fig:ccc1} between evaluation and the dynamics of linear recurrent neurons. From now on, $\cm{e}$ is shorthand for $\cm{e}(\rd{0})$.
\vspace{.4em}
\[
\cm{\iter{\tt}{x}{\ite{\ul{x}}{\ul{\ff}}{\ul{\tt}}}{2}} = \cm{\tt} \neq \cm{\ff}
\]
\vspace{-.3em}

\noindent This correspondence holds for any closed\footnote{A closed program has no free variables.} program of type $\nb{\N}$ or $\nb{\Bool}$. Notably, this includes programs with nested iteration or higher-order functions. These correspond to more complex neural network architectures that we do not explore empirically---hierarchical recurrent neural networks and recurrent hypernetworks \cite{pascanu2013construct, ha2016hypernetworks}.

\vspace{.7em}
\begin{bthm}{(Compiler preserves program behavior)}{}
\refstepcounter{thm_ctr}\label{thm:compiler-correct}
\vspace{.25em}
\t{If }$\p{\emp}{e}{\t{base}\bl{(}\ty\bl{)}}$, \t{ then }
\vspace{.4em}
\begin{flalign*}
&\begin{aligned}
(a) \; & \step{e}{v}  \implies \cm{e}=\cm{v} \\
(b) \; & \nb{e} \;\cancel{\bstep}\; \nb{v} \implies \cm{e}\neq\cm{v}\\
\end{aligned}&&
\end{flalign*}    
\end{bthm}
\vspace{.3em}

\textit{Proof sketch.} The argument requires showing that the compiler preserves what programs will $(a)$ and won't $(b)$ do. These correspond to soundness and adequacy of a denotational semantics. Appendix B.4 contains a detailed proof. In short, the argument for $(a)$ is by induction on evaluation. Then $(b)$ follows as a corollary of $(a)$ and Theorem \ref{thm:type-sound}. 

For $(a)$, the induction on evaluation proceeds on the following lemma, which generalizes the type. This gives a sufficient induction hypothesis.

\vspace{.5em}
\begin{blem}{\textsf{(Compiler preserves what programs will do)}}{}
\vspace{.25em}
\t{If }$\p{\emp}{e}{\ty}$ \t{ and }$\step{e}{v}$ \t{, then } $\cm{e} = \cm{v}$  
\end{blem}
\vspace{.5em}

It is worth considering the iteration case in detail $\step{\iter{e_1}{y}{e_2}{e_3}}{v}$, where $\nb{e_3}$ evaluates to $\nb{\suc{v_3}}$. It captures the essential reasoning happening across cases in the proof. We must show $\cm{\iter{e_1}{y}{e_2}{e_3}} = \cm{v}$. \footnote{For functions $=$ is extensional equality.}

\[
\scalebox{.8}{$
\begin{aligned}
      &\,\,\;\;\;\,\cm{\iter{e_1}{y}{e_2}{e_3}}\\[.5em]
      &=\rd{\sum_{n\in\N}}\,\cm{e_3}(\rd{n})\,\rd{\.}\,\cm{e_2}^\rd{n}(\cm{e_1}) && \t{By compiling}\\
         &= \rd{\sum_{n\in\N}}\,\cm{e_3}(\rd{n})\,\rd{\.}\,\cm{e_2}(\cm{e_2}^\rd{(n-1)}(\cm{e_1})) && \t{Because }\rd{f}^\rd{n}(\rd{v})=\rd{f}(\rd{f}^{\rd{(n-1)}}(\rd{v}))\\
         &= \rd{\sum_{n\in\N}}\,\cm{\suc{v_3}}(\rd{n})\,\rd{\.}\,\cm{e_2}(\cm{e_2}^\rd{(n-1)}(\cm{e_1})) && \t{By induction}\\
     &= \cm{\suc{v_3}}(\rd{0})\,\rd{\.}\,\cm{e_2}^\rd{0}(\cm{e_1}) \rp \rd{\sum_{n\in\N\setminus0}}\,\cm{{\suc{v_3}}}(\rd{n})\,\rd{\.}\,\cm{e_2}(\cm{e_2}^\rd{(n-1)}(\cm{e_1})) && \t{Because }\rd{\sum_{n\in\N}}\,\rd{f}(\rd{n})=\rd{f}(\rd{0})\rp \rd{\sum_{n\in\N\setminus0}}\,\rd{f}(\rd{n})\\
      &= \rd{0}\,\rd{\.}\,\cm{e_2}^\rd{0}(\cm{e_1}) \rp \rd{\sum_{n\in\N\setminus0}}\,\cm{{v_3}}(\rd{n-1})\,\rd{\.}\,\cm{e_2}(\cm{e_2}^\rd{(n-1)}(\cm{e_1})) && \t{By compiling}\\
          &= \rd{\sum_{n\in\N\setminus0}}\,\cm{{v_3}}(\rd{n-1})\,\rd{\.}\,\cm{e_2}(\cm{e_2}^\rd{(n-1)}(\cm{e_1})) && \t{By scalar multiplication}\\ 
         &= \rd{\sum_{n\in\N\setminus0}}\,\cm{e_2}(\cm{{v_3}}(\rd{n-1})\,\rd{\.}\,\cm{e_2}^{\rd{(n-1)}}(\cm{e_1})) && \t{Compiling maps to multilinear maps}\\ 
         &= \cm{e_2}\Bigl(\rd{\sum_{n\in\N\setminus0}}\,\cm{{v_3}}(\rd{n-1})\,\rd{\.}\,\cm{e_2}^{\rd{(n-1)}}(\cm{e_1})\Bigr) && \t{Compiling maps to multilinear maps}\\
         &= \cm{e_2}\Bigl(\,\rd{\sum_{n\in\N}}\,\cm{{v_3}}(\rd{n})\,\rd{\.}\,\cm{e_2}^{\rd{n}}(\cm{e_1})\Bigr)&& \t{Because }\rd{\sum_{n\in\N\setminus0}\rd{f}}(\rd{n-1})=\rd{\sum_{n\in\N}\rd{f}}(\rd{n})\\[.5em]
        &= \cm{e_2}(\cm{\iter{e_1}{y}{e_2}{v_3}}) && \t{By compiling}\\
        &= \cm{e_2}(\cm{v_\t{n}}) && \t{By induction}\\
        &= \cm{\{y \map v_\t{n}\}(e_2)} && \t{Compiling commutes with substitution}\\
        &= \cm{v} && \t{By induction}
\end{aligned}$}
\]
\vspace{1em}

The argument introduces two key lemmas, used across cases. The first states our compiler produces multilinear maps, as indicated in Fig. \ref{fig:compiling-programs}. 

\vspace{.5em}
\begin{blem}{\textsf{(Compiler maps programs to multilinear maps)}}{}
\vspace{.25em}
\t{If }$\p{\D,\bind{x}{\ty_1}}{e}{\ty_2}$, \t{ then }
\vspace{.2em}
\[
\cm{e}(\svec,\a1\rt\xv1\rp\a2\rt\xv2)=\a1\rt\cm{e}(\svec,\xv1)\rp \a2\rt\cm{e}(\svec,\xv2)
\vspace{.2em}
\]
\end{blem}
\vspace{.8em}

\noindent The second is a substitution lemma, typical of compiler correctness proofs. Both are shown by structural induction on $\nb{e}$. Appendix B.7/B.10 contain detailed proofs.

\vspace{.5em}
\begin{blem}{\textsf{(Compiler commutes with substitution )}}{}
\vspace{.25em}
\t{If }$\p{\D,\bind{x}{\ty_1}}{e}{\ty_2}$, \t{ then } $\cm{\{x \map v\}(e)}(\svec)=\cm{e}(\svec,\cm{v})$
\end{blem}
\vspace{.8em}

These are the essential arguments in establishing the correctness of our compiler. Note how \textit{linearity} is critical to the argument. We compile to \textit{linear} neural networks.\footnote{These can still be \textit{linked} against nonlinear neural networks, as we do in the experiments.} These are essentially linear maps whose matrices are easy to compute. Without linearity, there would be no effective procedure for determining the coefficients associated with the matrices inside nonlinear neural networks. There is a rich history studying this problem. It teaches us that determining such coefficients for various architectures and classes of functions is typically NP-complete \cite{judd1990neural}. 

Now we address the empirical question, as to whether the correctness of our compiler yields any benefit to differentiable programming.

\section{Experiments}
\label{sec:experiments}

Our experiments test whether programming neural networks with $\cj${\scriptsize $(\nb{\lto}, \nb{\Bool}, \nb{\N})$} helps them learn tasks with discrete structure, and whether the correctness theorem we proved is actually driving this benefit. Programming a neural network with $\cj${\scriptsize $(\nb{\lto}, \nb{\Bool}, \nb{\N})$} means that we specify \textit{part} of the neural network using an \textit{open}\footnote{Open programs have free variables.} $\cj${\scriptsize $(\nb{\lto}, \nb{\Bool}, \nb{\N})$} program, which compiles to a $\nb{k}$-linear map implemented in a differentiable programming language. When programming neural networks this way, the equations defining their behavior can look as follows---a composition of matrices, nonlinear functions, and compiled $\nb{k}$-linear maps.
$$\rd{\eta}(\rd{x}) = \rd{W_2} \o \rd{\t{relu}} \o \cm{e} \o \rd{\t{relu}} \o \rd{W_1}$$
\vspace{-.6em}

Our implementation of $\cj${\scriptsize $(\nb{\lto}, \nb{\Bool}, \nb{\N})$} is a simple prototype built in \t{PyTorch}, a domain-specific language for programming neural networks in \t{Python} \cite{paszke2019pytorch}. It extends \t{PyTorch} with the primitives in $\cj${\scriptsize $(\nb{\lto}, \nb{\Bool}, \nb{\N})$}, enriching it with first-class iteration and conditionals. The compiler, which implements the specification in Fig. \ref{fig:compiling-programs}, is roughly 200 lines of \t{Python}.

The experiments study two related image transformation tasks. The first is the iterative image transformation task from Fig. \ref{fig:task}, where a neural network is programmed to solve it using first-class iteration. The second experiment is a conditional and iterative image transformation---if the input is an even digit, apply vertical blur iteratively, else return the original image. A neural network is programmed to solve it using first-class iteration \textit{and} conditionals. 

These experiments illustrate, but do not exhaust, possible approaches to these tasks. They are meant to highlight the benefits and challenges inherent to differentiable programming with first-class discrete structure. Specifically, they are designed to test our specification: whether a compiler can be implemented according to it, and whether the compiler correctness theorem we identify is driving positive impacts on learning. They are not engineering benchmarks; they test \textit{the theory}.

To briefly summarize our results, we find that programming neural networks with first-class discrete structure helps them learn faster and with greater data-efficiency relative to a neural network programmed without first-class iteration. These benefits persist even when using the compiler in an alternative \textit{type-preserving} way---returning instead a random $\nb{k}$-linear map from $\cm{\D}$ to $\cm{\ty}$. The data also reveal an interesting limitation of our specification. Linking with vectors of large \textit{norm} can destabilize learning dynamics. Yet it is allowed by our specification, which only demands that these vectors are in $\cm{\D}$. By bounding the norm of linked vectors, a compiler could feasibly prevent this instability.

Appendix C.1-4 contains 4 additional experiments, which replicate experiments from prior work on differentiable programming with first-class conditionals \cite{jvg26linear}. Since our specification and implementation differ, these data were helpful to establish consistency with prior work.

\subsection{Experiment 1: Iterative Image Transformation}
\label{sec:exp1}

\subsubsection{Task} Our first experiment studies the iterative image transformation task from Fig. \ref{fig:task}.

\setlist[itemize]{topsep=8pt, itemsep=1pt, parsep=2pt, leftmargin=1.5em, rightmargin=1.5em}
\begin{itemize}[label=$\triangleright$]
\item If $\nb{x}$ is the digit $\nb{0}$, return $\nb{x}$
\item If $\nb{x}$ is the digit $\nb{n}$, apply vertical blur $\nb{n}$ times to $\nb{x}$
\end{itemize}

\subsubsection{Models} Each model is a neural network implemented in \t{PyTorch}, and trained to solve the iterative image transformation. The experiments test whether a compiler satisfying our specification yields positive effects on learning dynamics. Our models are controls for measuring this effect. To justify our experimental design, we first give intuition for our choice of models. Then we review the specific details of each model.
\begin{itemize}
\item \t{Model D} is produced using our semantics-preserving compiler (Fig. \ref{fig:compiling-programs}). If \t{Model D} succeeds on the task, it could be that our compiler correctness theorem is stronger than necessary to obtain positive effects on learning dynamics. Therefore, we test a model produced from a compiler satisfying a weaker property (\t{Model T}). 

\item \t{Model T} is produced using our compiler, but the coefficients of its compiled linear map are randomized. This breaks the term translation but \textit{preserves the type translation}. If \t{Model T} succeeds, it could be that even type preservation is stronger than necessary. Therefore, we test a model produced without any consideration of compiler correctness (\t{Model I}).

\item \t{Model I} is a convolutional neural network baseline. Since \t{Model D} and \t{Model T} are composed of comparably simple subnetworks, we control for architectural complexity in \t{Model I} so that observed gains can be more confidently attributed to the compiler, which is wiring together simple neural components.
\end{itemize}

\t{Model D} is a neural network similar to Fig. \ref{fig:compiling-iteration}. The iterative structure is \textit{directly} programmed using our compiler.
\vspace{-.1em}
\[
\cm{\iter{x}{y}{\_}{n}}(\xvec:\rd{\R^{784}}, \,\rd{\vec{n}}:\rd{\R^{10}})
\]

\noindent After filling the hole $\nb{\_}$ with a differentiable map $\rd{\t{vblur}}:\rd{\R^{784}}\to\rd{\R^{784}}$, the program compiles according to our formal specification. We use a one-holed context
so that the iterative transformation $\rd{\t{vblur}}$ may be learned. $\rd{\t{vblur}}$ is a linear convolution whose kernel weights are determined by a 3-layer convolutional neural network.
Additionally, the inputs $(\xvec, \rd{\vec{n}})$ are determined by neural networks. $\xvec$ is the base case of iteration, and is simply the original image. $\rd{\vec{n}}$ controls how many times to iterate, up to 10 steps.\footnote{During training it is typical to assume a maximum number of steps the recurrent neural network takes through time.} It is also determined by a 3-layer convolutional neural network.

\t{Model T} is similar to \t{Model D}. But it compiles a simpler program $\cm{\iter{x}{y}{y}{n}}(\rd{\xvec}, \rd{\vec{n}})$ and replaces its $\nb{2}$-linear map with a random matrix $\rd{W}$.\footnote{The exclusion of $\rd{\t{vblur}}$ is necessary, otherwise the dimensionality explodes.}
\vspace{.1em}
\[
\rd{W}(\xvec \otimes \rd{\vec{n}})
\]
\vspace{-1.4em}

\noindent Observe that the tensor product is used to form a linear map from a $\nb{2}$-linear map. It is \textit{type-preserving} because it keeps the type translation from our compiler. However, the random matrix $\rd{W}$ does not generally satisfy the compiler correctness theorem. 

\t{Model I} is a 3-layer convolutional neural network, a simple baseline from which to compare the impacts of programming with and without first-class iteration. It is not directly programmed with any iterative structure, thus we say it is programmed \textit{indirectly} to solve the task.

Further details on model parameters are available in the artifact.

\begin{figure}[h]
  \includegraphics[scale=.43]{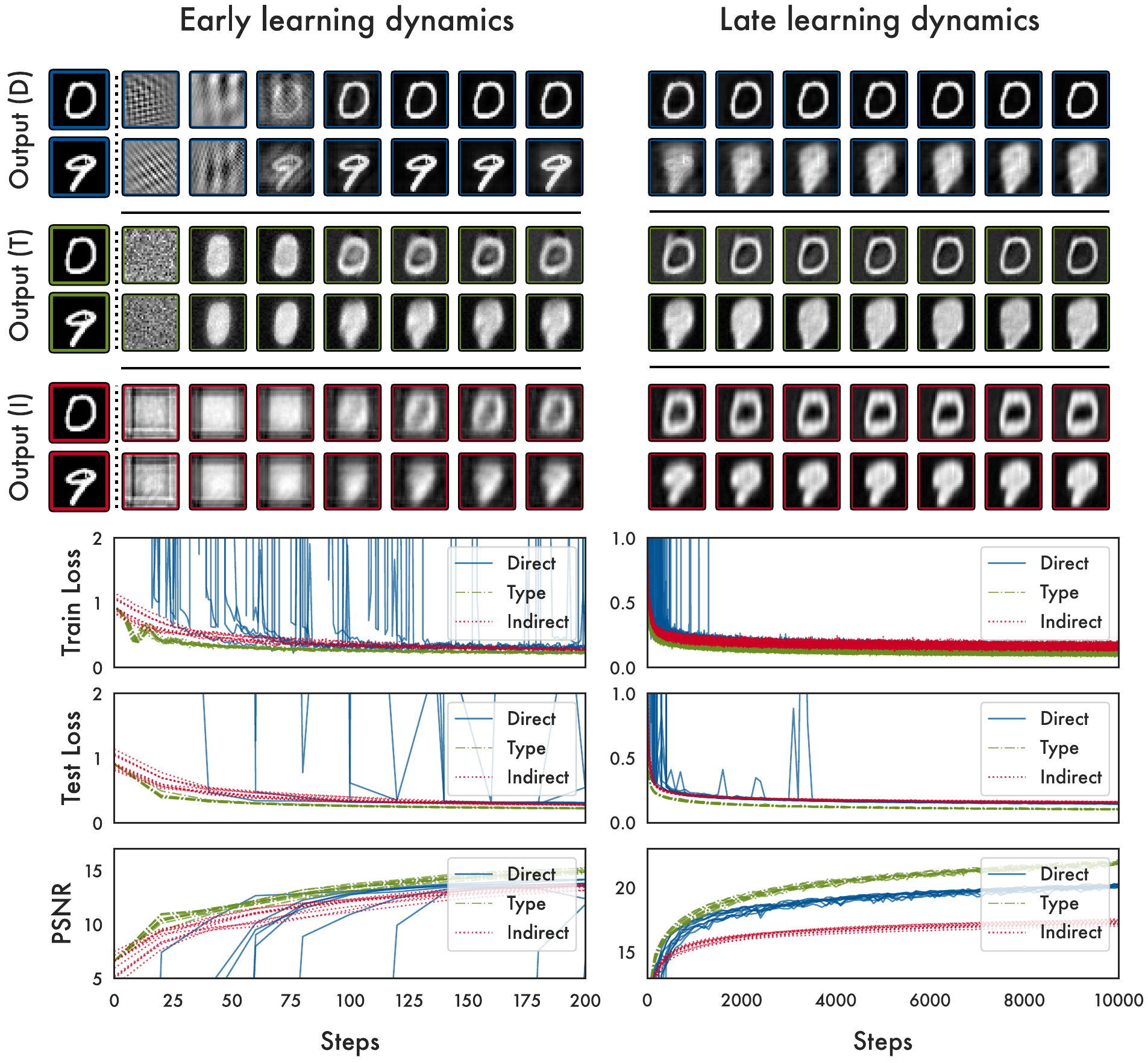}
  \centering

  \vspace{1.6em}
  \caption{Prototypical learning dynamics for Experiment 1}
  \label{fig:local-dynamics}
\end{figure}

\subsubsection{Measures}

The experiments take several measures of each model to help distinguish their learning dynamics. These feature prominently in Figs. \ref{fig:local-dynamics}, \ref{fig:global-dynamics}, \ref{fig:global-dynamics2}, and \ref{fig:local-dynamics2}.

The first measure is the model loss on the training and test set. These models learn by iterative adjustments of their parameters in order to minimize a loss function. The model loss helps us see if model training converges, and to what extent this convergence is stable. However, models can have similar loss curves while behaving in distinct ways. Therefore we use an additional measure that helps distinguish model behavior.

The second measure of learning dynamics is the peak signal-to-noise ratio (PSNR) between a model output and the target output specified in the test set. It is a kind of image distance, and is a useful measure for understanding actual model behavior \cite{hore2010image}. Qualitatively distinct behaviors correlate well with distinct PSNR curves. Higher PSNR indicates two images are \textit{more} similar.

\subsubsection{Results}

Each model was trained to approximate the image transformation specified in a training dataset of 60,000 image inputs and corresponding outputs. They were then tested on a distinct dataset of 10,000 image pairs. Using \t{L1} loss, the \t{Adam} optimizer computed parameter updates to each model \cite{kingma2014adam}. During training, the previously described measures were recorded at regular intervals. This was done for 10 random seeds, 3 learning rates, and 3 training batch sizes. 

Fig. \ref{fig:local-dynamics} presents the typical observed dynamics of learning for each model. On the $x$-axis, a step indicates a parameter update to the models. The $y$-axis contains previously discussed measures, in addition to model outputs across training time. For model outputs, the first image on the left is the input. The data for each seed is shown. To understand both early and late training dynamics, the left-half of the figure focuses on the first $200$ training steps, and the right-half on the first 10,000.

These results show that \t{Model D} and \t{Model T} learn to iteratively apply the image transformation. But \t{Model I} does not. It instead applies a more uniform blur to digits. For reference, the outputs specified in the test set for these inputs are in Fig. \ref{fig:task}. 

At a finer level, \t{Model D} and \t{Model T} learn comparably well, observed by their outputs and PSNR. However, they arrive at similar solutions in distinct ways. \t{Model D} first learns to reconstruct the original image, and then to iteratively apply blur. \t{Model T} instead slowly learns to iteratively blur the original image without passing through a stage where it reconstructs the original image. Interestingly, \t{Model D} is less stable in its early training dynamics. When \rd{\t{vblur}} has a large norm, its iterative application explodes model outputs. Before learning to produce appropriate values of $\rd{\t{vblur}}$, \t{Model T} encounters these unstable inputs early in training.

\begin{figure}[t]
  \includegraphics[scale=.4]{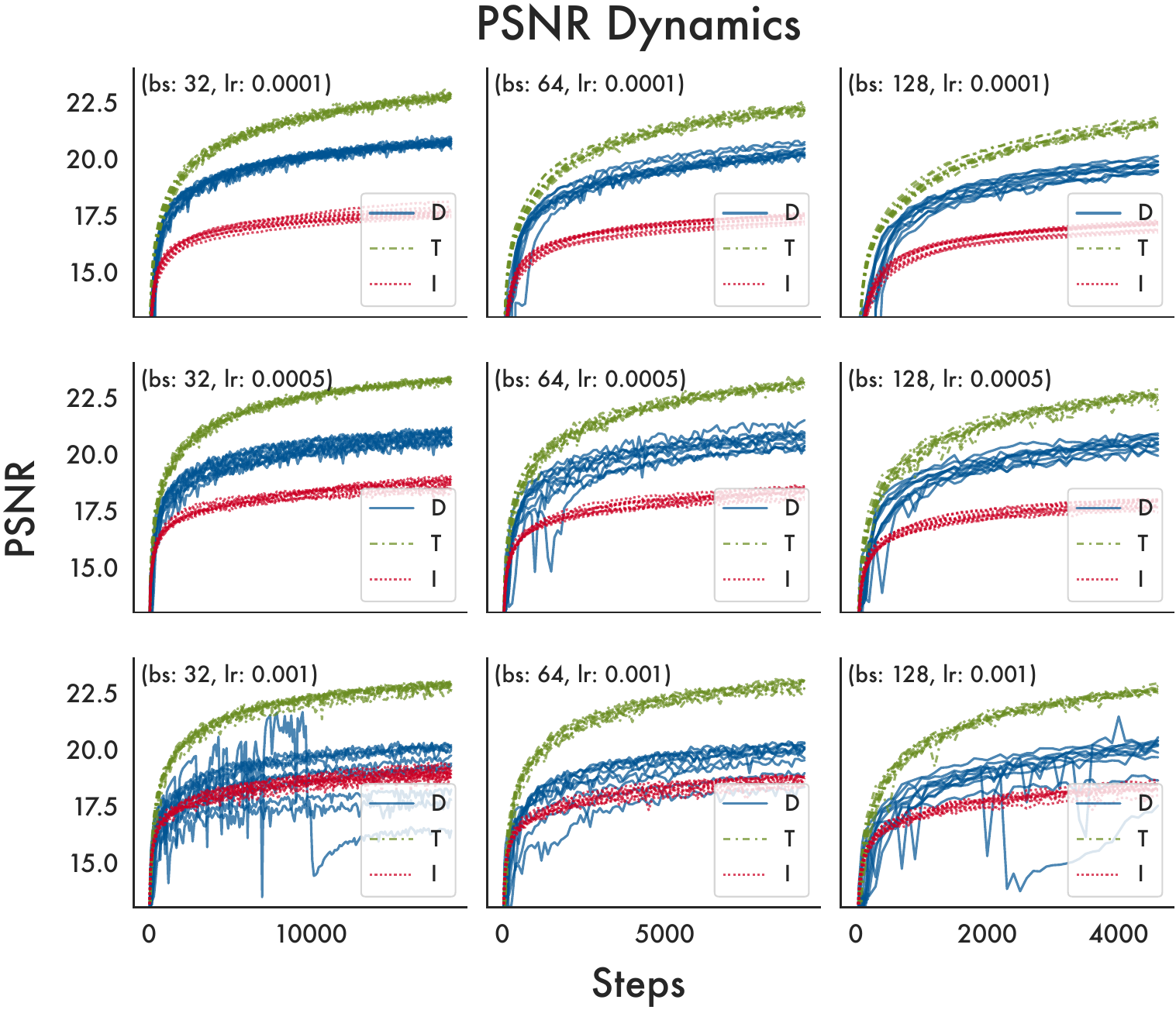}
  \centering
  \vspace{.6em}

  \caption{Summary of dynamics across configurations for Experiment 1}
  \vspace{-.5em}
  \label{fig:global-dynamics}
\end{figure}

Overall, model performance is generally consistent across learning rate and training batch size. Fig. \ref{fig:global-dynamics} shows this using PSNR, the measure which best correlates with model performance.

\subsection{Experiment 2: Conditional and Iterative Image Transformation}

\subsubsection{Task} Our second experiment studies a conditional and iterative image transformation task similar to Fig. \ref{fig:task}. The only difference is that the iterative image transform only applies if the input digit \textit{is odd}.
\vspace{0em}
\setlist[itemize]{topsep=8pt, itemsep=1pt, parsep=2pt, leftmargin=1.5em, rightmargin=1.5em}
\begin{itemize}[label=$\triangleright$]
\item If $\nb{x}$ is an even digit, return $\nb{x}$
\item If $\nb{x}$ is an odd digit $\nb{n}$, apply vertical blur $\nb{n}$ times to $\nb{x}$
\end{itemize}

\begin{figure}[t]
  \includegraphics[scale=.43]{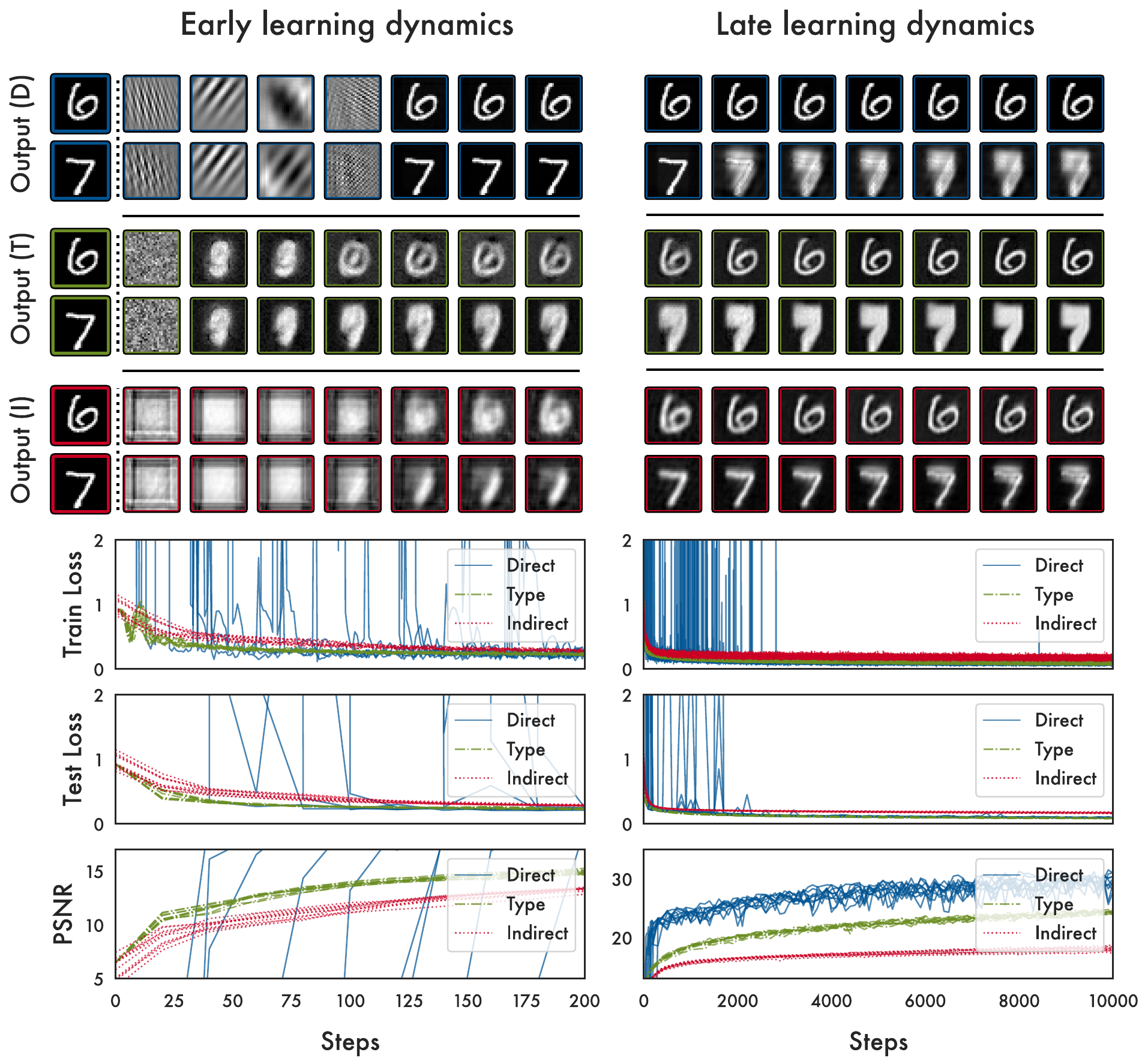}
  \centering

  \vspace{1.3em}
  \caption{Prototypical learning dynamics for Experiment 2}
  \label{fig:local-dynamics2}
\end{figure}

\subsubsection{Models} The models are similar to the first experiment. \t{Model D} is a neural network with conditional and iterative structure \textit{directly} programmed using our compiler.
Note that then-branch uses an identity iterator so that the variable 
$\nb{n}$ is used in each branch, a requirement of linear typing.

\vspace{-1em}
\[
\cm{\ite{\ul{b}}{\ul{\iter{x}{y}{y}{n}}}{\ul{\iter{x}{y}{\_}{n}}}}(\rd{\vec{b}}:\rd{\R^{2}}, \xvec:\rd{\R^{784}}, \,\rd{\vec{n}}:\rd{\R^{10}})
\]

\t{Model T} is similar to \t{Model D}. But compiles $\cm{\ite{\ul{b}}{\ul{x}}{\iter{x}{y}{y}{n}}}(\rd{\vec{b}}, \xvec, \rd{\vec{n}})$ and replaces its $\nb{3}$-linear map with a random matrix $\rd{W}$.
\vspace{0em}
\[
\rd{W}(\rd{\vec{b}} \otimes \xvec \otimes \rd{\vec{n}})
\]
\vspace{-1.2em}

\t{Model I} is the same model from the first experiment.

\subsubsection{Measures} The measures are unchanged from the first experiment.

\subsubsection{Results} The model training and data collection mirror the first experiment, using new training and test datasets (of same size) which specify the conditional and iterative image transformation.

\begin{figure}[t]
\vspace{2em}
  \includegraphics[scale=.4]{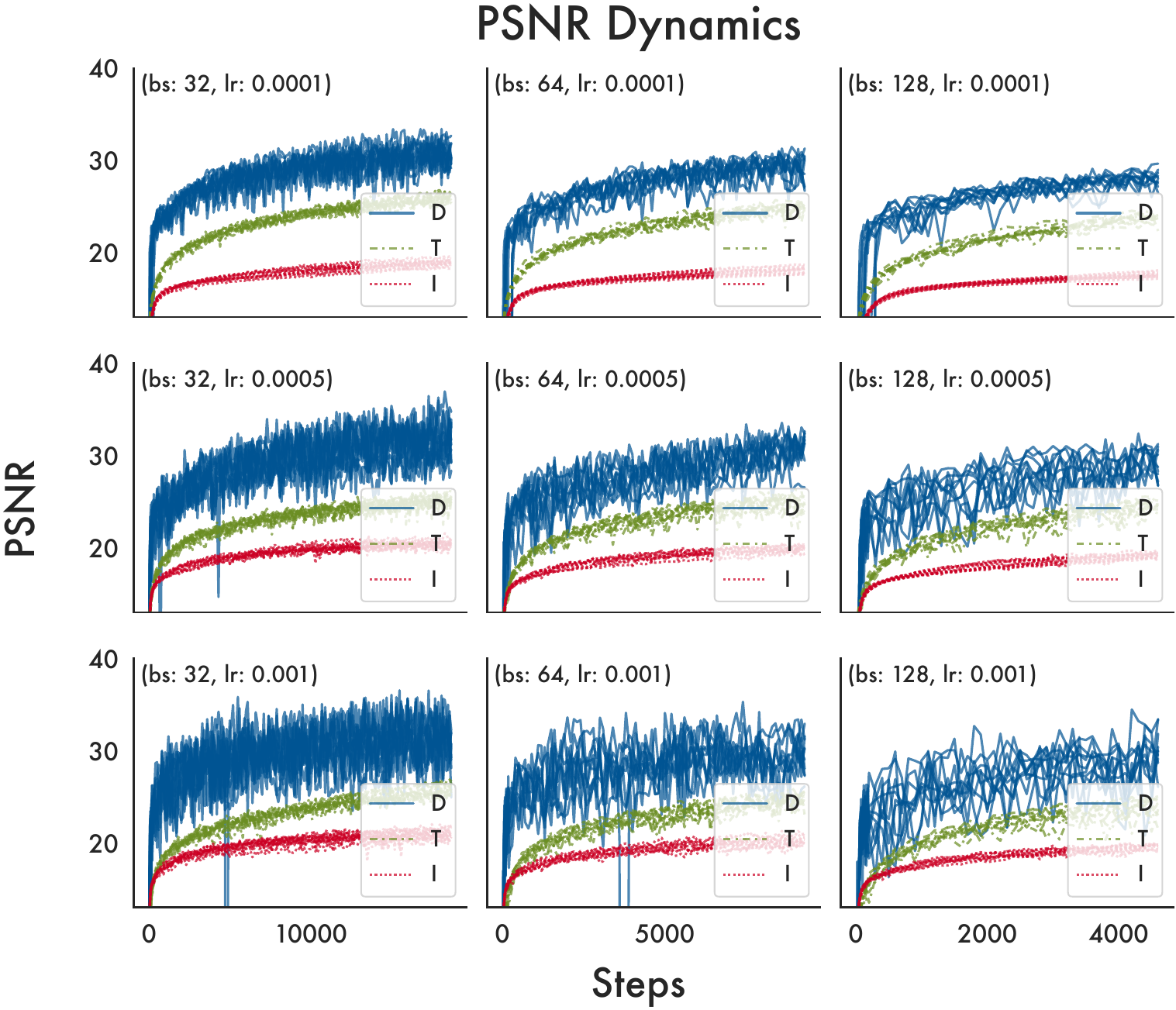}
  \centering
  \vspace{.6em}
    
  \caption{Summary of dynamics across configurations for Experiment 2}
  \vspace{0.0em}
  \label{fig:global-dynamics2}
\end{figure}

Fig. \ref{fig:global-dynamics2} and \ref{fig:local-dynamics2} also mirror the analyses from the first experiment. The trends in the data are also similar. However, \t{Model D} now exhibits the best PSNR. Because it programs in the identity for when an input image is even, if the network can just learn to classify even digits, it will produce correct outputs on many test cases. This drives up its PSNR measure. The quality of its blurring is similar to the first experiment.

\section{Related Work}
\label{sec:related}

$\cj${\scriptsize $(\nb{\lto}, \nb{\Bool}, \nb{\N})$} was deeply inspired by research in neurosymbolic programming, denotational semantics, and differentiable programming. We highlight their key insights while also taking the chance to identify where our work differs.

\subsection{Neurosymbolic Programming}
Neurosymbolic programs combine neural networks with discrete components. There are many approaches, but we focus on those which compile discrete components into differentiable form.

\t{Scallop}, \t{TerpreT}, and \t{Differentiable Forth} are programming languages where discrete components are compiled into differentiable form \cite{li2023scallop, bovsnjak2017programming, gaunt2017differentiable}. Their programming paradigms differ---across logic, imperative, and probabilistic programming. In general, their programs blend neural networks solving perceptual or linguistic tasks with discrete components performing logical tasks, mediated by the compiler as in $\cj${\scriptsize $(\nb{\lto}, \nb{\Bool}, \nb{\N})$}. For example, \t{Scallop} was recently used to write probabilistic logic programs whose inputs may derive from querying a large language model \cite{li2024relational}.  These languages established that this kind of programming is possible for a rich array of tasks, and that it has positive impacts on learning dynamics. Our first intuitions on compiling discrete structure to differentiable form came from these languages.

However, they lack a theory of compiler correctness. No compiler correctness theorems and proofs are available for these languages. As a result, some principles underlying their design are difficult to understand or fix. For example, these languages lack first-class iteration. Before completing our work, it was unclear whether its absence was fundamental. $\cj${\scriptsize $(\nb{\lto}, \nb{\Bool}, \nb{\N})$} offers a simple setting to isolate and resolve these basic language design issues, not only for iteration but also other discrete structures (Section \ref{sec:discussion}).

\subsection{Denotational Semantics}

Linear types originate from linear logic, whose denotational semantics exposed a connection between linear programs and linear maps. These semantics span \textit{decades}. Of special relevance are semantics which explicitly show the connection by computing denotations of programs as linear maps.

Our semantic approach is inspired by work from Valiron and Zdancewic, on categorical models of lambda calculi \cite{valiron2014finite}. They denote programs by linear maps between finite-dimensional vector spaces over finite fields, and their translation is proved sound and adequate. Several examples show programs with discrete structure denoted as matrices---though not over $\rd{\R}$-vector spaces. From these examples, we realized that if we could develop translations to $\rd{\R}$-vector spaces, then we could associate differentiable forms to discrete programs. But it was not clear whether this was possible. Fortunately, we encountered a denotational semantics of linear logic to Köthe sequence spaces \cite{ehrhard2002kothe}. These semantics denote programs by linear maps between \textit{topological} vector spaces over $\rd{\R}$, notably including infinite-dimensional spaces.

However, their motivations differ substantially from ours. As a result, their translation inherits commitments that complicate their use in our setting. For example, Köthe sequence spaces are topological vector spaces. Such spaces are suitable for their aims, but not necessary here. Instead, our work builds on prior efforts in simpler vector spaces, where \cj\ was first introduced to study first-class conditionals in a differentiable programming language \cite{jvg26linear}. We add primitive recursion over natural numbers, and explore a functional---rather than relational---specification of the compiler. Unlike prior work with \cj, we deal with infinite-dimensional vector spaces, and relate these spaces to discrete-time dynamical systems, introducing recurrence (i.e., \textit{time}) as a key mechanism for encoding infinite-dimensional data in neural networks.

\subsection{Differentiable Programming}

\t{PyTorch} and \t{JAX} are differentiable programming languages widely used to build and train neural networks \cite{paszke2019pytorch, jax2018github}. Automatic differentiation in these languages frees programmers from manually implementing derivatives necessary for gradient-based learning in neural networks. They also make it easier to train neural networks at scale; they support specialized hardware which greatly accelerate training. There is growing interest in developing their programming language theory---especially on automatic differentiation \cite{abadi2019simple, elliott2018simple, krawiec2022provably}. These efforts focus on proving that the derivatives computed for real-valued functions via automatic differentiation are correct, and what that means for functions which may internally use conditionals. But they do not study how to associate derivatives to functions whose \textit{inputs} may be discrete structures. Recent work introduced a branch developing this theory, studying the prospect of first-class discrete structure in differentiable programming \cite{jvg26linear}. They show conditionals can be first-class, but defer on more complicated discrete structure. With $\cj${\scriptsize $(\nb{\lto}, \nb{\Bool}, \nb{\N})$}, it is now clear iteration can be first-class. 

Contrasting a language-based approach, there are also specialized neural network architectures for learning discrete algorithms. These contrast the present work, which concerns itself with compiling these algorithms into a neural network \textit{prior to training}. Examples of these architectures include Neural Turing Machines and Neural Programmer Interpreters \cite{graves2014neural, reed2015neural}. They constrain the learning dynamics of networks in such a way that they more easily learn discrete algorithms, but do not leverage program specifications to generate an architecture for a particular task. 

\section{Discussion}
\label{sec:discussion}

Discrete structure is not \textit{overtly} differentiable. However, compilers can translate discrete structure to differentiable form---broadening the class of differentiable algorithms we can express. Our work shows $\cj${\scriptsize $(\nb{\lto}, \nb{\Bool}, \nb{\N})$} can be used to develop these compilers. Its formal specification and metatheory are tractable, and yield a simple implementation for empirical studies on the interplay between learning and the discrete structures of ordinary programming. Our experimental results show that discrete structure can benefit learning, and also expose an important limitation of our specification---it permits compilers which are not \textit{necessarily} stable with respect to learning. We expect stronger specifications which address these issues are possible, and may require including \textit{normalization} layers in the target neurons. Beyond this, our efforts expose an exciting possibility, on whether algebraic data like lists and trees could be first-class in differentiable programming.

We suspect that these structures could be compiled to differentiable form by compiling to linear \textit{recursive} neurons \cite{irsoy2014deep}. With recurrent neurons, their recurrent structure is simple and mirrors the natural numbers. But recursive neurons exhibit hierarchical recurrent structure---similar to the hierarchical structure found in algebraic data like trees. Establishing their formal correspondence would enable differentiable programming with first-class primitive recursion over algebraic data, a powerful programming abstraction. It is our next step.

\section*{Data-Availability Statement}
The artifact supporting this paper is archived on Zenodo \cite{zenodo}.

\bibliographystyle{ACM-Reference-Format}
\bibliography{bibliography}
\clearpage 

\appendix
\newtheorem{definitionx}{Definition}[section] 

\newenvironment{ndef}[1]{%
  \begin{definitionx}[#1]%
}{%
  \end{definitionx}%
}

\section{Definitions}

\subsection{Context Splitting}
\label{app:split}
\vspace{1em}
\begin{center}
\bottomAlignProof
\AxiomC{$\nb{\D} = \nb{\D_1} \,\cup\, \nb{\D_2}$}
\AxiomC{$\nb{\emp} = \nb{\t{dom}}{(\nb{\D_1})} \,\cap\, \nb{\t{dom}}{(\nb{\D_2})}$}
\BinaryInfC{$\nb{\D} = \nb{\D_1} \o \nb{\D_2}$}
\DisplayProof
\end{center}
\vspace{1em}

\subsection{Length of a context}
$$\nb{\t{len}}(\nb{\_}):\nb{\t{Context}}\to\nb{\N}$$\\[-1em]
$$
\begin{aligned}
	&\nb{\t{len}}(\nb{\emp})=\nb{0}\\
	&\nb{\t{len}}(\nb{\D,\bind{x}{\ty}})=\nb{1}\,\nb{+}\,\nb{\t{len}}(\nb{\D})\\
\end{aligned}
$$
\vspace{1em}

\subsection{Index set of context}
$$\nb{\iota}(\nb{\_}):\nb{\t{Context}}\to\{\nb{\N}\}$$\\[-1em]
$$
\nb{\iota}(\nb{\D})=\{\nb{1},...,\nb{\t{len}}(\nb{\D})\}
$$
\vspace{1em}

\subsection{Index of variable in context}
$$\nb{\iota_\_}(\nb{\_}):\nb{\t{Var}}\x \nb{\t{Context}}\to\nb{\N}$$\\[-1em]
$$
\begin{aligned}
&\nb{\iota_x}(\nb{\emp})=\nb{0}\\
&\nb{\iota_x}(\nb{\D,\bind{y}{\ty}})=
  \begin{cases} 
    \t{\nb{len}}(\nb{\D,\bind{y}{\ty}}) & \t{ if }\nb{x=y}\\ 
    \nb{\iota_x}(\nb{\D})& \t{ otherwise}
  \end{cases}
\end{aligned}
$$
\vspace{1em}

\subsection{Context concatenation}
$$\nb{\_ \mid\mid\_}: \nb{\t{Context}} \x \nb{\t{Context}} \to \nb{\t{Context}}$$\\[-1em]
$$
\begin{aligned}
	&\nb{\D_1} \,\nb{\mid\mid}\, (\nb{\emp}) = \nb{\D_1}\\
	&\nb{\D_1} \,\nb{\mid\mid}\, (\nb{\D_2,\bind{x}{\ty}}) = (\nb{\D_1 \mid\mid \D_2}),\nb{\bind{x}{\ty}}
\end{aligned}
$$
\vspace{1em}

\subsection{Restricting a source environment}
\t{If $\nb{\sig}\in\Nv{\Ds1}$, then $(\nb{\d})_\Ds2$ is the restriction of $\nb{\d}$ to the domain of $\nb{\Ds2}$,}\vspace{1em}
$$
\begin{aligned}
& (\nb{\d})_\nb{\emp}=\nb{\emp}\\
& (\nb{\d})_{\nb{\Ds2,x:\ty}} = 
    \begin{cases}
    (\nb{\d})_\Ds2\nb{\{x \map \d(x)\}} & \t{if }\nb{x} \in \nb{\t{dom}}(\nb{\s}) \\ 
    (\nb{\d})_\Ds2 & \t{otherwise}
    \end{cases}
\end{aligned}
$$
\vspace{1em}

\subsection{Restricting a target environment}
\t{If $\svec \in \cm{\D_1}$ and $\nb{\t{set}}(\Ds2) \subseteq \nb{\t{set}}(\Ds1)$, then $(\svec)^{\Ds1}_{\Ds2} \in \cm{\D_2}$ is the restructuring of $\svec$ according to $\cm{\D_2}$. When the source context $\Ds1$ is clear from context we use the notation $\svec_\Ds2$ as shorthand.}\vspace{1em}
$$
\begin{aligned}
& (\svec)^{\Ds1}_{\nb{\emp}} = \rd{\{0\}}\\
& (\svec)^{\Ds1}_{\Ds2,\nb{x:\ty}} = (\svec)^{\Ds1}_\Ds2 \,\rd{\x}\, \rd{\pi}_{\nb{\iota_x}(\Ds1)}(\svec)
\end{aligned}
$$
\vspace{1em}

\section{Proofs}

\subsection{Theorem (Programs evaluate to values)}
\vspace{.7em}
$$\p{\emp}{e}{\ty} \implies \exists \nb{v},\step{e}{v}$$
\begin{proof}\leavevmode\\
\t{
Because programs can be logically typed $\lp{\emp}{e}{\ty}$\\
By empty substitution and logical typing $\nb{\emp(e)}=\nb{e}\in\E{\ty}$\\
Because logically typed programs terminate $\exists \nb{v},\step{e}{v}$ 
}
\end{proof}
\vspace{1em}

\subsection{Lemma (Programs can be logically typed)}
\vspace{.7em}
$$\p{\D}{e}{\ty} \implies \lp{\D}{e}{\ty}$$
\begin{proof}\leavevmode\\
\t{By induction on $\p{\D}{e}{\ty}$,
\setlist[itemize]{topsep=4pt, itemsep=4pt, parsep=2pt, leftmargin=1.5em, rightmargin=1.5em}
\begin{itemize}[label=$\triangleright$]
\item \textbf{Case} $\p{\emp}{\z}{\N}$\\[.25em]
	\textbf{Show} $\nb{\z} \in \E{\N}$\\
	Consider each condition,
    \begin{itemize}[label=$\triangleright$]
	   \item \textbf{Condition} $\exists \nb{v},\step{\z}{v}$\\[.25em]
		  Let $\exists \nb{v}=\nb{\z}$\\
		By evaluation $\nb{\z} \bstep \nb{\z}$
	   \item \textbf{Condition} $\nb{\z} \in \V{\N}$\\[.25em]
		By logical relation $\nb{\z} \in \V{\N}$
    \end{itemize}
\item \textbf{Case} $\p{\D}{\suc{e}}{\N}$\\[.25em]
	\textbf{Show} $\nb{\suc{\d(e)}} \in \E{\N}$\\
	Consider each condition,
    \begin{itemize}[label=$\triangleright$]
	   \item \textbf{Condition} $\exists \nb{v}, \step{\suc{\d(e)}}{v}$\\[.25em]
		By assumption $\p{\D}{e}{\N}$\\
		By induction $\lp{\D}{e}{\N}$\\
		By logical relation $\step{\d(e)}{v}$ where $\nb{v} \in \V{\N}$\\
		Let $\exists \nb{v}=\suc{v}$\\
		By evaluation $\step{\suc{\d(e)}}{\suc{v}}$ 
	   \item \textbf{Condition} $\suc{v} \in \V{\ty_2}$\\[.25em]
		Recall by logical relation $\step{\d(e)}{v}$ where $\nb{v} \in \V{\N}$\\
		By logical relation $\suc{v} \in \V{\N}$
    \end{itemize}
\item \textbf{Case} $\p{\bind{x}{\ty}}{x}{\ty}$\\[.25em]
	\textbf{Show} $\nb{\{x \map v\}(x)}\in\E{\ty}$ where $\nb{\{x \map v\}}\in\Nv{\bind{x}{\ty}}$ \\
	Consider each condition,
    \begin{itemize}[label=$\triangleright$]
	\item \textbf{Condition} $\exists \nb{v_1}, \step{\{x \map v\}(x)}{v_1}$\\[.25em]
		By substitution $\nb{\{x \map v\}(x)}=\nb{v}$ \\
		Let $\exists \nb{v_1}=\nb{v}$ \\
		By evaluation $\step{v}{v}$ 
	\item \textbf{Condition} $\nb{v} \in \V{\ty}$ \\[.25em]
		By logical relation $\nb{\{x \map v\}}\in\Nv{\bind{x}{\ty}}$ where $\nb{v} \in \V{\ty}$ 
    \end{itemize}
\item \textbf{Case} $\p{\D}{\lam{y}{e}}{\ty_1 \lto \ty_2}$\\[.25em]
	\textbf{Show} $\nb{\lam{y}{\d(e)}}\in\E{\ty_1 \lto \ty_2}$ where $\nb{\d}\in\Nv{\D}$ \\
	Consider each condition,
    \begin{itemize}[label=$\triangleright$]
	\item \textbf{Condition} $\exists \nb{v}, \step{\lam{y}{\d(e)}}{v}$\\[.25em]
		Let $\exists \nb{v}=\nb{\lam{y}{\d(e)}}$ \\
		By evaluation $\step{\lam{y}{\d(e)}}{\lam{y}{\d(e)}}$
	\item \textbf{Condition} $\nb{\lam{y}{\d(e)}}\in\V{\ty_1\lto\ty_2}$\\[.25em]
		\textbf{Show} $\nb{(\lam{y}{\d(e)})v_1} \in \E{\ty_2}$ \\
		By assumption $\nb{v_1} \in \V{\ty_1}$\\
		Consider each condition,
        \begin{itemize}[label=$\triangleright$]
		\item \textbf{Condition} $\exists \nb{v}, \step{(\lam{y}{\d(e)})v_1}{v}$ \\[.25em]
			By assumption $\p{\D,\bind{y}{\ty_1}}{e}{\ty_2}$ \\
			By induction $\lp{\D,\bind{y}{\ty_1}}{e}{\ty_2}$ i.e. $\nb{\d\{y \map v_1\}(e)} \in \E{\ty_2}$ \\
			By logical relation $\step{\d\{y \map v_1\}(e)}{v}$ where $\nb{v} \in \V{\ty_2}$ \\
			Because substitutions commute $\nb{\d\{y \map v_1\}(e)}=\nb{\{y \map v_1\}(\d(e))}$\\ 
			Let $\exists \nb{v}=\nb{v}$ \\
			By evaluation $\step{(\lam{y}{\d(e)})v_1}{v}$ 
		\item \textbf{Condition} $\nb{v} \in \V{\ty_2}$ \\[.25em]
			By logical relation $\step{\{y \map v_1\}(\d(e))}{v}$ where $\nb{v} \in \V{\ty_2}$
        \end{itemize}
    \end{itemize}
\item \textbf{Case} $\p{\D_1 \o \D_2}{e_1e_2}{\ty_2}$\\[.25em]
	\textbf{Show} $\nb{\d(e_1)\d(e_2)} \in \E{\ty_2}$\\
	Consider each condition,
    \begin{itemize}[label=$\triangleright$]
	\item \textbf{Condition} $\exists \nb{v_\bullet},\step{\d(e_1)\d(e_2)}{v_\bullet}$ \\[.25em]
		By induction $\lp{\D_1}{e_1}{\ty_1 \lto \ty_2}$\\
		By induction $\lp{\D_2}{e_2}{\ty_1}$ \\
		By logical relation $\step{\d_\Ds1(e_1)}{\lam{x}{e}}$ where $\nb{\lam{x}{e}}\in\V{\ty_1 \lto \ty_2}$\\
		By logical relation $\step{\d_\Ds2(e_2)}{v_2}$ where $\nb{v_2}\in\V{\ty_1}$ \\
		By logical relation $\step{(\lam{x}{e})v_2}{v}$ where $\nb{v}\in\V{\ty_2}$ \\
		Note that $\nb{\d_\Ds1}(\ess1)=\nb{\d}(\ess1)$ and $\nb{\d_\Ds2}(\ess2)=\nb{\d}(\ess2)$ \\
		Let $\exists \nb{v_\bullet}=\nb{v}$ \\
		By evaluation $\step{\d(e_1)\d(e_2)}{v}$
	\item \textbf{Condition} $\nb{v} \in\V{\ty_2}$ \\[.25em]
		By logical relation $\step{(\lam{x}{e})v_2}{v}$ where $\nb{v}\in\V{\ty_2}$
    \end{itemize}
\item \textbf{Case} $\p{\D_1 \o \D_2}{\ite{e_1}{e_2}{e_3}}{\ty}$\\[.25em]
	\textbf{Show} $\nb{\ite{\d(e_1)}{\d(e_2)}{\d(e_3)}} \in \E{\ty_2}$\\
	Consider each condition,
    \begin{itemize}[label=$\triangleright$]
	\item \textbf{Condition} $\exists \nb{v_\bullet},\step{\nb{\ite{\d(e_1)}{\d(e_2)}{\d(e_3)}}}{v_\bullet}$\\[.25em]
		By induction $\lp{\D_1}{e_1}{\Bool}$\\
		By induction $\lp{\D_2}{e_2}{\ty}$ and $\lp{\D_2}{e_3}{\ty}$ \\
		By logical relation $\step{\d_\Ds1(e_1)}{v}$ where $\nb{v}\in\V{\Bool}=\{\tt,\ff\}$ \\
		Consider whether $\nb{v}=\tt$ or $\nb{v}=\ff$,
        \begin{itemize}[label=$\triangleright$]
		\item \textbf{Case} $\nb{v}=\tt$\\[.25em]
			By logical relation $\step{\d_\Ds2(e_2)}{v_2}$ where $\nb{v_2}\in\V{\ty}$ \\
			Note that $\nb{\d_\Ds1}(\ess1)=\nb{\d}(\ess1)$ and $\nb{\d_\Ds2}(\ess2)=\nb{\d}(\ess2)$ \\
			Let $\exists \nb{v_\bullet}=\vs2$ \\
			By evaluation $\step{\nb{\ite{\d(e_1)}{\d(e_2)}{\d(e_3)}}}{v_2}$\\
			\item \textbf{Condition} $\nb{v_2} \in\V{\ty}$ \\[.25em]
				Recall by logical relation $\nb{v_2}\in\V{\ty}$ 
		\item \textbf{Case} $\nb{v}=\ff$\\[.25em]
			Similar to previous case	
        \end{itemize}
    \end{itemize}
\item \textbf{Case} $\p{\D_1 \o \D_3}{\iter{e_1}{y}{e_2}{e_3}}{\ty}$\\[.25em]
	\textbf{Show} $\nb{\iter{\d(e_1)}{y}{e_2}{\d(e_3)}} \in \E{\ty}$ \\
	By assumption $\p{\D_3}{e_3}{\N}$ \\
	By induction $\lp{\D_3}{e_3}{\N}$ \\
	By logical relation $\step{\d_\Ds3(e_3)}{v_3}$ where $\nb{v_3}\in\V{\N}=\{\z\} \cup \{ \suc{v_\bullet} \mid \nb{v_\bullet} \in \V{\N}\}$\\
	It suffices to show $\nb{\iter{\d(e_1)}{y}{e_2}{v_3}} \in \E{\ty}$ by backward closure of logical relation\\
	By induction on $\nb{v_3}$ where,
    \begin{itemize}[label=$\triangleright$]
	\item \textbf{Case} $\nb{v_3} =\z$\\[.25em]
		Consider each condition,
        \begin{itemize}[label=$\triangleright$]
		\item \textbf{Condition} $\exists \nb{v}, \step{\iter{\d(e_1)}{y}{e_2}{\z}}{v}$\\[.25em]
			By assumption $\p{\bind{y}{\ty}}{e_2}{\ty}$ and $\p{\D_1}{e_1}{\ty}$ \\
			By induction $\lp{\bind{y}{\ty}}{e_2}{\ty}$ and $\lp{\D_1}{e_1}{\ty}$ \\
			By logical relation $\step{\sig_\Ds1(e_1)}{v_1}$ where $\nb{v_1}\in\V{\ty}$ \\
			Note that $\nb{\d_\Ds1}(\ess1)=\nb{\d}(\ess1)$\\
			Let $\exists \nb{v}=\nb{v_1}$ \\
			By evaluation $\step{\iter{\d(e_1)}{y}{e_2}{\z}}{v_1}$ \\
		\item \textbf{Condition} $\nb{v_1}\in\V{\ty}$ \\[.25em]
			Recall by logical relation $\step{\d(e_1)}{v_1}$ where $\nb{v_1}\in\V{\ty}$ 
        \end{itemize}
	\item \textbf{Case} $\nb{v_3} = \suc{v_\bullet}$\\[.25em]
		Consider each condition,
        \begin{itemize}[label=$\triangleright$]        
		\item \textbf{Condition} $\exists \nb{v}, \step{\iter{\d(e_1)}{y}{e_2}{v_3}}{v}$\\[.25em]
			By assumption $\nb{v_\bullet}\in \V{\N}$ \\
			By induction $\lp{\D_1}{e_1}{\ty}$ and $\lp{\bind{y}{\ty}}{e_2}{\ty}$\\
			By induction $\nb{\iter{\d(e_1)}{y}{e_2}{v_\bullet}} \in \E{\ty}$\\
			By logical relation $\step{\nb{\iter{\d_\Ds1(e_1)}{y}{e_2}{v_3}}}{v_\t{n}}$ \\
			By logical relation $\step{\{y \map v_\t{n}\}(e_2)}{v_2}$ where $\nb{v_2} \in \V{\ty}$ \\
			By logical relation $\step{\d_1(e_1)}{v_1}$ where $\nb{v_1} \in \V{\ty}$\\
			Note that $\nb{\d_\Ds1}(\ess1)=\nb{\d}(\ess1)$ and $\nb{\d_\Ds3}(\ess3)=\nb{\d}(\ess3)$\\
			Let $\exists \nb{v} = \nb{v_2}$\\
			By evaluation $\step{\iter{\sig(e_1)}{y}{e_2}{\sig(e_3)}}{v_2}$ 
		\item \textbf{Condition} $\nb{v} \in \V{\ty}$\\[.25em]
			Recall by logical relation $\step{\{y \map v_\t{n}\}(e_2)}{v_2}$ where $\nb{v_2} \in \V{\ty}$
        \end{itemize}
    \end{itemize}
\end{itemize}
}\end{proof}

\subsection{Lemma (Logically typed programs evaluate to values)}
\t{Where $\nb{\d}\in\Nv{\D}$,
$$\lp{\D}{e}{\ty} \implies \exists \nb{v},\step{\d(e)}{v}$$}
\begin{proof}\leavevmode\\
\t{By logical typing $\nb{\d(e)}\in\E{\ty}$ \\
By logical relation $\exists \nb{v},\step{\d(e)}{v}$}
\end{proof}
\vspace{1em}

\subsection{Theorem (Compiler preserves program behavior)}
\label{app:ccc}
\t{If $\p{\emp}{e}{\nb{\t{base}}(\nb{\ty})}$, then
\begin{flalign*}
&\begin{aligned}
(a) \; & \step{e}{v}  \implies \cm{e}=\cm{v} \\
(b) \; & \nb{e} \;\cancel{\bstep}\; \nb{v} \implies \cm{e}\neq\cm{v}\\
\end{aligned}&&
\end{flalign*} }

\begin{proof}\leavevmode\\
\t{
Consider each condition,
\setlist[itemize]{topsep=4pt, itemsep=4pt, parsep=2pt, leftmargin=1.5em, rightmargin=1.5em}
\begin{itemize}[label=$\triangleright$]
\item \textbf{Condition} $(a)$,\\[.25em]
	Because compiler preserves what programs will do $\cm{e}=\cm{v}$
\item \textbf{Condition} $(b)$,\\[.25em]
	Because compiler preserves what programs won't do $\cm{e} \neq \cm{v}$
\end{itemize}}
\end{proof}
\vspace{1em}

\subsection{Lemma (Compiler preserves what programs will do)}
\vspace{.7em}
$$\p{\emp}{e}{\ty} \,\t{ and }\, \step{e}{v} \implies \cm{e}=\cm{v}$$
\begin{proof}\leavevmode\\
\t{By induction on $\step{e}{v}$,
\setlist[itemize]{topsep=4pt, itemsep=4pt, parsep=2pt, leftmargin=1.5em, rightmargin=1.5em}
\begin{itemize}[label=$\triangleright$]
\item \textbf{Case} $\step{v}{v}$\\[.25em]
	Because compiler is deterministic $\c{\nb{v}}=\c{\nb{v}}$
\item \textbf{Case} $\step{\suc{e}}{\suc{v}}$\\[.25em]
	By assumption $\typ{\suc{e}}{\N}$\\
	By inversion on typing $\typ{e}{\N}$\\
	 \[\begin{aligned}
	 &\,\,\;\;\; \c{\suc{e}}\\
	 &= \rd{n} \map \begin{cases}\rd{0} & \t{if } \rd{n=0}\\ \c{\nb{e}}(\rd{n-1}) & \t{otherwise}\end{cases} && \t{By compiling}\\ 
	 &= \rd{n} \map \begin{cases}\rd{0} & \t{if } \rd{n=0}\\ \c{\nb{v}}(\rd{n-1}) & \t{otherwise}\end{cases} && \t{By induction} \\ 
	 &= \c{\suc{v}} && \t{By compiling}
	 \end{aligned}\]
\item \textbf{Case} $\step{e_1e_2}{v}$\\[.25em]
	By assumption $\step{e_1}{\lam{x}{e}}$ and $\step{e_2}{v_2}$ and $\step{\{x \map v_2\}(e)}{v}$\\
	By assumption $\typ{e_1e_2}{\ty_2}$\\
	By inversion on typing $\typ{e_1}{\ty_1 \lto \ty_2}$ and $\typ{e_2}{\ty_1}$ \\
	Because evaluation preserves typing $\typ{\lam{x}{e}}{\ty_1\lto\ty_2}$ and $\typ{v_2}{\ty_1}$\\
	By inversion on typing $\p{\bind{x}{\ty_1}}{e}{\ty_2}$ \\
	Because closing substitutions preserve typing $\typ{\{x \map v_2\}(e)}{\ty_2}$ \\
	$$\begin{aligned}
					&\,\,\;\;\;\, \cm{e_1e_2} \\
		            &= \cm{e_1}(\cm{e_2}) && \t{By compiling}\\
				    &= \cm{\lam{x}{e}}(\cm{v_2}) &&\t{By induction}\\
				    &= (\rd{\vec{x}} \map \cm{e}{(\rd{\vec{x}})})(\cm{v_2}) && \t{By compiling}\\
				    &= \cm{e}({\cm{v_2}}) && \t{By function application}\\
				    &= \cm{\{x \map v_2\}(e)} && \t{Compiling commutes with substitution}\\
				    &= \cm{v} && \t{By induction}
	\end{aligned}$$\\
\item \textbf{Case} $\step{\ite{e_1}{e_2}{e_3}}{v_2}$\\[.25em]
	By assumption $\step{e_1}{\tt}$ and $\step{e_2}{v_2}$\\
	By assumption $\typ{\ite{e_1}{e_2}{e_3}}{\ty}$\\
	By inversion on typing $\typ{e_1}{\Bool}$ and $\typ{e_2}{\ty}$\\
	$$\begin{aligned}
					&\,\,\;\;\;\, \cm{\ite{e_1}{e_2}{e_3}} \\
		            &= \rd{\pi_1}(\cm{e_1})\rt\cm{e_2}\rp\rd{\pi_2}(\cm{e_1})\rt\cm{e_3} && \t{By compiling}\\
		            &= \rd{\pi_1}(\cm{\tt})\rt\cm{v_2}\rp\rd{\pi_2}(\cm{\tt})\rt\cm{e_3} && \t{By induction}\\
		            &= \rd{1}\rt\cm{v_2}\rp\rd{0}\rt\cm{e_3} && \t{By compiling}\\
				    &= \cm{v_2} &&\t{By scalar multiplication}\\
	\end{aligned}$$\\
\item \textbf{Case} $\step{\ite{e_1}{e_2}{e_3}}{v_2}$\\[.25em]
	Similar to previous case\\
\item \textbf{Case} $\step{\iter{e_1}{y}{e_2}{e_3}}{v_1}$\\[.25em]
	By assumption $\step{e_1}{v_1}$ and $\step{e_3}{\z}$\\
	By assumption $\typ{\iter{e_1}{y}{e_2}{e_3}}{\ty}$ \\
	By inversion on typing $\typ{e_1}{\ty}$ and $\typ{e_3}{\N}$\\[-1em]
    \[\scalebox{.8}{$\begin{aligned}
			 &\,\,\;\;\;\, \cm{\iter{e_1}{y}{e_2}{e_3}}\\[.5em]
			 &=\rd{\sum_{n\in\N}}\,\cm{e_3}(\rd{n})\,\rd{\.}\,\cm{e_2}^\rd{n}(\cm{e_1}) && \t{By compiling}\\
	 &= \cm{e_3}(\rd{0})\,\rd{\.}\,\cm{e_2}^\rd{0}(\cm{e_1}) \rp \rd{\sum_{n\in\N\setminus0}}\,\cm{e_3}(\rd{n})\,\rd{\.}\,\cm{e_2}^\rd{n}(\cm{e_1}) && \t{Because }\rd{\sum_{n\in\N}}\,\rd{f}(\rd{n})=\rd{f}(\rd{0})\rp \rd{\sum_{n\in\N\setminus0}}\,\rd{f}(\rd{n})\\
     &= \cm{\z}(\rd{0})\,\rd{\.}\,\cm{e_2}^\rd{0}(\cm{v_1}) \rp \rd{\sum_{n\in\N\setminus0}}\,\cm{\z}(\rd{n})\,\rd{\.}\,\cm{e_2}^\rd{n}(\cm{v_1}) && \t{By induction}\\
			 &= \rd{1}\,\rd{\.}\,\cm{e_2}^\rd{0}(\cm{v_1}) \rp \rd{\sum_{n\in\N\setminus0}}\,\cm{\z}(\rd{n})\,\rd{\.}\,\cm{e_2}^\rd{n}(\cm{v_1}) &&\t{By compiling}\\
			 &= \rd{1}\,\rd{\.}\,\cm{v_1} \rp \rd{\sum_{n\in\N\setminus0}}\,\cm{\z}(\rd{n})\,\rd{\.}\,\cm{e_2}^\rd{n}(\cm{v_1}) &&\t{Because }\rd{f}^\rd{0}(\rd{v})=\rd{v}\\
			 &= \cm{v_1} \rp \rd{\sum_{n\in\N\setminus0}}\,\rd{0}\,\rd{\.}\,\cm{e_2}^\rd{n}(\cm{e_1}) && \t{Because }\cm{\z}(\rd{n})=\rd{0} \t{ if }\rd{n}\neq \rd{0} \\
			 &= \cm{v_1} &&\t{By addition}
	\end{aligned}$}\]\\
\item \textbf{Case} $\step{\iter{e_1}{y}{e_2}{e_3}}{v}$\\[.25em]
	By assumption $\step{\iter{e_1}{y}{e_2}{v_3}}{v_\t{n}}$ and $\step{\{y \map v_\t{n}\}(e_2)}{v}$ and $\step{e_3}{\suc{v_3}}$\\
	By assumption $\typ{\iter{e_1}{y}{e_2}{e_3}}{\ty}$ \\
	By inversion on typing $\typ{e_1}{\ty}$ and $\p{\bind{y}{\ty}}{e_2}{\ty}$ and $\typ{e_3}{\N}$ \\
	Because evaluation preserves typing $\typ{\suc{v_3}}{\N}$ \\
	By inversion on typing $\typ{v_3}{\N}$\\
	By typing $\typ{\iter{e_1}{y}{e_2}{v_3}}{\ty}$ \\
	Because evaluation preserves typing $\typ{v_\t{n}}{\ty}$ \\
	Because substitutions preserve typing $\typ{\{y \map v_\t{n}\}(e_2)}{\ty}$\\
	\[\scalebox{.8}{$\begin{aligned}
	  &\,\,\;\;\;\,\cm{\iter{e_1}{y}{e_2}{e_3}}\\[.5em]
	  &=\rd{\sum_{n\in\N}}\,\cm{e_3}(\rd{n})\,\rd{\.}\,\cm{e_2}^\rd{n}(\cm{e_1}) && \t{By compiling}\\
		 &= \rd{\sum_{n\in\N}}\,\cm{e_3}(\rd{n})\,\rd{\.}\,\cm{e_2}(\cm{e_2}^\rd{(n-1)}(\cm{e_1})) && \t{Because }\rd{f}^\rd{n}(\rd{v})=\rd{f}(\rd{f}^{\rd{(n-1)}}(\rd{v}))\\
		 &= \rd{\sum_{n\in\N}}\,\cm{\suc{v_3}}(\rd{n})\,\rd{\.}\,\cm{e_2}(\cm{e_2}^\rd{(n-1)}(\cm{e_1})) && \t{By induction}\\
	 &= \cm{\suc{v_3}}(\rd{0})\,\rd{\.}\,\cm{e_2}^\rd{0}(\cm{e_1}) \rp \rd{\sum_{n\in\N\setminus0}}\,\cm{{\suc{v_3}}}(\rd{n})\,\rd{\.}\,\cm{e_2}(\cm{e_2}^\rd{(n-1)}(\cm{e_1})) && \t{Because }\rd{\sum_{n\in\N}}\,\rd{f}(\rd{n})=\rd{f}(\rd{0})\rp \rd{\sum_{n\in\N\setminus0}}\,\rd{f}(\rd{n})\\
	  &= \rd{0}\,\rd{\.}\,\cm{e_2}^\rd{0}(\cm{e_1}) \rp \rd{\sum_{n\in\N\setminus0}}\,\cm{{v_3}}(\rd{n-1})\,\rd{\.}\,\cm{e_2}(\cm{e_2}^\rd{(n-1)}(\cm{e_1})) && \t{By compiling}\\
		  &= \rd{\sum_{n\in\N\setminus0}}\,\cm{{v_3}}(\rd{n-1})\,\rd{\.}\,\cm{e_2}(\cm{e_2}^\rd{(n-1)}(\cm{e_1})) && \t{By scalar multiplication}\\ 
		 &= \rd{\sum_{n\in\N\setminus0}}\,\cm{e_2}(\cm{{v_3}}(\rd{n-1})\,\rd{\.}\,\cm{e_2}^{\rd{(n-1)}}(\cm{e_1})) && \t{Compiling maps to multilinear maps}\\ 
		 &= \cm{e_2}(\rd{\sum_{n\in\N\setminus0}}\,\cm{{v_3}}(\rd{n-1})\,\rd{\.}\,\cm{e_2}^{\rd{(n-1)}}(\cm{e_1})) && \t{Compiling maps to multilinear maps}\\
		 &= \cm{e_2}\left(\,\rd{\sum_{n\in\N}}\,\cm{{v_3}}(\rd{n})\,\rd{\.}\,\cm{e_2}^{\rd{n}}(\cm{e_1})\right)&& \t{Because }\rd{\sum_{n\in\N\setminus0}\rd{f}}(\rd{n-1})=\rd{\sum_{n\in\N}\rd{f}}(\rd{n})\\[.5em]
		&= \cm{e_2}(\cm{\iter{e_1}{y}{e_2}{v_3}}) && \t{By compiling}\\
		&= \cm{e_2}(\cm{v_\t{n}}) && \t{By induction}\\
		&= \cm{\{y \map v_\t{n}\}(e_2)} && \t{Compiling commutes with substitution}\\
		&= \cm{v} && \t{By induction}
	\end{aligned}$}\]
\end{itemize}
}\end{proof}
\vspace{1em}

\subsection{Lemma (Compiler preserves what programs won't do)}
\vspace{.7em}
$$\p{\emp}{e}{\nb{\t{base}}\bl{(}\ty\bl{)}} \,\t{ and }\,  \nb{e} \;\cancel{\bstep}\; \nb{v} \implies \cm{e}\neq\cm{v}$$
\begin{proof}\leavevmode\\
\t{Closed programs produce values, thus $\step{e}{v_1}$ where $\nb{v_1} \neq \nb{v}$. Now consider the boolean case. Because of canonical forms $\nb{v}$ and $\nb{v_1}$ are either $\nb{\tt}$ or $\nb{\ff}$, but cannot be equal. Without loss of generality, suppose $\nb{v}=\nb{\ff}$ and $\nb{v_1}=\nb{\tt}$. By negation $\cm{e}=\cm{v}$, but the compiler preserves what programs do. Therefore we have a contradiction. 
$$\cm{e}=\cm{\tt}=\ttv=\cm{\ff}=\ffv\vspace{.3em}$$
A similar argument holds for the natural numbers case.
}
\end{proof}
\vspace{1em}

\subsection{Lemma (Compiler maps programs to multilinear maps)}
\vspace{.7em}
$$\p{\D,\bind{x}{\ty_1}}{e}{\ty_2} \implies \cm{e}(\svec,\a1\rt\xv1\rp\a2\rt\xv2)=\a1\rt\cm{e}(\svec,\xv1)\rp \a2\rt\cm{e}(\svec,\xv2)$$
\begin{proof}\leavevmode\\
\t{Immediate from compiler mapping programs to additive and homogeneous maps.}
\end{proof}

\subsection{Lemma (Compiler maps programs to additive maps)}
\vspace{.7em}
$$\p{\D,\bind{x}{\ty_1}}{e}{\ty_2} \implies \cm{e}(\svec,\xv1\rp\xv2)=\cm{e}(\svec,\xv1)\rp\cm{e}(\svec,\xv2)$$
\begin{proof}\leavevmode\\
\t{By induction on $\nb{e}$,
\begin{itemize}[label=$\triangleright$]
\item \textbf{Case} $\ys$\\[.25em]
	Consider whether $\xs=\ys$,
    \begin{itemize}[label=$\triangleright$]
	\item \textbf{Case} $\xs=\ys$\\[.25em]
		By inversion $\nb{\D}=\nb{\emp}$
		$$\begin{aligned}
			&\;\;\;\;\;\cm{x}(\xvec\rp\yvec) && \\
			&= \xvec \rp \yvec && \t{By compiling}\\
			&= \cm{x}(\xvec)\rp\cm{y}(\yvec) && \t{By compiling}
		\end{aligned}$$
	\item \textbf{Case} $\xs \neq \ys$\\[.25em]
		Vacuous because we assume a contradiction $\p{\D,\bind{x}{\ty_1}}{y}{\ty_2}$ 
    \end{itemize}
\item \textbf{Case} $\tt$\\[.25em]
	Vacuous because we assume $\p{\D,\bind{x}{\ty_1}}{\tt}{\ty_2}$ 
\item \textbf{Case} $\ff$\\[.25em]
	Vacuous because we assume $\p{\D,\bind{x}{\ty_1}}{\ff}{\ty_2}$ 
\item \textbf{Case} $\z$\\[.25em]
	Vacuous because we assume $\p{\D,\bind{x}{\ty_1}}{\z}{\ty_2}$ 
\item \textbf{Case} $\suc{e}$\\[.25em]
	By inversion $\p{\D,\bind{x}{\ty}}{e}{\N}$ 
			\[\scalebox{.8}{$\begin{aligned}
				& \,\,\,\,\,\,\,\, \cm{\suc{e}}(\sv1\rp\sv2) && \\
				&= \rd{n} \map
				  \begin{cases}
					  \rd{0} & \t{if } \rd{n=0}\\
					  \cm{e}(\svec,\xv1\rp\sv2)(\rd{n-1}) & \t{otherwise}
				  \end{cases} && \t{By compiling}\\
				&= \rd{n} \map
				  \begin{cases}
					  \rd{0} & \t{if } \rd{n=0}\\
					  \cm{e}(\svec,\xv1)(\rd{n-1})\rp\cm{e}(\svec,\xv2)(\rd{n-1}) & \t{otherwise}
				  \end{cases} && \t{By induction}\\
				&= \Biggl(\rd{n} \map
				  \begin{cases}
					  \rd{0} & \t{if } \rd{n=0}\\
					  \cm{e}(\sv1)(\rd{n-1}) & \t{otherwise}
				  \end{cases}\;\Biggr)
				  \rp
				  \Biggl(\rd{n} \map
				  \begin{cases}
					  \rd{0} & \t{if } \rd{n=0}\\
					  \cm{e}(\sv2)(\rd{n-1}) & \t{otherwise}
				  \end{cases}\;\Biggr) && \t{By addition of functions}\\
				&= \cm{\suc{e}}(\sv1)\rp\cm{\suc{e}}(\sv2) && \t{By compiling}
			\end{aligned}$}\]
\item \textbf{Case} $\nb{\lam{y}{e}}$\\[.25em]
	Consider whether $\xs=\ys$,
    \begin{itemize}[label=$\triangleright$]
	\item \textbf{Case} $\xs = \ys$\\[.25em]
		Vacuous by Barendregt's convention: free and bound variables must be unique.
	\item \textbf{Case} $\xs \neq \ys$\\[.25em]
		By inversion on typing $\p{\D,\bind{x}{\ty},\bind{y}{\ty_1}}{e}{\ty_2}$\\
		\[\begin{aligned}
				& \,\,\,\,\,\,\,\,\cm{\lam{y}{e}}(\svec,\xv1\rp\xv2)\\
				&= \yvec \map\cm{e}(\svec,\xv1\rp\xv2,\yvec) && \t{By compiling}\\
				&= \yvec \map\cm{e}(\svec,\xv1\rp\xv2,\yvec)_{\nb{\D,y:\ty_1,x:\ty}} && \t{Compiling invariant to exchange}\\
				&= \yvec \map\cm{e}(\svec,\yvec,\xv1\rp\xv2) && \t{By restricting environments}\\
				&= \yvec \map\cm{e}(\svec,\yvec,\xv1) \rp \cm{e}(\svec,\yvec,\xv2) && \t{By induction}\\
				&= \yvec \map\cm{e}(\svec,\yvec,\xv1)_{\nb{\D,x:\ty,y:\ty_1}} \rp \cm{e}(\svec,\yvec,\xv2)_{\nb{\D,x:\ty,y:\ty_1}} && \t{Compiling invariant to exchange}\\
				&= \yvec \map\cm{e}(\svec,\xv1,\yvec) \rp \cm{e}(\svec,\xv2,\yvec) && \t{By restricting environments}\\
				&= \bigl(\yvec \map\cm{e}(\svec,\xv1,\yvec)\bigr) \rp \bigl(\yvec \map\cm{e}(\svec,\xv2,\yvec)\bigr) && \t{By addition of functions}\\
				&= \cm{\lam{y}{e}}(\svec,\xv1) \rp \cm{\lam{y}{e}}(\svec,\xv2) && \t{By compiling}
			\end{aligned}\]
    \end{itemize}
\item \textbf{Case} $\nb{\ite{e_1}{e_2}{e_3}}$\\[.25em]
	By inversion on typing $\nb{\D,\bind{x}{\ty_1}}=\nb{\D_1 \o \D_2}$ and $\p{\D_1}{e_1}{\Bool}$ and $\p{\D_2}{e_2}{\ty}$ and $\p{\D_2}{e_3}{\ty}$ \\
	Consider whether $\xs \in \fv{\ess1}$ or $(\xs \in \fv{\ess2} \land\xs \in \fv{\ess3})$,
    \begin{itemize}[label=$\triangleright$]
	\item \textbf{Case} $\xs \in \fv{e_1}$\\[.25em]
		Because contexts permit exchange $\p{\Ds{11},\bind{x}{\ty_1}}{e_1}{\Bool}$
		\[\scalebox{.8}{$\begin{aligned}
				& \,\,\,\,\,\,\,\,\cm{\ite{e_1}{e_2}{e_3}}(\svec,\xv1 \rp \xv2)\\ 
				&= \rd{\pi_1}(\cm{e_1}(\svec,\xv1\rp\xv2)_\Ds1)\rt\cm{e_2}(\svec,\xv1\rp\xv2)_\Ds2\\
				& \rp \rd{\pi_2}(\cm{e_1}(\svec,\xv1\rp\xv2)_\Ds1)\rt\cm{e_3}(\svec,\xv1\rp\xv2)_\Ds2 && \t{By compiling}\\
				&= \rd{\pi_1}(\cm{e_1}((\svec,\xv1\rp\xv2)_\Ds1)_\nb{\Ds{11},x:\ty_1})\rt\cm{e_2}(\svec,\xv1\rp\xv2)_\Ds2 \\
				&\rp \rd{\pi_2}(\cm{e_1}((\svec,\xv1\rp\xv2)_\Ds1)_\nb{\Ds{11},x:\ty_1})\rt\cm{e_3}(\svec,\xv1\rp\xv2)_\Ds2 && \t{Compiling invariant to exchange}\\
				&= \rd{\pi_1}(\cm{e_1}(\svec,\xv1\rp\xv2)_\nb{\Ds{11},x:\ty_1})\rt\cm{e_2}(\svec,\xv1\rp\xv2)_\Ds2\\
				& \rp \rd{\pi_2}(\cm{e_1}(\svec,\xv1\rp\xv2)_\nb{\Ds{11},x:\ty_1})\rt\cm{e_3}(\svec,\xv1\rp\xv2)_\Ds2 && \t{Because }((\svec,\xv1\rp\xv2)_\Ds1)_\nb{\Ds{11},x:\ty_1}=(\svec,\xv1\rp\xv2)_\nb{\Ds{11},x:\ty_1}\\
				&= \rd{\pi_1}(\cm{e_1}(\svec_\Ds{11},\xv1\rp\xv2))\rt\cm{e_2}(\svec,\xv1\rp\xv2)_\Ds2\\
				& \rp \rd{\pi_2}(\cm{e_1}(\svec_\Ds{11},\xv1\rp\xv2))\rt\cm{e_3}(\svec,\xv1\rp\xv2)_\Ds2 && \t{By restricting environments}\\
				&= \rd{\pi_1}(\cm{e_1}(\svec_\Ds{11},\xv1)\rp \cm{e_1}(\svec_\Ds{11},\xv2))\rt\cm{e_2}(\svec,\xv1\rp\xv2)_\Ds2\\
				& \rp \rd{\pi_2}(\cm{e_1}(\svec_\Ds{11},\xv1) \rp\cm{e_1}(\svec_\Ds{11},\xv2))\rt\cm{e_3}(\svec,\xv1\rp\xv2)_\Ds2 && \t{By induction}\\
				&= (\rd{\pi_1}(\cm{e_1}(\svec_\Ds{11},\xv1))\rp \rd{\pi_1}(\cm{e_1}(\svec_\Ds{11},\xv2))) \rt\cm{e_2}(\svec,\xv1\rp\xv2)_\Ds2\\
				& \rp (\rd{\pi_2}(\cm{e_1}(\svec_\Ds{11},\xv1)) \rp \rd{\pi_2}(\cm{e_1}(\svec_\Ds{11},\xv2)))\rt\cm{e_3}(\svec,\xv1\rp\xv2)_\Ds2 && \t{Because }\rd{\pi_i} \t{ is linear}\\
				&= (\rd{\pi_1}(\cm{e_1}(\svec_\Ds{11},\xv1)_\Ds1)\rp \rd{\pi_1}(\cm{e_1}(\svec_\Ds{11},\xv2)_\Ds1)) \rt\cm{e_2}(\svec,\xv1\rp\xv2)_\Ds2\\
				& \rp (\rd{\pi_2}(\cm{e_1}(\svec_\Ds{11},\xv1)_\Ds1) \rp \rd{\pi_2}(\cm{e_1}(\svec_\Ds{11},\xv2)_\Ds1))\rt\cm{e_3}(\svec,\xv1\rp\xv2)_\Ds2 && \t{Compiling invariant to exchange}\\
				&= (\rd{\pi_1}(\cm{e_1}(\svec,\xv1)_\Ds1)\rp \rd{\pi_1}(\cm{e_1}(\svec,\xv2)_\Ds1)) \rt\cm{e_2}(\svec,\xv1\rp\xv2)_\Ds2\\
				& \rp (\rd{\pi_2}(\cm{e_1}(\svec,\xv1)_\Ds1) \rp \rd{\pi_2}(\cm{e_1}(\svec,\xv2)_\Ds1))\rt\cm{e_3}(\svec,\xv1\rp\xv2)_\Ds2 && \t{Because }(\svec_\Ds{11},\xv{i})_\Ds1=(\svec,\xv{i})_\Ds1\\
				&= \rd{\pi_1}(\cm{e_1}(\svec,\xv1)_\Ds1)\rt\cm{e_2}(\svec,\xv1\rp\xv2)_\Ds2\\
				& \rp \rd{\pi_2}(\cm{e_1}(\svec,\xv1)_\Ds1)\rt\cm{e_3}(\svec,\xv1\rp\xv2)_\Ds2 \\
				& \rp \rd{\pi_1}(\cm{e_1}(\svec,\xv2)_\Ds1)\rt\cm{e_2}(\svec,\xv1\rp\xv2)_\Ds2\\
				& \rp \rd{\pi_2}(\cm{e_1}(\svec,\xv2)_\Ds1)\rt\cm{e_3}(\svec,\xv1\rp\xv2)_\Ds2 && \t{Because }\rt\t{ distributes over }\rp\\
				&= \rd{\pi_1}(\cm{e_1}(\svec,\xv1)_\Ds1)\rt\cm{e_2}(\svec,\xv1)_\Ds2\\
				& \rp \rd{\pi_2}(\cm{e_1}(\svec,\xv1)_\Ds1)\rt\cm{e_3}(\svec,\xv1)_\Ds2 \\
				& \rp \rd{\pi_1}(\cm{e_1}(\svec,\xv2)_\Ds1)\rt\cm{e_2}(\svec,\xv2)_\Ds2\\
				& \rp \rd{\pi_2}(\cm{e_1}(\svec,\xv2)_\Ds1)\rt\cm{e_3}(\svec,\xv2)_\Ds2 &&\t{Because }(\svec,\xv1\rp\xv2)_\Ds2=(\svec,\xv1)_\Ds2 = (\svec,\xv2)_\Ds2\\
				&= \cm{\ite{e_1}{e_2}{e_3}}(\svec,\xv1) \rp \cm{\ite{e_1}{e_2}{e_3}}(\svec,\xv2) &&\t{By compiling}
			\end{aligned}$}\]
	\item \textbf{Case} $\xs \in \fv{e_2} \land \xs \in \fv{\ess3}$ \\[.25em]
		Similar to previous case
    \end{itemize}
\item \textbf{Case} $\nb{\iter{e_1}{y}{e_2}{e_3}}$\\[.25em]
	By inversion on typing $\nb{\D,\bind{x}{\ty_1}}=\nb{\D_1 \o \D_3}$ and $\p{\D_1}{e_1}{\ty}$ and $\p{\bind{y}{\ty}}{e_2}{\ty}$ and $\p{\D_3}{e_3}{\N}$\\
	Consider whether $\xs \in \fv{\ess1}$ or $\xs \in \fv{e_3}$,
    \begin{itemize}[label=$\triangleright$]
	\item \textbf{Case} $\xs \in \fv{e_1}$\\[.25em]
		Because contexts permit exchange $\p{\Ds{11},\bind{x}{\ty_1}}{e_1}{\ty}$\\
		\[\scalebox{.7}{$\begin{aligned}
				& \,\,\,\,\,\,\,\,\cm{\iter{e_1}{y}{e_2}{e_3}}(\svec,\xv1\rp\xv2)\\ 
				&= \rd{\sum_{n \in \N}}\,\cm{\nb{e_3}}{(\svec,\xv1\rp\xv2)_\Ds{3}}(\rd{n})\,\rd{\.}\,(\cm{\nb{e_2}}{)}^\rd{n}(\cm{\nb{e_1}}{(\svec,\xv1\rp\xv2)_\Ds{1})} && \t{By compiling}\\
				&= \rd{\sum_{n \in \N}}\,\cm{\nb{e_3}}{(\svec,\xv1\rp\xv2)_\Ds{3}}(\rd{n})\,\rd{\.}\,(\cm{\nb{e_2}}{)}^\rd{n}(\cm{\nb{e_1}}{((\svec,\xv1\rp\xv2)_\Ds{1})_{\nb{\Ds{11},x:\ty_1}}}) && \t{Compiling invariant to exchange}\\
				&= \rd{\sum_{n \in \N}}\,\cm{\nb{e_3}}{(\svec,\xv1\rp\xv2)_\Ds{3}}(\rd{n})\,\rd{\.}\,(\cm{\nb{e_2}}{)}^\rd{n}(\cm{\nb{e_1}}{(\svec,\xv1\rp\xv2)_{\nb{\Ds{11},x:\ty_1}})} && \t{Because }((\svec,\xv1\rp\xv2)_\Ds1)_{\nb{\Ds{11},x:\ty_1}}=(\svec,\xv1\rp\xv2)_\nb{\Ds{11},x:\ty_1}\\
				&= \rd{\sum_{n \in \N}}\,\cm{\nb{e_3}}{(\svec,\xv1\rp\xv2)_\Ds{3}}(\rd{n})\,\rd{\.}\,(\cm{\nb{e_2}}{)}^\rd{n}(\cm{\nb{e_1}}{(\svec_\Ds{11},\xv1\rp\xv2))} && \t{By restricting environments}\\
				&= \rd{\sum_{n \in \N}}\,\cm{\nb{e_3}}{(\svec,\xv1\rp\xv2)_\Ds{3}}(\rd{n})\,\rd{\.}\,(\cm{\nb{e_2}}{)}^\rd{n}(\cm{\nb{e_1}}{(\svec_\Ds{11},\xv1)\rp\cm{e_1}(\svec_\Ds{11},\xv2))} && \t{By induction}\\
				&= \rd{\sum_{n \in \N}}\,\cm{\nb{e_3}}{(\svec,\xv1\rp\xv2)_\Ds{3}}(\rd{n})\,\rd{\.}\,(\cm{\nb{e_2}}{)}^\rd{n}(\cm{\nb{e_1}}{(\svec_\Ds{11},\xv1)_\Ds1\rp\cm{e_1}(\svec_\Ds{11},\xv2)_\Ds1)} && \t{Compiling invariant to exchange}\\
				&= \rd{\sum_{n \in \N}}\,\cm{\nb{e_3}}{(\svec,\xv1\rp\xv2)_\Ds{3}}(\rd{n})\,\rd{\.}\,(\cm{\nb{e_2}}{)}^\rd{n}(\cm{\nb{e_1}}{(\svec,\xv1)_\Ds1\rp\cm{e_1}(\svec,\xv2)_\Ds1)} && \t{Because }(\svec_\Ds{11},\xv1)_\Ds1=(\svec,\xv1)_\Ds1 \t{ and } (\svec_\Ds{11},\xv2)_\Ds1=(\svec,\xv2)_\Ds1 \\
				&= \rd{\sum_{n \in \N}}\,\cm{\nb{e_3}}{(\svec,\xv1\rp\xv2)_\Ds{3}}(\rd{n})\,\rd{\.}\,(\cm{\nb{e_2}}{)}^\rd{n}(\cm{\nb{e_1}}{(\svec,\xv1)_\Ds{1})} && \t{If}\,\rd{f} \t{ additive},\\
				&\;\;\;\;\;\;\rp 
				 \cm{\nb{e_3}}{(\svec,\xv1\rp\xv2)_\Ds{3}}(\rd{n})\,\rd{\.}\,(\cm{\nb{e_2}}{)}^\rd{n}(\cm{\nb{e_1}}{(\svec,\xv2)_\Ds{1})} && \rd{a\.f^n}(\rd{v+w})=\rd{a\,\.}\,(\rd{f^n}(\rd{v})\,\rd{\rp f^n}(\rd{w}))=\rd{a\.f^n}(\rd{v})\rd{\rp a\.f^n}(\rd{w})\\[.8em]
				&= \rd{\sum_{n \in \N}}\,\cm{\nb{e_3}}{(\svec,\xv1\rp\xv2)_\Ds{3}}(\rd{n})\,\rd{\.}\,(\cm{\nb{e_2}}{)}^\rd{n}(\cm{\nb{e_1}}{(\svec,\xv1)_\Ds{1})} && \\
				&\rp  
				 \rd{\sum_{n \in \N}}\,\cm{\nb{e_3}}{(\svec,\xv1\rp\xv2)_\Ds{3}}(\rd{n})\,\rd{\.}\,(\cm{\nb{e_2}}{)}^\rd{n}(\cm{\nb{e_1}}{(\svec,\xv2)_\Ds{1})} && \t{Because } \rd{\sum_{n \in \N}}\,\rd{f}(\rd{n})\rp\rd{g}(\rd{n})=\rd{\sum_{n \in \N}}\,\rd{f}(\rd{n}) \rp \rd{\sum_{n \in \N}}\,\rd{g}(\rd{n})\\
				&= \rd{\sum_{n \in \N}}\,\cm{\nb{e_3}}{(\svec,\xv1)_\Ds{3}}(\rd{n})\,\rd{\.}\,(\cm{\nb{e_2}}{)}^\rd{n}(\cm{\nb{e_1}}{(\svec,\xv1)_\Ds{1})}\\
				&\rp 
				 \rd{\sum_{n \in \N}}\,\cm{\nb{e_3}}{(\svec,\xv2)_\Ds{3}}(\rd{n})\,\rd{\.}\,(\cm{\nb{e_2}}{)}^\rd{n}(\cm{\nb{e_1}}{(\svec,\xv2)_\Ds{1})} && \t{Because }(\svec,\xv1)_\Ds3=(\svec,\xv2)_\Ds3=(\svec,\xv1\rp\xv2)_\Ds3\\
				&= \cm{\iter{e_1}{y}{e_2}{e_3}}(\svec,\xv1) \rp \cm{\iter{e_1}{y}{e_2}{e_3}}(\svec,\xv2) &&\t{By compiling}
			\end{aligned}$}\]\vspace{.3em}
	\item \textbf{Case} $\xs \in \fv{e_3}$\\[.25em]
		Similar to previous case
    \end{itemize}
\end{itemize}}
\end{proof}

\subsection{Lemma (Compiler maps programs to homogeneous maps)}
\vspace{.7em}
$$\p{\D,\bind{x}{\ty_1}}{e}{\ty_2} \implies \rd{\alpha} \rt \cm{e}(\svec,\xvec)$$
\begin{proof}\leavevmode\\
\t{By induction on $\nb{e}$, similar to showing the compiler maps programs to additive maps.}
\end{proof}
\vspace{1em}

\subsection{Lemma (Compiler commutes with substitution)}
\vspace{.7em}
$$\p{\D,\bind{x}{\ty_1}}{e}{\ty_2} \implies \cm{\{x \map v\}(e)}(\svec)=\cm{e}(\svec,\cm{v})$$
\begin{proof}\leavevmode\\
\t{By induction on $\nb{e}$,
\begin{itemize}[label=$\triangleright$]
\item \textbf{Case} $\nb{y}$\\[.25em]
	Consider whether $\nb{x} = \nb{y}$
    \begin{itemize}[label=$\triangleright$]
	\item \textbf{Case} $\nb{x} = \nb{y}$\\[.25em]
		By inversion on typing $\D=\nb{\emp}$\\
		$$\begin{aligned}
			&\;\;\;\;\;\cm{\{x \map v\}(x)} && \\
			&= \cm{v} && \t{By substitution}\\
			&= \cm{x}(\cm{v}) && \t{By compiling} 
		\end{aligned}$$
	\item \textbf{Case} $\nb{x}\neq\nb{y}$\\[.25em]
		Vacuous because we assume $\p{\D,\bind{x}{\ty_1}}{y}{\ty_2}$ 
    \end{itemize}
\item \textbf{Case} $\tt$\\[.25em]
	Vacuous because we assume $\p{\D,\bind{x}{\ty_1}}{\tt}{\ty_2}$ 
\item \textbf{Case} $\ff$\\[.25em]
	Vacuous because we assume $\p{\D,\bind{x}{\ty_1}}{\ff}{\ty_2}$ 
\item \textbf{Case} $\z$\\[.25em]
	Vacuous because we assume $\p{\D,\bind{x}{\ty_1}}{\z}{\ty_2}$ 
\item \textbf{Case} $\suc{e}$\\[.25em]
	By inversion on typing $\p{\D,\bind{x}{\ty}}{e}{\N}$\\
		$$\begin{aligned}
				& \,\,\,\,\,\,\,\, \cm{\{x \map v\}(\suc{e})}(\svec) && \\
				&= \cm{\suc{\{x \map v\}(e)}}(\svec) && \t{By substitution}\\
				&= \rd{n} \map
				  \begin{cases}
					  \rd{0} & \t{if } \rd{n=0}\\
					  \cm{\{x \map v\}(e)}(\rd{\vec{\sig}})(\rd{n-1}) & \t{otherwise}
				  \end{cases} && \t{By compiling}\\
				&= \rd{n} \map
				  \begin{cases}
					  \rd{0} & \t{if } \rd{n=0}\\
					  \cm{e}(\rd{\vec{\sig}}, \cm{v})(\rd{n-1}) & \t{otherwise}
				  \end{cases} && \t{By induction}\\
				&= \cm{\suc{e}}(\svec,\cm{v}) && \t{By compiling}
			\end{aligned}$$
\item \textbf{Case} $\nb{\lam{y}{e}}$\\[.25em]
	Consider whether $\nb{x}\neq \nb{y}$,
    \begin{itemize}[label=$\triangleright$]
	\item \textbf{Case} $\nb{x} \neq \nb{y}$\\[.25em]
		By inversion on typing $\p{\D,\bind{x}{\ty},\bind{y}{\ty_1}}{e}{\ty_2}$\\
		Because contexts permit exchange $\p{\D,\bind{y}{\ty_1},\bind{x}{\ty}}{e}{\ty_2}$ \\
		Because substitution preserves typing $\p{\D,\bind{y}{\ty_1}}{\{x \map v\}(e)}{\ty_2}$\\
		$$\begin{aligned}
				& \,\,\,\,\,\,\,\,\cm{\{x \map v\}(\lam{y}{e})}(\svec)\\
				&= \cm{\lam{y}{\{x \map v\}(e)}}(\svec) && \t{By substitution}\\
				&= \yvec \map\cm{\{x \map v\}(e)}(\svec,\yvec) && \t{By compiling}\\
				&= \yvec \map\cm{e}(\svec,\yvec,\cm{v}) && \t{By induction}\\
				&= \yvec \map\cm{e}(\svec,\yvec,\cm{v})_{\nb{\D,x:\ty,y:\ty_1}} && \t{Compiling is exchange invariant }\\
				&= \yvec \map\cm{e}(\svec,\cm{v},\yvec) && \t{By restricting environments}\\
				&= \cm{\lam{y}{e}}(\svec,\cm{v}) && \t{By compiling}
			\end{aligned}$$\\
	\item \textbf{Case} $\xs = \ys$\\[.25em]
		Vacuous by Barendregt's convention: free and bound variables must be distinct.
    \end{itemize}
\item \textbf{Case} $\nb{e_1e_2}$\\[.25em]
	By typing $\xs \in \fv{e_1e_2}$\\
	Consider whether $\xs \in \fv{e_1}$ or $\xs \in \fv{e_2}$,
    \begin{itemize}[label=$\triangleright$]
	\item \textbf{Case} $\xs \in \fv{e_1}$\\[.25em]
		By assumption $(\svec,\cm{v}) \in \cm{\D,\bind{x}{\ty}}$ \\
		By inversion $\nb{\D}=\nb{\D_{11} \o \D_{12}}$ and $\p{\D_{11}}{\{x \map v\}(e_1)}{\ty_1 \lto \ty_2}$ and $\p{\D_{12}}{e_2}{\ty_1}$ 	\\
		By inversion $\nb{\D,\bind{x}{\ty}}=\nb{\D_{21} \o \D_{22}}$ and $\p{\D_{21}}{e_1}{\ty_1 \lto \ty_2}$ and $\p{\D_{22}}{e_2}{\ty_1}$ \\
		Because contexts permit exchange $\p{\D_{11},\bind{x}{\ty}}{e_1}{\ty_1 \lto \ty_2}$\\
		$$\begin{aligned}
				& \,\,\,\,\,\,\,\,\cm{\{x \map v\}(e_1e_2)}(\svec)\\ 
				&= \cm{\{x \map v\}(e_1)e_2}(\svec) && \t{By substitution and }\nb{x} \in \nb{\t{fv}}(\nb{e
				_1})\\ 
				&= \cm{\{x \map v\}(e_1)}(\svec_\Ds{11})(\cm{e_2}(\svec_\Ds{12})) && \t{By compiling}\\
				&= \cm{e_1}(\svec_\Ds{11}, \cm{v})(\cm{e_2}(\svec_\Ds{12})) && \t{By induction}\\ 
				&= \cm{e_1}(\svec_\Ds{11}, \cm{v})_{\nb{\D_{21}}}(\cm{e_2}(\svec_\Ds{12})_{\nb{\D_{22}}}) && \t{Compiling invariant to exchange}\\
				&= \cm{e_1}(\svec, \cm{v})_{\nb{\D_{21}}}(\cm{e_2}(\svec_\Ds{12})_{\nb{\D_{22}}}) && \t{Because }(\svec_\Ds{11},\cm{v})_{\nb{\D_{21}}}=(\svec,\cm{v})_{\nb{\D_{21}}}\\ 
				&= \cm{e_1}(\svec, \cm{v})_{\nb{\D_{21}}}(\cm{e_2}(\svec, \cm{v})_{\nb{\D_{22}}}) && \t{Because }(\svec_\Ds{12})_{\nb{\D_{22}}}=(\svec,\cm{v})_{\nb{\D_{22}}}\\ 
				&= \cm{e_1e_2}(\svec,\cm{v}) &&\t{By compiling}
			\end{aligned}$$\\
	\item \textbf{Case} $\xs \in \fv{e_2}$\\[.25em]
		Similar to previous case
    \end{itemize}
\item \textbf{Case} $\nb{\ite{e_1}{e_2}{e_3}}$\\[.25em]
	By inversion on typing $\nb{\D}=\nb{\D_{11} \o \D_{12}}$ and $\p{\D_{11}}{\{x \map v\}(e_1)}{\Bool}$ and $\p{\D_{12}}{e_2}{\ty}$ and $\p{\D_{12}}{e_3}{\ty}$ 	\\
	By inversion on typing $\nb{\D,\bind{x}{\ty_1}}=\nb{\D_{21} \o \D_{22}}$ and $\p{\D_{21}}{e_1}{\Bool}$ and $\p{\D_{22}}{e_2}{\ty}$ and $\p{\D_{22}}{e_3}{\ty}$ \\
	Consider whether $\xs \in \fv{\ess1}$ or $(\xs \in \fv{\ess2} \land\xs \in \fv{\ess3})$,
    \begin{itemize}[label=$\triangleright$]
	\item \textbf{Case} $\xs \in \fv{e_1}$\\[.25em]
		Because contexts permit exchange $\p{\Ds{11},\bind{x}{\ty_1}}{e_1}{\Bool}$
        \vspace{-.3em}
		\[\scalebox{.75}{$\begin{aligned}
				& \,\,\,\,\,\,\,\,\cm{\{x \map v\}(\ite{e_1}{e_2}{e_3})}(\svec)\\ 
				&= \cm{\ite{\{x \map v\}(e_1)}{e_2}{e_3}}(\svec) && \t{By substitution and }\xs \in \fv{\ess1}\\ 
				&= \rd{\pi_1}(\cm{\{x \map v\}(e_1)}(\svec_\Ds{11}))\rt\cm{e_2}(\svec)_\Ds{12}\\
				& \rp \rd{\pi_2}(\cm{\{x \map v\}(e_1)}(\svec_\Ds{11}))\rt\cm{e_3}(\svec)_\Ds{12} && \t{By compiling}\\
				&= \rd{\pi_1}(\cm{e_1}(\svec_\Ds{11},\cm{v}))\rt\cm{e_2}(\svec)_\Ds{12}\\
				& \rp \rd{\pi_2}(\cm{e_1}(\svec_\Ds{11},\cm{v}))\rt\cm{e_3}(\svec)_\Ds{12} && \t{By induction}\\
				&= \rd{\pi_1}(\cm{e_1}(\svec_\Ds{11},\cm{v})_\Ds{21})\rt\cm{e_2}((\svec)_\Ds{12})_\Ds{22}\\
				& \rp \rd{\pi_2}(\cm{e_1}(\svec_\Ds{11},\cm{v})_\Ds{21})\rt\cm{e_3}((\svec)_\Ds{12})_\Ds{22} && \t{Compiling invariant to exchange}\\
				&= \rd{\pi_1}(\cm{e_1}(\svec,\cm{v})_\Ds{21})\rt\cm{e_2}((\svec)_\Ds{12})_\Ds{22}\\
				& \rp \rd{\pi_2}(\cm{e_1}(\svec,\cm{v})_\Ds{21})\rt\cm{e_3}((\svec)_\Ds{12})_\Ds{22} && \t{Because }(\svec_\Ds{11},\cm{v})_\Ds{21} = (\svec,\cm{v})_\Ds{21}\\
				&= \rd{\pi_1}(\cm{e_1}(\svec,\cm{v})_\Ds{21})\rt\cm{e_2}(\svec,\cm{v})_\Ds{22}\\
				& \rp \rd{\pi_2}(\cm{e_1}(\svec,\cm{v})_\Ds{21})\rt\cm{e_3}(\svec,\cm{v})_\Ds{22} && \t{Because }((\svec)_\Ds{12})_\Ds{22}=(\svec,\cm{v})_\Ds{22}\\
				&= \cm{\ite{e_1}{e_2}{e_3}}(\svec,\cm{v}) &&\t{By compiling}
			\end{aligned}$}\]
	\item \textbf{Case} $\xs \in \fv{e_2} \land \xs \in \fv{\ess3}$\\[.25em]
		Similar to previous case
    \end{itemize}
\item \textbf{Case} $\nb{\iter{e_1}{y}{e_2}{e_3}}$\\[.25em]
	By typing $\xs \in \fv{\iter{e_1}{y}{e_2}{e_3}}$ \\
	Consider whether $\xs \in \fv{e_1}$ or $\xs \in \fv{e_3}$,
    \begin{itemize}[label=$\triangleright$]
	\item \textbf{Case} $\xs \in \fv{e_1}$\\[.25em]
		By inversion on typing $\nb{\D} = \nb{\D_{11} \o \D_{13}}$ and $\p{\D_{11}}{\{x \map v\}(e_1)}{\ty}$ and $\p{\bind{y}{\ty}}{e_2}{\ty}$ and $\p{\D_{13}}{e_3}{\N}$ \\
		By inversion on typing $\nb{\D,\bind{x}{\ty}} = \nb{\D_{21} \o \D_{23}}$ and $\p{\D_{21}}{e_1}{\ty}$ and $\p{\bind{y}{\ty}}{e_2}{\ty}$ and $\p{\D_{23}}{e_3}{\N}$ \\
		Because contexts permit exchange $\p{\Ds{11},\bind{x}{\ty}}{e_1}{\ty}$\\
		\[\scalebox{.75}{$\begin{aligned}
				& \,\,\,\,\,\,\,\,\cm{\{x \map v\}(\iter{e_1}{y}{e_2}{e_3})}(\svec)\\ 
				&= \cm{\iter{\{x \map v\}(e_1)}{y}{e_2}{e_3}}(\svec) && \t{By substitution and }\nb{x} \in \nb{\t{fv}}(\nb{e
				_1})\\ 
				&= \rd{\sum_{n \in \N}}\,\cm{\nb{e_3}}{(\svec_\Ds{13})}(\rd{n})\,\rd{\.}\,\cm{\nb{e_2}}{}^\rd{n}(\cm{\nb{\{x \map v\}(e_1)}}{(\svec_\Ds{11})}) && \t{By compiling}\\
				&= \rd{\sum_{n \in \N}}\,\cm{\nb{e_3}}{(\svec_\Ds{13})}(\rd{n})\,\rd{\.}\,\cm{\nb{e_2}}{}^\rd{n}(\cm{\nb{e_1}}{(\svec_\Ds{11},\cm{v})}) && \t{By induction}\\ 
				&= \rd{\sum_{n \in \N}}\,\cm{\nb{e_3}}{(\svec_\Ds{13})_\Ds{23}}(\rd{n})\,\rd{\.}\,\cm{\nb{e_2}}{}^\rd{n}(\cm{\nb{e_1}}{(\svec_\Ds{11},\cm{v})_\Ds{21}}) && \t{Compiling invariant to exchange}\\
				&= \rd{\sum_{n \in \N}}\,\cm{\nb{e_3}}{(\svec_\Ds{13})_\Ds{23}}(\rd{n})\,\rd{\.}\,\cm{\nb{e_2}}{}^\rd{n}(\cm{\nb{e_1}}((\svec,\cm{v})_\Ds{21}) && \t{Because }(\svec_\Ds{11},\cm{v})_{\nb{\D_{21}}}=(\svec,\cm{v})_{\nb{\D_{21}}}\\
				&= \rd{\sum_{n \in \N}}\,\cm{\nb{e_3}}{(\svec,\cm{v})_{\nb{\D_{23}}}}(\rd{n})\,\rd{\.}\,\cm{\nb{e_2}}{}^\rd{n}(\cm{\nb{e_1}}((\svec,\cm{v})_\Ds{21}) && \t{Because }(\svec_\Ds{13})_{\nb{\D_{23}}}=(\svec,\cm{v})_{\nb{\D_{23}}}\\ 
				&= \cm{\iter{e_1}{y}{e_2}{e_3}}(\svec,\cm{v}) &&\t{By compiling}
			\end{aligned}$}\]\\
	\item \textbf{Case} $\xs \in \fv{e_3}$\\[.25em]
		Similar to previous case
    \end{itemize}
\end{itemize}}
\end{proof}
\vspace{1em}

\subsection{Lemma (Compiler is invariant to exchange)}
\vspace{.7em}
$$\p{\D_1}{e}{\ty} \;\t{ and }\;\p{\D_2}{e}{\ty}  \implies \cm{e}(\sv1)=\cm{e}(\sv1)_\Ds2$$
\begin{proof}\leavevmode\\
\t{By induction on $\p{\D_1}{e}{\ty}$,
\begin{itemize}[label=$\triangleright$]
\item \textbf{Case} $\p{\bind{x}{\ty}}{x}{\ty}$\\[.25em]
	By assumption $\Ds1=\Ds2=\nb{\bind{x}{\ty}}$  \\
	By assumption $\xvec \in \cm{\bind{x}{\ty}}$\\
		$$\begin{aligned}
			&\;\;\;\;\;\cm{x}(\xvec) && \\
			&= \cm{x}(\xvec_{\nb{x:\ty}}) && \t{By restricting environments} 
		\end{aligned}$$\\
\item \textbf{Case} $\p{\emp}{\tt}{\Bool}$\\[.25em]
	By inversion on typing $\Ds1=\Ds2=\nb{\emp}$\\
		$$\begin{aligned}
			&\;\;\;\;\;\cm{\tt}(\rd{0}) && \\
			&= \cm{\tt}(\rd{0}) && \t{By reflexivity} 
		\end{aligned}$$\\
\item \textbf{Case} $\p{\emp}{\ff}{\Bool}$\\[.25em]
	Similar to previous case
\item \textbf{Case} $\p{\emp}{\z}{\ty}$\\[.25em]
	By assumption $\Ds1 = \Ds2 = \nb{\emp}$\\
	$$ \begin{aligned}
			&\;\;\;\;\;\cm{\z}(\rd{0}) && \\
			&= \cm{\z}(\rd{0}_{\nb{\emp}}) && \t{By restricting environments} 
		\end{aligned}$$\\
\item \textbf{Case} $\p{\D_1}{\suc{e}}{\N}$ \\[.25em]
	By inversion on typing $\p{\D_2}{e}{\N}$\\
	By assumption $\sv1 \in \cm{\D_1}$\\ 
		$$\begin{aligned}
				& \,\,\,\,\,\,\,\, \cm{\suc{e}}(\sv1) && \\
				&= \rd{n} \map
				  \begin{cases}
					  \rd{0} & \t{if } \rd{n=0}\\
					  \cm{e}(\sv1)(\rd{n-1}) & \t{otherwise}
				  \end{cases} && \t{By compiling}\\
				&= \rd{n} \map
				  \begin{cases}
					  \rd{0} & \t{if } \rd{n=0}\\
					  \cm{e}(\sv1)_\Ds2(\rd{n-1}) & \t{otherwise}
				  \end{cases} && \t{By induction}\\
				&= \cm{\suc{e}}(\sv1)_\Ds2 && \t{By compiling}
			\end{aligned}$$\\
\item \textbf{Case} $\p{\D_1}{\lam{x}{e}}{\ty_1 \lto \ty_2}$\\[.25em]
	By assumption and inversion on typing $\p{\D_2}{\lam{x}{e}}{\ty_1 \lto \ty_2}$ and $\p{\D_2,\bind{x}{\ty_1}}{e}{\ty_2}$\\
	By assumption and compiling environments $(\sv1,\xvec) \in \cm{\D_1,\bind{x}{\ty}}$\\
		$$\begin{aligned}
				& \,\,\,\,\,\,\,\,\cm{\lam{x}{e}}(\sv1)\\
				&= \xvec \map\cm{e}(\sv1,\xvec) && \t{By compiling}\\
				&= \xvec \map\cm{e}(\sv1,\xvec)_{\nb{\Ds{2},x:\ty_1}} && \t{By induction}\\
				&= \xvec \map\cm{e}((\sv1)_\Ds2,\xvec) && \t{By restricting environments}\\
				&= \cm{\lam{x}{e}}(\sv1)_{\nb{\Ds{2}}} && \t{By compiling}
			\end{aligned}$$\\
\item \textbf{Case} $\p{\D_1}{e_1e_2}{\ty_2}$\\[.25em]
	By assumption $\nb{\D_1}=\nb{\Ds{11} \o \Ds{12}}$ and $\p{\Ds{11}}{e_1}{\ty_1 \lto \ty_2}$ and $\p{\Ds{12}}{e_2}{\ty_1}$ \\
	By assumption and inversion on typing $\p{\D_2}{e_1e_2}{\ty_2}$ and $\nb{\D_2}=\nb{\Ds{21}\o\Ds{22}}$ and $\p{\Ds{21}}{e_1}{\ty_1 \lto \ty_2}$ and $\p{\Ds{22}}{e_2}{\ty_1}$ \\
	Because typing only admits exchange of contexts $\nb{\D_1}=\nb{\Ds{11} \o \Ds{12}}=\nb{\Ds{21}\o\Ds{22}}$\\
		$$\begin{aligned}
				& \,\,\,\,\,\,\,\,\cm{e_1e_2}(\sv1)\\ 
				&= \cm{e_1}(\sv1)_\Ds{11}(\cm{e_2}(\sv1)_\Ds{12}) && \t{By compiling}\\
				&= \cm{e_1}((\sv1)_\Ds{11})_\Ds{21}(\cm{e_2}((\sv1)_\Ds{12})_\Ds{22}) && \t{By induction}\\ 
				&= \cm{e_1}(\sv1)_\Ds{21}(\cm{e_2}((\sv1)_\Ds{12})_\Ds{22}) && \t{Because }((\sv1)_\Ds{11})_{\nb{\D_{21}}}=(\sv1)_{\nb{\D_{21}}}\\
				&= \cm{e_1}(\sv1)_\Ds{21}(\cm{e_2}(\sv1)_\Ds{22}) && \t{Because }((\sv1)_\Ds{12})_{\nb{\D_{22}}}=(\sv1)_{\nb{\D_{22}}}\\
				&= \cm{e_1e_2}(\sv1)_\Ds2 &&\t{By compiling}
			\end{aligned}$$
\item \textbf{Case} $\p{\D_1}{\ite{e_1}{e_2}{e_3}}{\ty}$\\[.25em]
	By assumption $\nb{\D}=\nb{\D_{11} \o \D_{12}}$ and $\p{\D_{11}}{\{x \map v\}(e_1)}{\Bool}$ and $\p{\D_{12}}{e_2}{\ty}$ and $\p{\D_{12}}{e_3}{\ty}$ 	\\
	By inversion $\nb{\D,\bind{x}{\ty_1}}=\nb{\D_{21} \o \D_{22}}$ and $\p{\D_{21}}{e_1}{\Bool}$ and $\p{\D_{22}}{e_2}{\ty}$ and $\p{\D_{22}}{e_3}{\ty}$ \\
		$$\begin{aligned}
				& \,\,\,\,\,\,\,\,\cm{\ite{e_1}{e_2}{e_3}}(\svec)\\ 
				&= \rd{\pi_1}(\cm{e_1}(\svec_\Ds{11}))\rt\cm{e_2}(\svec)_\Ds{12}\\
				& \rp \rd{\pi_2}(\cm{e_1}(\svec_\Ds{11}))\rt\cm{e_3}(\svec)_\Ds{12} && \t{By compiling}\\
				&= \rd{\pi_1}(\cm{e_1}(\svec_\Ds{11})_\Ds{21})\rt\cm{e_2}((\svec)_\Ds{12})_\Ds{22}\\
				& \rp \rd{\pi_2}(\cm{e_1}(\svec_\Ds{11})_\Ds{21})\rt\cm{e_3}((\svec)_\Ds{12})_\Ds{22} && \t{By induction}\\
				&= \rd{\pi_1}(\cm{e_1}(\svec_\Ds{21}))\rt\cm{e_2}((\svec)_\Ds{12})_\Ds{22}\\
				& \rp \rd{\pi_2}(\cm{e_1}(\svec_\Ds{21}))\rt\cm{e_3}((\svec)_\Ds{12})_\Ds{22} && \t{Because }(\svec_\Ds{11})_\Ds{21}=(\svec_{\Ds{21}})\\
				&= \rd{\pi_1}(\cm{e_1}(\svec_\Ds{21}))\rt\cm{e_2}(\svec)_\Ds{22}\\
				& \rp \rd{\pi_2}(\cm{e_1}(\svec_\Ds{21}))\rt\cm{e_3}(\svec)_\Ds{22} && \t{Because }((\svec)_\Ds{12})_\Ds{22}=(\svec)_\Ds{22}\\
				&= \cm{\ite{e_1}{e_2}{e_3}}(\svec)_\Ds{2} &&\t{By compiling}
			\end{aligned}$$\\
\item \textbf{Case} $\p{\D_1}{\iter{e_1}{y}{e_2}{e_3}}{\ty}$\\[.25em]
	By assumption $\nb{\D_1}=\nb{\Ds{11} \o \Ds{13}}$ and $\p{\Ds{11}}{e_1}{\ty}$ and $\p{\bind{y}{\ty}}{e_2}{\ty}$ and $\p{\Ds{13}}{e_3}{\N}$\\
	By assumption $\p{\D_2}{\iter{e_1}{y}{e_2}{e_3}}{\ty}$\\
    By inversion on typing $\nb{\D_2}=\nb{\Ds{21}\o\Ds{23}}$ and $\p{\Ds{21}}{e_1}{\ty}$ and $\p{\bind{y}{\ty}}{e_2}{\ty}$ and $\p{\Ds{23}}{e_3}{\N}$\\
		$$\begin{aligned}
				& \,\,\,\,\,\,\,\,\cm{\iter{e_1}{y}{e_2}{e_3}}(\sv1)\\ 
				&= \rd{\sum_{n \in \N}}\,\cm{\nb{e_3}}{((\sv1)_\Ds{13})}(\rd{n})\,\rd{\.}\,\cm{\nb{e_2}}{}^\rd{n}(\cm{\nb{e_1}}{((\sv1)_\Ds{11})}) && \t{By compiling}\\
				&= \rd{\sum_{n \in \N}}\,\cm{\nb{e_3}}{(((\sv1)_\Ds{13})_{\Ds{23}})}(\rd{n})\,\rd{\.}\,\cm{\nb{e_2}}{}^\rd{n}(\cm{\nb{e_1}}{(((\sv1)_\Ds{11})_\Ds{21}})) && \t{By induction}\\
				&= \rd{\sum_{n \in \N}}\,\cm{\nb{e_3}}{((\sv1)_\Ds{23})}(\rd{n})\,\rd{\.}\,\cm{\nb{e_2}}{}^\rd{n}(\cm{\nb{e_1}}{(((\sv1)_\Ds{11})_\Ds{21}})) && \t{Because }((\sv1)_\Ds{13})_{\nb{\D_{23}}}=(\sv1)_{\nb{\D_{23}}}\\
				&= \rd{\sum_{n \in \N}}\,\cm{\nb{e_3}}{((\sv1)_\Ds{23})}(\rd{n})\,\rd{\.}\,\cm{\nb{e_2}}{}^\rd{n}(\cm{\nb{e_1}}{((\sv1)_\Ds{21})}) && \t{Because }((\sv1)_\Ds{11})_{\nb{\D_{21}}}=(\sv1)_{\nb{\D_{21}}}\\
				&= \cm{\iter{e_1}{y}{e_2}{e_3}}(\sv1)_\Ds2 &&\t{By compiling}
			\end{aligned}$$
\end{itemize}}
\end{proof}
\vspace{1em}

\clearpage
\section{Experiments}
\label{app:experiments}

The following experiments replicate data from prior work on differentiable programming with first-class conditionals. Since our specification and implementation differ, these data were helpful to establish consistency with prior work. They are binary classification tasks with explicit conditional structure. Our findings are consistent with prior work, showing that programming neural networks with first-class conditionals helps them solve these tasks. Detailed specifications of model parameters are available in the artifact; they are consistent with prior work.

\subsection{Conditional classification: XOR}

\subsubsection{Task}

\begin{wrapfigure}{ht}{0.35\textwidth}
    \centering
    \vspace{-1.4em}
    \includegraphics[width=.3\textwidth]{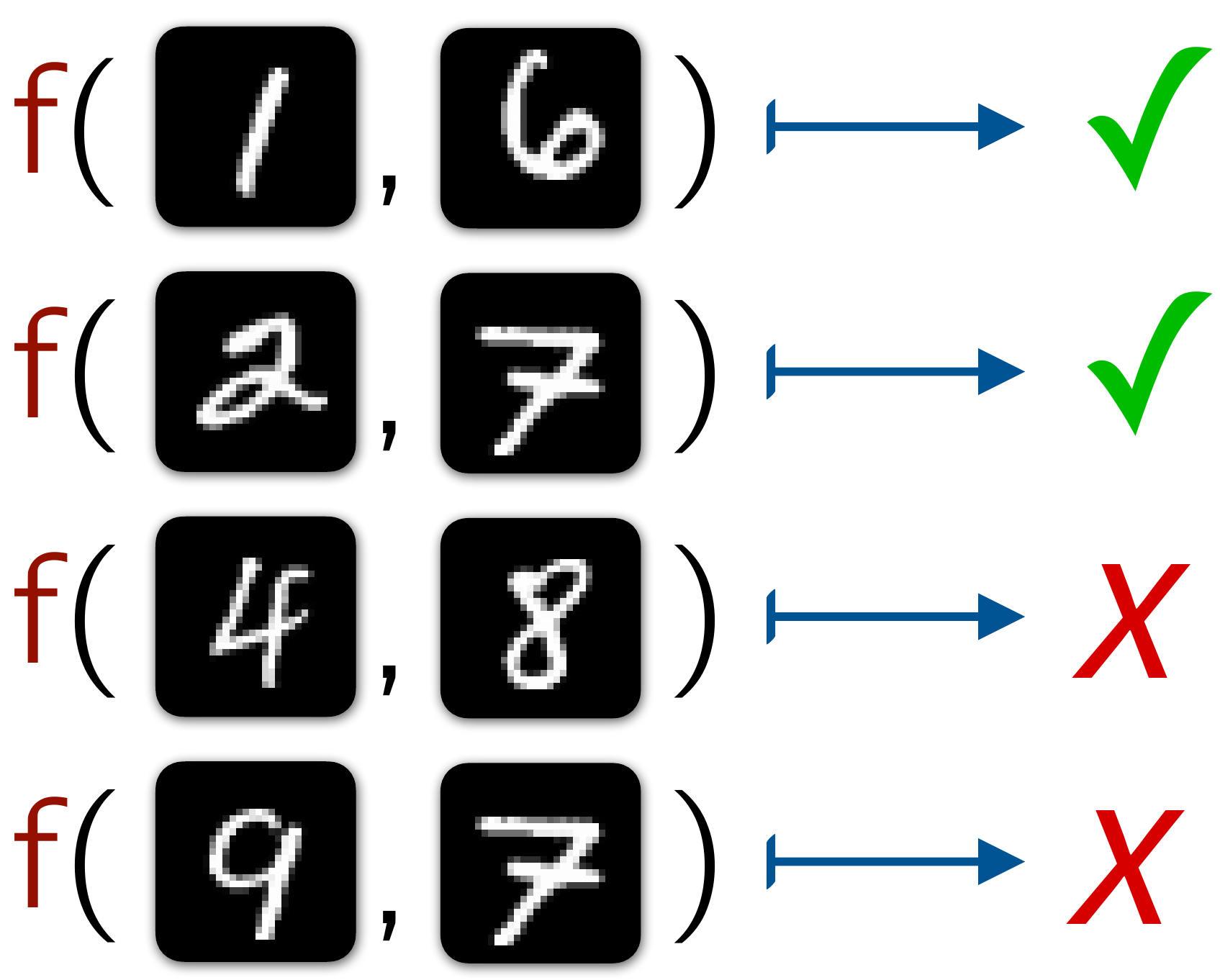}
    \vspace{.6em}
  \caption{XOR task}
  \label{fig:task2xor}
  \vspace{-2.5em}
\end{wrapfigure}

This experiment studies how neural networks learn a binary classification involving \t{XOR} logic, shown in Fig. \ref{fig:task2xor}. The input is two images $x_1$ and $x_2$:
\setlist[itemize]{topsep=4pt, itemsep=1pt, parsep=2pt, leftmargin=1.5em, rightmargin=1.5em}
\begin{itemize}[label=$\triangleright$]
\item If $x_1$ is even and $x_2$ is odd, return \textcolor{ForestGreen}{\cmark}
\item If $x_1$ is odd and $x_2$ is even, return \textcolor{ForestGreen}{\cmark}
\item Otherwise, return \textcolor{red}{\xmark}
\end{itemize}

\subsubsection{Models}
\t{Model D} is a neural network programmed using $\cj{\scriptstyle(\nb{\lto},\, \nb{\Bool},\, \nb{\mathbb{N}})}$, the \t{XOR} structure is \textit{directly} programmed using our compiler. 
\vspace{.5em}
\[
\cm{\ite{\ul{x}}{(\ite{\ul{y}}{\ul{\ff}}{\ul{\tt}})}{(\ite{\ul{y}}{\ul{\tt}}{\ul{\ff}})}}(\xvec, \yvec)
\]

\vspace{.4em}
\t{Model T} replaces this $\nb{2}$-linear map with random matrix $\rd{W}$, similar to our iterative image transform experiments. \t{Model I} is a simple 3-layer feedforward neural network.

\vspace{1em}
\begin{figure}[ht]
  \includegraphics[scale=.39]{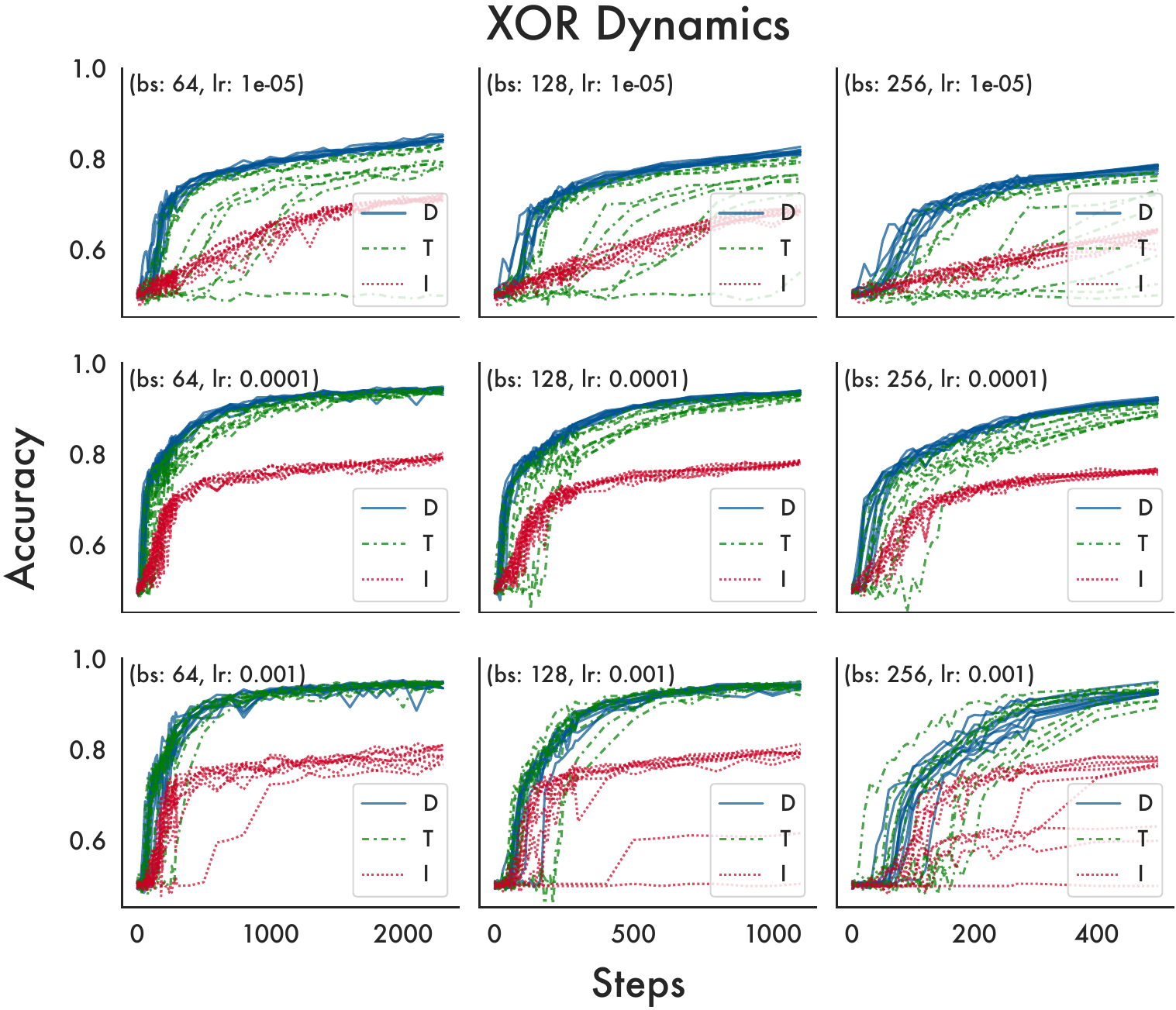}
  \centering
  \vspace{.4em}    
  \caption{Summary of dynamics across configurations}
\end{figure}

\clearpage

\subsection{Conditional classification: EQ}

\subsubsection{Task}

\begin{wrapfigure}{ht}{0.35\textwidth}
    \centering
    \vspace{-1.4em}
    \includegraphics[width=.3\textwidth]{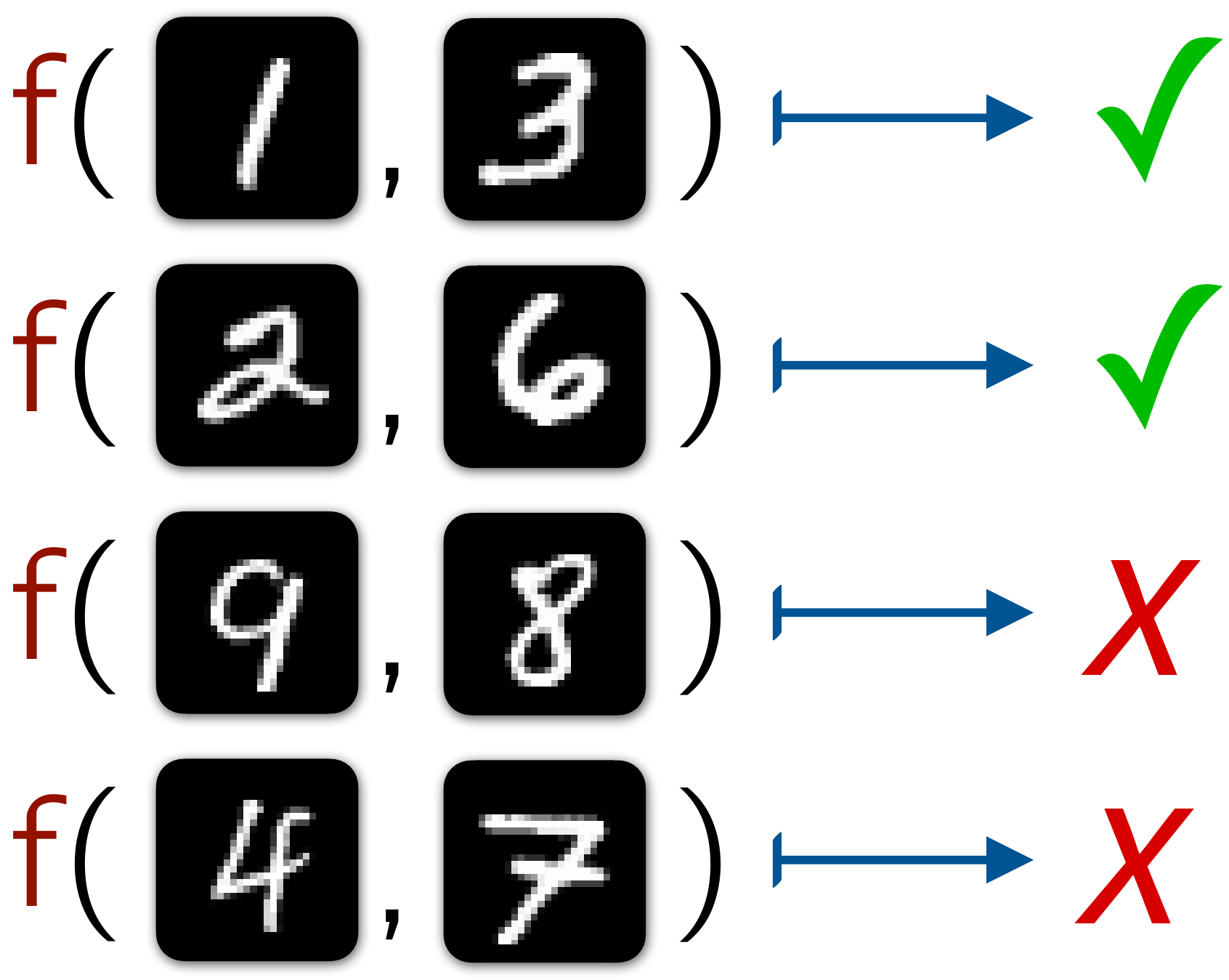}
    \vspace{.6em}
  \caption{EQ task}
  \label{fig:task2eq}
  \vspace{-2.5em}
\end{wrapfigure}

This experiment studies how neural networks learn a binary classification involving \t{EQ} logic, shown in Fig. \ref{fig:task2eq}. The input is two images $x_1$ and $x_2$:
\setlist[itemize]{topsep=4pt, itemsep=1pt, parsep=2pt, leftmargin=1.5em, rightmargin=1.5em}
\begin{itemize}[label=$\triangleright$]
\item If $x_1$ and $x_2$ are even, return \textcolor{ForestGreen}{\cmark}
\item If $x_1$ and $x_2$ are odd, return \textcolor{ForestGreen}{\cmark}
\item Otherwise, return \textcolor{red}{\xmark}
\end{itemize}

\subsubsection{Models}
\t{Model D} is a neural network programmed using $\cj{\scriptstyle(\nb{\lto},\, \nb{\Bool},\, \nb{\mathbb{N}})}$, the \t{EQ} structure is \textit{directly} programmed using our compiler. 
\vspace{.5em}
\[
\cm{\ite{\ul{x}}{(\ite{\ul{y}}{\ul{\tt}}{\ul{\ff}})}{(\ite{\ul{y}}{\ul{\ff}}{\ul{\tt}})}}(\xvec, \yvec)
\]

\vspace{.4em}
\t{Model T} replaces this $\nb{2}$-linear map with random matrix $\rd{W}$, similar to our iterative image transform experiments. \t{Model I} is a simple 3-layer feedforward neural network.

\vspace{1em}
\begin{figure}[ht]
  \includegraphics[scale=.45]{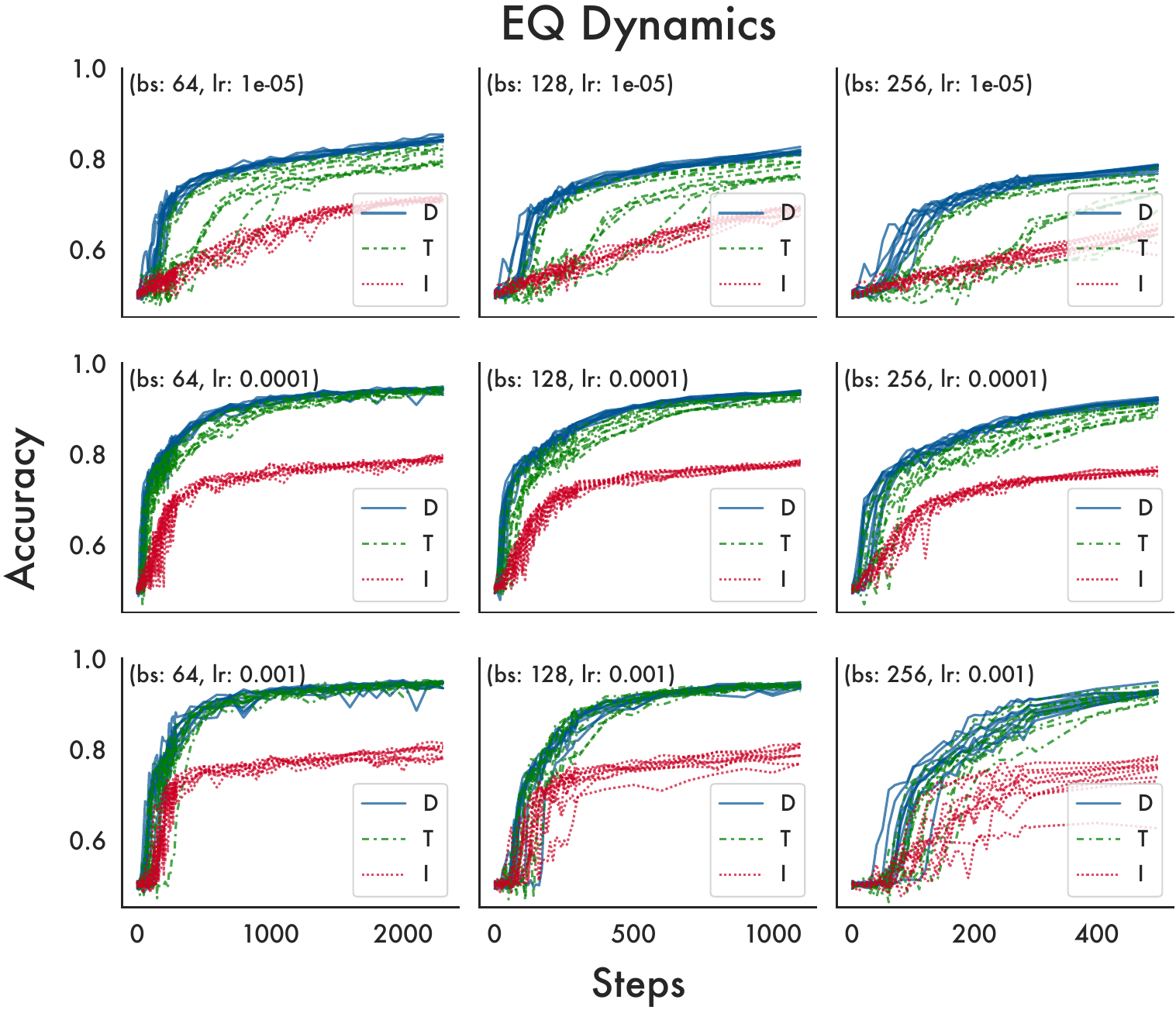}
  \centering
  \vspace{1.6em}    
  \caption{Summary of dynamics across configurations}
\end{figure}

\clearpage

\subsection{Conditional classification: AND}

\subsubsection{Task}

\begin{wrapfigure}{ht}{0.35\textwidth}
    \centering
    \vspace{-1.4em}
    \includegraphics[width=.3\textwidth]{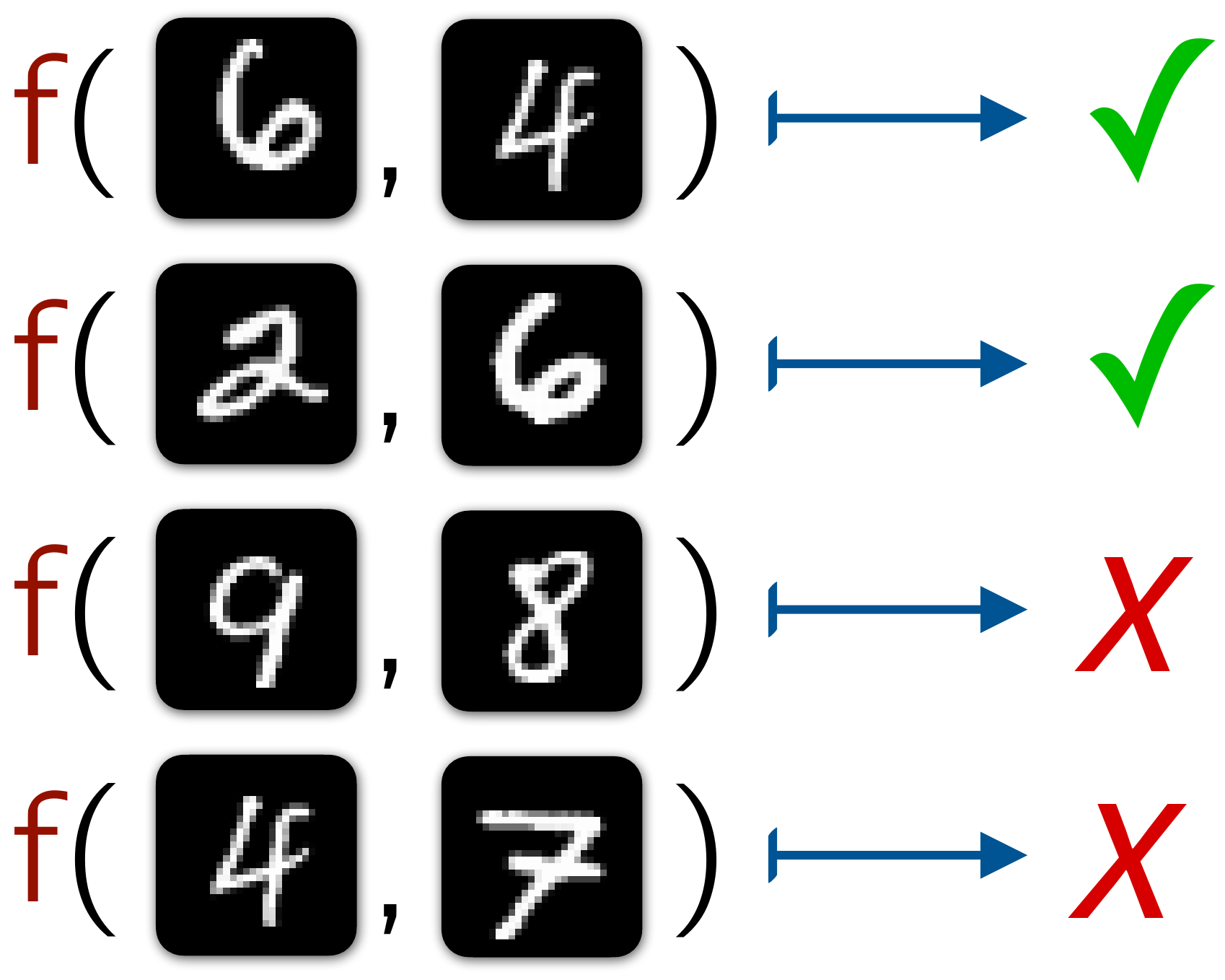}
    \vspace{.6em}
  \caption{AND task}
  \label{fig:task2and}
  \vspace{-2.5em}
\end{wrapfigure}

This experiment studies how neural networks learn a binary classification involving \t{AND} logic, shown in Fig. \ref{fig:task2and}. The input is two images $x_1$ and $x_2$:
\setlist[itemize]{topsep=4pt, itemsep=1pt, parsep=2pt, leftmargin=1.5em, rightmargin=1.5em}
\begin{itemize}[label=$\triangleright$]
\item If $x_1$ is even and $x_2$ is even, return \textcolor{ForestGreen}{\cmark}
\item Otherwise, return \textcolor{red}{\xmark}
\end{itemize}

\subsubsection{Models}
\t{Model D} is a neural network programmed using $\cj{\scriptstyle(\nb{\lto},\, \nb{\Bool},\, \nb{\mathbb{N}})}$, the \t{AND} structure is \textit{directly} programmed using our compiler. 
\vspace{1.2em}
\[
\cm{\ite{\ul{x}}{(\ite{\ul{y}}{\ul{\tt}}{\ul{\ff}})}{(\ite{\ul{y}}{\ul{\ff}}{\ul{\ff}})}}(\xvec, \yvec)
\]

\vspace{.4em}
\t{Model T} replaces this $\nb{2}$-linear map with random matrix $\rd{W}$, similar to our iterative image transform experiments. \t{Model I} is a simple 3-layer feedforward neural network.

\vspace{1em}
\begin{figure}[h]
  \includegraphics[scale=.45]{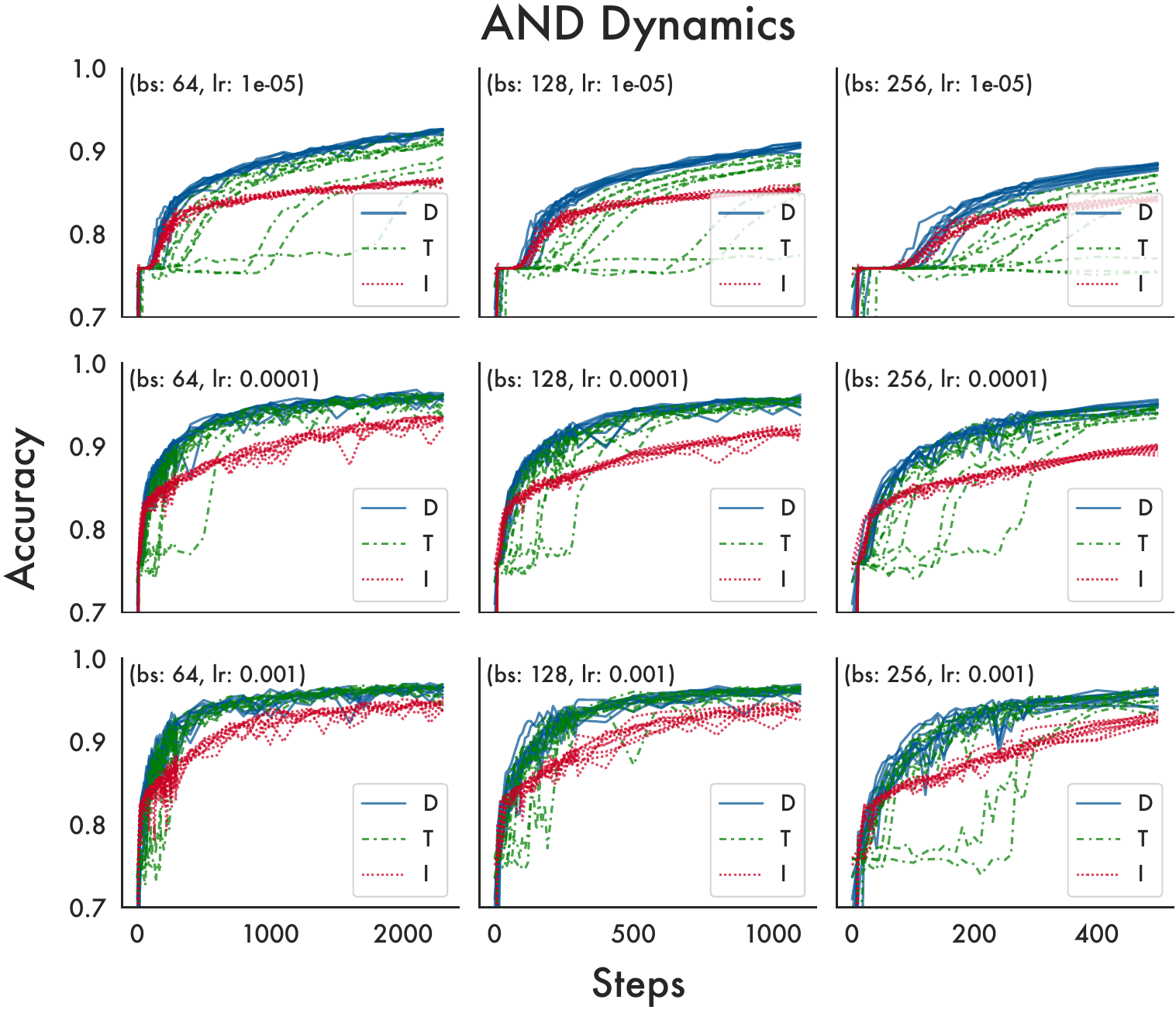}
  \centering
  \vspace{1.6em}    
  \caption{Summary of dynamics across configurations}
\end{figure}

\clearpage

\subsection{Conditional classification: OR}

\subsubsection{Task}

\begin{wrapfigure}{ht}{0.35\textwidth}
    \centering
    \vspace{-1.4em}
    \includegraphics[width=.3\textwidth]{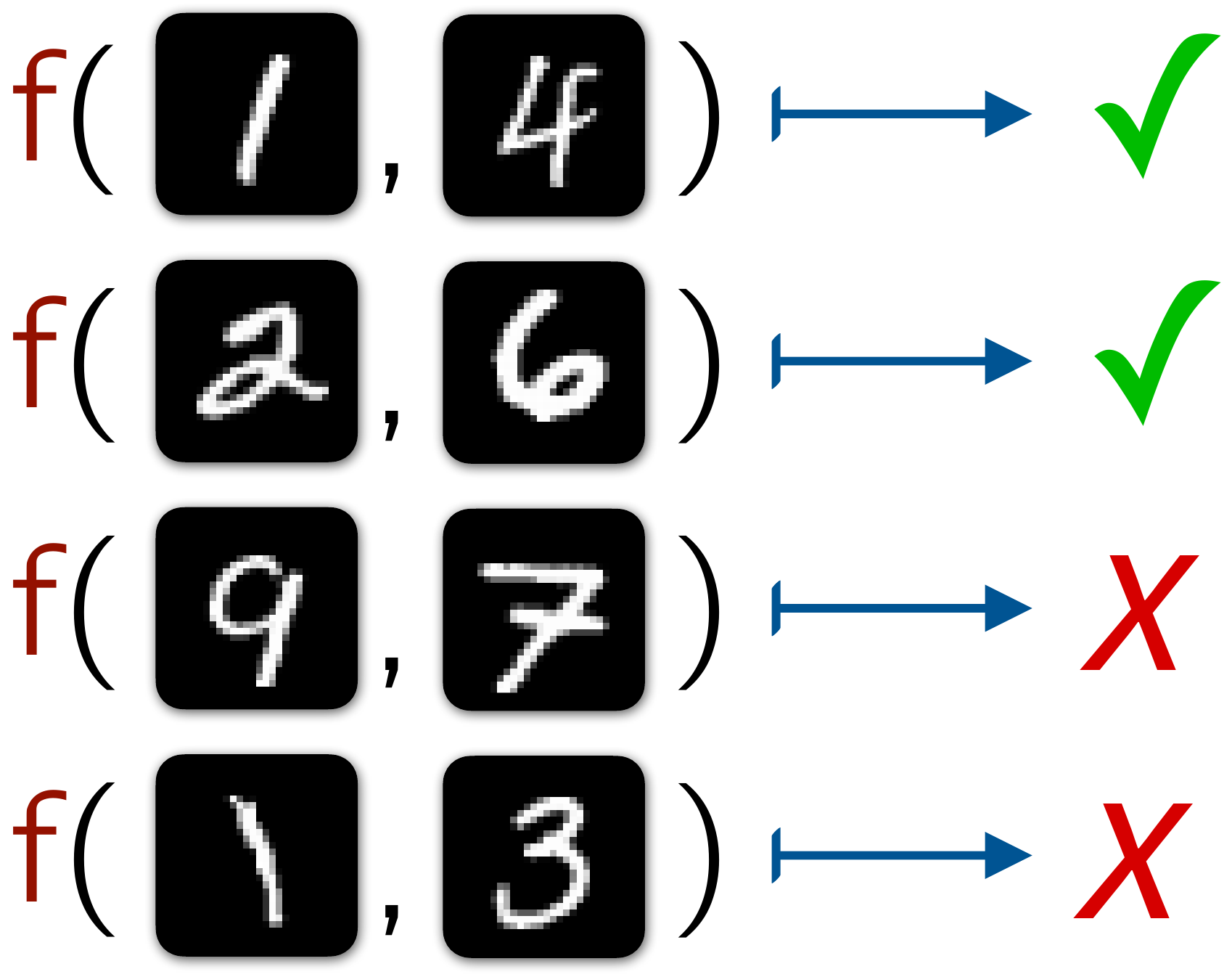}
    \vspace{.6em}
  \caption{OR Task}
  \label{fig:task2or}
  \vspace{-2.5em}
\end{wrapfigure}

This experiment studies how neural networks learn a binary classification involving \t{OR} logic, shown in Fig. \ref{fig:task2or}. The input is two images $x_1$ and $x_2$:
\setlist[itemize]{topsep=4pt, itemsep=1pt, parsep=2pt, leftmargin=1.5em, rightmargin=1.5em}
\begin{itemize}[label=$\triangleright$]
\item If $x_1$ is even or $x_2$ is odd, return \textcolor{ForestGreen}{\cmark}
\item Otherwise, return \textcolor{red}{\xmark}
\end{itemize}

\subsubsection{Models}
\t{Model D} is a neural network programmed using $\cj{\scriptstyle(\nb{\lto},\, \nb{\Bool},\, \nb{\mathbb{N}})}$, the \t{OR} structure is \textit{directly} programmed using our compiler. 
\vspace{1.2em}
\[
\cm{\ite{\ul{x}}{(\ite{\ul{y}}{\ul{\tt}}{\ul{\tt}})}{(\ite{\ul{y}}{\ul{\tt}}{\ul{\ff}})}}(\xvec, \yvec)
\]

\vspace{.4em}
\t{Model T} replaces this $\nb{2}$-linear map with random matrix $\rd{W}$, similar to our iterative image transform experiments. \t{Model I} is a simple 3-layer feedforward neural network.

\vspace{1em}
\begin{figure}[h]
  \includegraphics[scale=.45]{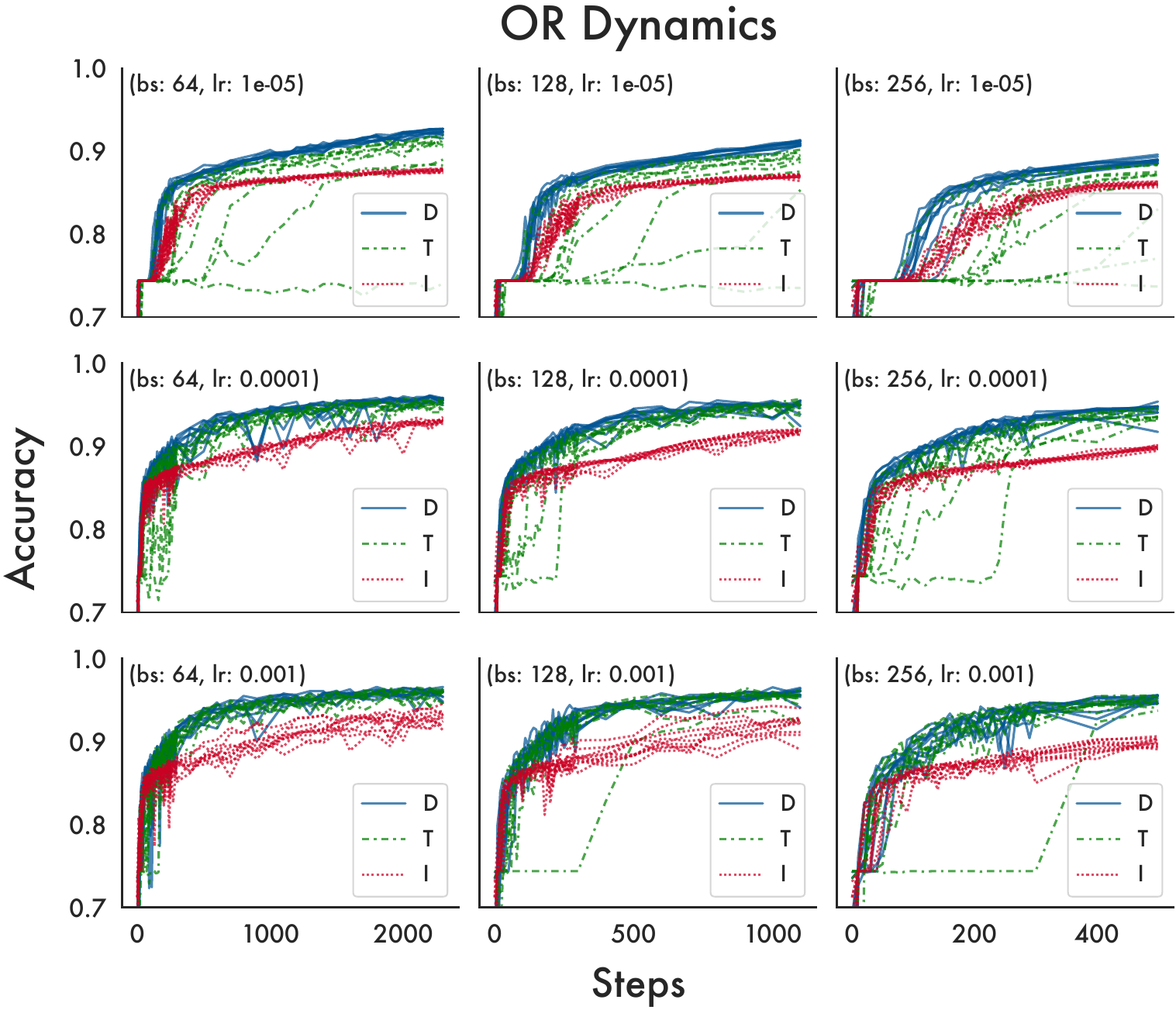}
  \centering
  \vspace{1.6em}    
  \caption{Summary of dynamics across configurations}
\end{figure}

\clearpage

\end{document}